%% file: AdS-diagrams.tex
\documentclass[12pt]{article}

\usepackage[left=2.51cm,right=2.51cm,top=2.54cm,bottom=2.54cm]{geometry}
\usepackage[english=british]{csquotes}
\usepackage{amsfonts}
\usepackage{graphicx}
\usepackage{amsmath,amssymb, amsfonts}
\usepackage{latexsym}
\usepackage{mathrsfs}
\usepackage{bbold}
\usepackage{color}
\usepackage{slashed}
\usepackage{cancel}
\usepackage{soul}
\usepackage{ulem} %
\usepackage{setspace} %
\usepackage{comment}
\usepackage{booktabs}
\usepackage{multirow}
\usepackage{tikz-feynman}
\tikzfeynmanset{compat=1.0.0}
\usepackage{cite}
\usepackage[small]{caption}
\usepackage{epsfig}
\usepackage{slashed}
\usepackage{epstopdf}
\usepackage{latexsym}
\usepackage{graphicx}
\usepackage{subcaption}
\usepackage{tikz}
\usepackage{rotating}
\usepackage{bbm}
\usepackage{bbold}
\usepackage{enumitem}
\usepackage{duckuments}
\def\checkmark{\tikz\fill[scale=0.4](0,.35) -- (.25,0) -- (1,.7) -- (.25,.15) -- cycle;}

\usepackage{empheq}
\usepackage[colorlinks,linkcolor=black,citecolor=blue,urlcolor=blue,linktocpage,pagebackref]{hyperref}

\numberwithin{equation}{section}
\begin{document}

\thispagestyle{empty}

\begin{center}

	\vspace*{-.6cm}

	\begin{center}

		\vspace*{1.1cm}

		{\hspace{-0.3cm}\centering \Large\textbf{Dressing and Screening in Anti-de Sitter}}

	\end{center}

	\vspace{0.8cm}
	{\hspace{-0.8cm}\bf Ankur$^{a,b,c}$, Lorenzo~Di~Pietro$^{a,b}$, Victor~Gorbenko$^{d}$, Shota~Komatsu$^{e}$,
	Veronica~Sacchi$^{d}$}

	\vspace{1.cm}
	
	${}^a\!\!$
	{ Dipartimento di Fisica, Universit\`a di Trieste, \\ Strada Costiera 11, I-34151 Trieste, Italy}
		
	\vspace{.3cm}

	${}^b\!\!$
	{ INFN, Sezione di Trieste, Via Valerio 2, I-34127 Trieste, Italy}

    \vspace{.3cm}

	${}^c\!\!$
	{Kavli Institute for Theoretical Sciences,
University of Chinese Academy of Sciences, Beijing, 100190, China}

	\vspace{.3cm}

	${}^d\!\!$
	{ Laboratory for Theoretical Fundamental Physics, Institute of Physics,
Ecole Polytechnique F\'{e}d\'{e}rale de Lausanne, Switzerland}

	\vspace{.3cm}

 ${}^e\!\!$
	{ CERN, Theoretical Physics Department, CH-1211 Geneva 23, Switzerland}

	\vspace{.3cm}

\end{center}

\vspace{1cm}

\centerline{\bf Abstract}
\vspace{2 mm}
\begin{quote}
Motivated by the question of defining gauge-invariant observables in  cosmology and by the close connection between perturbation theory in de Sitter (dS) and Anti–de Sitter (AdS), we study scalar electrodynamics in AdS in setups that are largely unexplored but relevant for dS physics. For photons with standard (Dirichlet) boundary conditions, we analyze charged scalars whose boundary conditions break the 
$U(1)$ symmetry. This leads to a nonstandard Higgs mechanism in which the gauge field acquires a one-loop mass without a classical vacuum expectation value. Using recent advances in perturbation theory in AdS, we compute this mass explicitly and evaluate charged-scalar four-point functions. We also provide an alternative derivation based on boundary Ward identities. For photons with alternate (Neumann) boundary conditions, where local charged operators are not gauge invariant, we construct physical observables by dressing charged fields with geodesic Wilson lines. These dressed operators have well-behaved conformal  properties and unphysical photon modes decouple from their correlation functions. Explicit one-loop computations further reveal the decoupling of the boundary field strength, for which we provide a nonperturbative argument based on higher-form symmetry. Along the way, we explain the physical consequences of spontaneous breaking of higher-form symmetry in AdS, including the role of the tilt operator, the relation between one-form symmetry and endpoints of Wilson lines at the boundary, and a generalized-symmetry interpretation of conserved currents dual to bulk gauge fields.
\end{quote}

\newpage

\tableofcontents

\input{newcommands}

\section{Introduction}
\label{sec:introduction}
\input{introduction.tex}


\section{Scalar QED Lagrangian and spin 1 AdS propagators}
\label{sec:propagators}
\input{propagators.tex}

\section{Dirichlet photon exchange}
\label{sec:dirichlet-photon-exchange}
\input{Dirichlet-tree-level.tex}

\section{Neumann photon exchange and dressing}
\label{sec:neumann-photon-exchange}
\input{Neumann-tree-level.tex}
\section{One loop Higgs mechanism, or screening}
\label{sec:1-loop-self-energy}
\input{1-loop-self-energy.tex}

\section{SSB and Higgs mechanism in AdS}
\label{sec:SecHiggs}
\input{SSB-and-Higgs-mech-in-AdS.tex}

\section{Implications of higher-form symmetry}
\label{sec:one-form-symmetries}
\input{1-form-symmetries.tex}

\section{Conclusions}
\label{sec:conclusions}
\input{conclusions.tex}

\subsection*{Acknowledgment}
We thank Jonáš Dujava and Manuel Loparco for discussions, collaboration on related topics and helpful comments on the manuscript. We also thank Giovanni Galati, Petr Kravchuk, Mehrdad Mirbabayi, Guilherme Pimentel, Fedor Popov, Massimo Porrati for discussions.
VG and VS are partially supported by  the SNF starting grant “The Fundamental Description of the Expanding Universe”. VG is partially supported by the Simons Foundation grant 994310
(Simons Collaboration on Confinement and QCD Strings). VS acknowledges support from the NCCR SwissMAP. The work of LD~is partially supported by INFN Iniziativa Specifica ST\&FI. LD~would like to thank the Isaac Newton Institute for Mathematical Sciences, Cambridge, for support and hospitality during the program ``Quantum field theory with boundaries, impurities, and defects", where work on this paper was undertaken. LD~and SK~thank the Yukawa Institute for Theoretical Physics and the organizers of the workshop ``Progress on Theoretical Bootstrap" for hospitality during the completion of this work. LD thanks the organizers of the workshop ``Constraining Effective Field Theories without Lorentz'' at IFPU, Trieste, for the opportunity to present this work. A would like to thank the Abdus Salam International Centre for Theoretical Physics (ICTP), Trieste, for the PhD grant and for hospitality during a post-PhD visit. 

\appendix

\section{Details for derivation of spin 1 AdS propagators}
\label{sec:propagators-details}
\input{propagators-details.tex}

\section{Field strength 2--point function}
\label{app:field-strength-2p-function}
\input{field-strength-2p-function.tex}


\section{Boundary limit of bulk to boundary propagators}
\label{app:boundary-limit-bulk-boundary-propagators}
\input{boundary-limit-bulk-to-boundary-prop.tex}

\section{Boundary expansion for parallel part of Dirichlet current exchange}
\label{app:parallel-current-boundary-expansion}
\input{parallel-current-bdy-expansion.tex}


\section{Dimreg of one-loop photon propagator}
\label{app:spin-1-simple-bubble}
\input{spin-1-simple-bubble.tex}


\input{tadpole.tex}




\bibliographystyle{utphys}
\bibliography{bibliography}

\end{document}

%% file: newcommands.tex
\definecolor{turquoise}{RGB}{0, 247, 230}
\definecolor{goldenyellow}{RGB}{255, 218, 66}
\definecolor{fuchsia}{RGB}{255, 0, 172}
\definecolor{petrolgreen}{RGB}{66, 144, 145}

\definecolor{blond}{RGB}{251,231,161}
\definecolor{skincolor}{RGB}{224,177,132}
\definecolor{idea}{RGB}{254,231,2} 	        

\newtheorem{theorem}{Theorem}[section]

\newtheorem{corollary}[theorem]{Corollary}

\newtheorem{lemma}[theorem]{Lemma}

\newtheorem{prop}[theorem]{Proposition}

\newcommand{\VS}[1]{\textcolor{fuchsia}{[VS: #1]}}
\newcommand{\LDP}[1]{{\color{blue}[LDP: #1]}} 
\newcommand{\SK}[1]{{\color{purple}[SK: #1]}} 
\newcommand{\VG}[1]{{\color{orange}[VG: #1]}}
\newcommand{\ANK}[1]{{\color{green}[ANK: #1]}}

\newcommand{\R}{{\mathbb{R}}}
\newcommand{\N}{{\mathbb{N}}}
\newcommand{\Z}{{\mathbb{Z}}}
\newcommand{\Tr}{\operatorname{Tr}}
\newcommand{\dd}{\mathrm{d}}
\newcommand{\sgn}{\operatorname{sgn}}
\newcommand{\ee}{\operatorname{e}}
\newcommand{\Res}{\operatorname{Res}}
\newcommand{\ResUnder}[1]{\mathop{\mathrm{Res}}\limits_{#1}}

\newcommand*\pFq[2]{{}_{#1}F_{#2}} 

\newcommand{\pprec}{\prec\mathrel{\mkern-5mu}\prec}

%% file: introduction.tex
The study of quantum field theories placed in a rigid AdS spacetime is interesting for multiple reasons. First, AdS space acts as a very symmetric IR regulator which makes the spectrum of the theory discrete and at the same time allows for a definition of asymptotic observables. These asymptotic observables, namely boundary correlation functions, are related to flat space asymptotic observables in the large-radius limit. Thus AdS correlation functions of a QFT give us a new tool to study its flat space S-matrix \cite{Paulos:2016fap,Komatsu:2020sag,vanRees:2022zmr}. If a QFT is strongly coupled in the infrared but weakly coupled or solvable in the UV, one can continuously connect strongly and weakly coupled phases by changing the AdS radius accordingly. This is particularly useful for understanding confining gauge theories, like QCD \cite{Callan:1989em,Aharony:2012jf,Ciccone:2024guw,Gabai:2025hwf}, as well as for the study of phase transitions \cite{Carmi:2018qzm,Copetti:2023sya}. Another motivation, of course, comes from holography, for example if there is a sector of a holographic theory in which the gravitational interactions are not important, or if one likes to study a bulk theory which is more general than those arising in known microscopic examples of AdS/CFT. Finally, AdS space is closely connected to its other maximally symmetric cousin -- dS space -- which is understood significantly worse. Direct calculations in dS space being  technically complicated, it is often convenient to perform a certain analytic continuation which allows us to compute dS observables in Euclidean AdS spacetime. The details of this analytic continuation depend on which dS observable one chooses to compute \cite{Maldacena:2002vr}. 

Of particular interest are correlation functions of operators located on a future asymptotic boundary of dS. At least in certain models these are similar to inflationary correlators directly measured in various cosmological experiments. In \cite{di2022analyticity}, building on earlier results of \cite{Sleight:2019hfp, Sleight:2020obc}, it was shown that for any theory of scalars in dS, at least perturbatively, one can associate a theory on EAdS with a doubled field content which has exactly the same boundary correlation functions. A doubled set of fields consists of fields of the same mass and spin, but different boundary conditions in AdS. For spinning fields the connection was established in \cite{Sleight:2020obc, Sleight:2021plv, Schaub:2023scu} and more recently for gauge fields in \cite{MdAbhishek:2025dhx,Sleight:2025dmt}.
 Ultimately, also in de Sitter space we are interested in gravitational theories. However, even a consistent definition of the asymptotic observables in the case of gauge theories is an unsolved problem. The basic obstacle is not even in non-perturbative effects, but it simply lies in definition of observables that are both well-defined, calculable and can be measured at least in Gedankenexperiments. See, for example, \cite{Chakraborty:2025izq} for a recent discussion.
 
 An analogous problem occurs also in abelian gauge theories in dS space. Since gauge fields and hence gauge transformation do not vanish on the future boundary, naively defined charged operators are not gauge-invariant. On the other hand, the perturbative intuition suggests that at least at small electric charge, it should be possible to define operators that approximate well local charged operator and also respect the de Sitter isometries. While the details of this construction are deferred to a future publication, it motivated us to study two phenomena specific to gauge theories in AdS. We believe that these phenomena possess independent interest and are important for understanding of the dynamics of QFT in AdS more generally.

The first phenomenon is associated with a photon with standard (Dirichlet) boundary conditions in AdS. As discussed in \cite{Porrati:2001db,Rattazzi:2009ux}, boundary conditions of matter fields can, and sometimes have to, break the symmetries that are gauged in the bulk of AdS. This leads to a somewhat unusual Higgs phenomenon, which gives mass to a gauge field at one loop, without a need for a classical vev for any of the charged fields. We study such Higgsing in detail focusing on scalar electrodynamics with scalar fields of generic mass. The phenomenon occurs when the boundary conditions for the scalars break the global $U(1)$ symmetry corresponding to the bulk gauge symmetry. This in turn leads to spontaneous symmetry breaking (SSB) in the bulk.  Recent advances in computational techniques for QFTs in AdS allow us to perform a very explicit calculation of the photon one-loop propagator, as well as of scalar four-point functions in which the photon is exchanged. We do so  by carefully renormalizing UV divergences using dimensional regularization and extract the finite piece of the photon mass in AdS$_4$ using a direct one-loop calculation. The physical effect of the mass is a faster decay of the electric field near the boundary. Being generated by a loop of the charged matter, we refer to this phenomenon as ``screening''.

The spontaneous breaking of a bulk gauge symmetry has an interesting interpretation in terms of boundary operators. Namely, at zero gauge coupling there are two protected operators: the vector current and the scalar tilt operator, the existence of which is equivalent to having spontaneous breaking. When the gauge coupling is turned on, the conservation of the current is broken by the tilt operator which leads to a multiplet recombination. A careful manipulation of Ward identities and of the bulk-to-boundary operator product expansion (bOPE) allows us to relate the anomalous dimension of the current to the VEV of the symmetry variation of the charged bulk operator, which serves as an order parameter for the symmetry.  This relation is completely general, and in our case it allows to calculate the photon mass much easier than and in agreement with the diagrammatic approach. This discussion, as well as the relevant references, are presented in section \ref{sec:SecHiggs}.

The second phenomenon is relevant for the photon with alternate (Neumann) boundary conditions. While this type of boundary conditions for the gauge fields is less studied -- see \cite{Compere:2008us,Giombi:2013yva} -- however, it has important applications, in particular, for the studies of confinement in non-Abelian theories, as recently discussed in \cite{Gabai:2025hwf}. With these boundary conditions correlation functions of charged boundary operators are not gauge-invariant, similarly to those in dS. In order to define operators that resemble as closely as possible the local ones, we implement a procedure that we refer to as ``dressing''. Namely, we attach geodesic Wilson lines to such charged operators. As we will see, again by a direct calculation, the correlation functions of such operators have some nice properties. For example, they transform under conformal transformations in the same way as correlation functions of local operators of corresponding dimension. At the same time, a potentially sick non-unitary mode of the photon decouples from our operators, as we see from the OPE expansion of the corresponding four-point function. While this mode is decoupled, the existence of a bulk gauge symmetry manifests itself in the existence of a protected boundary operator. As we explain in section \ref{sec:one-form-symmetries}, this operator is a conserved current and also a tilt, but now connected with the spontaneous breaking of the global higher-form magnetic bulk symmetry. We study the selection rules imposed by the current conservation and systematize the symmetries of the boundary theory for various boundary conditions and patterns of symmetry breaking in QED. In particular, we observe that even though an electric one-form symmetry is explicitly broken in the bulk, it is effectively restored near the boundary  even for massless matter. On the other hand, the magnetic symmetry explains the decoupling of the protected mode.

The discussion of sections \ref{sec:SecHiggs} and \ref{sec:one-form-symmetries} is general and can be read independently of more technical sections \ref{sec:dirichlet-photon-exchange},\ref{sec:neumann-photon-exchange},\ref{sec:1-loop-self-energy}.




%% file: propagators.tex
Throughout the document we will use a Lagrangian with a scalar sector slightly more general than the standard scalar QED. We will consider the usual unit-charge complex scalar \(\Phi = \frac{\varphi_{_1} + i\varphi_{_2}}{\sqrt{2}}\), coupled to a gauge field \(A_\mu\), with the (Euclidean) Lagrangian:
\begin{equation}
    \label{eq:full-lagrangian}
    \begin{aligned}
        \mathcal{L} &= \frac{1}{4}F_{\alpha\beta}F^{\alpha\beta} + \frac{1}{2\xi}\left(\nabla^{\alpha}A_{\alpha}\right)^2 \\
        & \quad + \nabla_{\alpha}\Phi\nabla^{\alpha}\Phi^* + m^2\Phi\Phi^* \\
        & \quad -ie A_{\alpha}(\Phi\nabla^{\alpha}\Phi^* - \Phi^*\nabla^{\alpha}\Phi) + e^2A_{\alpha}A^{\alpha}\Phi\Phi^* ~.
    \end{aligned}
\end{equation}
Let us also introduce parameters \(\nu_{1,2}\) to parametrize the boundary conditions of the real fields \(\varphi_1, \varphi_2\) in a slightly unusual way, motivated by the de Sitter conventions,
\begin{equation}
    \Delta_{1,2} = \frac{d}{2} + i\nu_{1,2},
\end{equation}
where unitarity requires $i\nu_{1,2}$ real and $\geq -1$, and 
\begin{equation}
    m^2 = \Delta_{1}(\Delta_{1} - d) = \Delta_{2}(\Delta_{2} - d) \implies \nu_1 = \pm \nu_2.
\end{equation}
The standard boundary condition is \(\nu_1 = \nu_2\),  i.e. the same boundary condition for both real scalar fields. Here we will also study the case where the two real degrees of freedom have opposite boundary conditions, and therefore break the $U(1)$ symmetry at the boundary.  As we will see, imposing such a boundary condition in AdS has various similarities with the spontaneous symmetry breaking of $U(1)$ symmetry in flat space.
 The main goal of this article is to understand the physics in AdS$_4$. However, in what follows, we present most of the results in generic boundary dimension $d$.\footnote{See \cite{Marolf:2006nd} for a discussion of the possible AdS boundary conditions for a free vector in general dimension. Note that the Neumann boundary condition for the photon always contains a boundary two-form operator $f_{\mu \nu}$ of dimension $2$. Such an operator saturates unitarity for $d=4$, meaning that it cannot couple to the bulk, and it violates unitarity for $d\geq 5$. Various subtle points arising for Neumann boundary conditions were discussed in \cite{Compere:2008us} for gravity  and also for higher spin gauge fields in \cite{Giombi:2013yva}.}

In the remainder of this section we review the propagator of a spin 1 gauge boson on EAdS\({}_{d + 1}\), in a generic \(\xi\) gauge, both with Dirichlet and Neumann boundary condition \cite{ankur2023scalar, Ciccone:2024guw}, employing the spectral representation for spinning AdS fields \cite{costa2014spinning}. We also discuss the physical meaning of the two conditions. The propagator in the Landau gauge $\xi = 1$ was derived in \cite{Allen:1985wd}. Other derivations of the gauge boson propagator that focus on the transverse part, or employ different types of gauge fixings can be found in \cite{DHoker:1998bqu, Liu:1998ty, DHoker:1999bve, Marotta:2024sce}. 

From the gauge-fixed (Euclidean) action
\begin{equation}
    S = \int_{AdS} \dd^{d + 1} x \sqrt{ g} \left(\frac{1}{4} F_{\mu \nu} F^{\mu \nu} + \frac{1}{2 \xi} \left( \nabla_\mu A^\mu \right)^2 \right).
    \label{eq:EAdS-action}
\end{equation}
we obtain the following equation for the propagator \(\Pi\) of the gauge field, using embedding coordinates in the conventions of \cite{costa2014spinning}
\begin{align}\label{eq:equation-of-motion-embedding}
\begin{split}
        \left[-\nabla_1^2 - d + \frac{1}{\frac{d - 1}{2}}\left(1 - \frac{1}{\xi}\right)(W_1\cdot\nabla_1)(K_1\cdot\nabla_1)\right] &\Pi(X_1, X_2; W_1, W_2)  \\
        & \quad = (W_1 \cdot W_2) \, \delta^{d + 1}(X_1, X_2)~,
\end{split}
\end{align}
where $\nabla_1$ denotes the covariant derivative with respect to $X_1$ and \(K\) is the differential operator to project polarization vectors \(W\), reported in equation \eqref{eq:K-index-projector-definition}.

In general, a spin 1 AdS two-point function admits two structures, which in embedding space can be chosen to be
\begin{equation}
    \label{eq:propagator-tensor-basis}
    \Pi(X_1, X_2; W_1, W_2) = F_0(u) (W_1\cdot W_2) + F_1(u) (X_1\cdot W_2)(X_2\cdot W_1)~.
\end{equation}
Hence, finding the propagator amounts to determining the two scalar functions $F_0$ and $F_1$ of the invariant
\begin{equation}
    u := \frac{(X_1 - X_2)^2}{2} = - (1 + X_1 \cdot X_2) ~.
\end{equation}

To fix completely the propagator we need to specify the boundary condition at the AdS boundary. Next, we consider both the Dirichlet and the Neumann boundary condition and give the corresponding expressions for the propagator.

\subsection{The Dirichlet propagator}
In appendix \ref{sec:propagators-details} we derive the propagators in spectral representation with the embedding formalism, and we find that a particular solution to the equation of motion is

\begin{equation}
    \label{eq:Dirichlet-propagator}
    \boxed{
    \begin{aligned}
        \Pi^{\mathcal{D}}_{d - 1}(X_1, X_2; W_1, W_2) &=
        \int_{-\infty}^{+\infty}\dd \lambda \, \frac{1}{\lambda^2 + \left(\frac{d}{2} - 1\right)^2} \, \Omega^{(1)}_{\lambda}(X_1, X_2; W_1, W_2) \\
        &\quad+\int_{-\infty}^{+\infty}\dd \lambda \, \frac{\xi}{\left(\lambda^2 + \frac{d^2}{4}\right)^2} \,  \left(W_1 \cdot \nabla_1\right)\left(W_2 \cdot \nabla_2\right) \Omega^{(0)}_{\lambda} (X_1, X_2).
    \end{aligned}
    }
\end{equation}
Notice that the transverse part is the same as the transverse part of a massive spinning propagator, whose mass squared in terms of its conformal dimension is \(M^2 = (\Delta - 1)(\Delta - d + 1)\).
Furthermore, the propagator may be written explicitly in a tensor basis as
\begin{equation}
    \label{eq:Dirichlet-propagator-tensor-basis}
    \Pi^{\mathcal{D}}_{d - 1}(X_1, X_2; W_1, W_2) = F_0^{\mathcal{D}}(u) (W_1\cdot W_2) + F_1^{\mathcal{D}}(u) (X_1\cdot W_2)(X_2\cdot W_1).
\end{equation}
Because of EAdS isometries, propagators will only be a function of the geodesic distance \(u\): the pull-back from the embedding space to AdS is
\begin{equation}
    \Pi^{\mathcal{D}}_{d - 1, \mu\nu} (x_1, x_2)
    = -F_0^{\mathcal{D}}(u) \frac{\partial^2 u}{\partial x_1^{\mu}\partial x_2^{\nu}} 
    + F_1^{\mathcal{D}}(u)\frac{\partial u}{\partial x_1^{\mu}}\frac{\partial u}{\partial x_2^{\nu}}.
\end{equation}
Again in appendix \ref{sec:propagators-details}, we derive the boundary limit of \(\Pi^{\mathcal{D}}_{d - 1, \mu\nu} (x_1, x_2)\) to be:
\begin{equation}
    \label{eq:Dirichlet-propagator-boundary-limit}
    \begin{aligned}
        F_0^{\mathcal{D}}(u) &\underset{u\to\infty}{\sim}
        \frac{\Gamma \left(\frac{d + 1}{2}\right)}
        {2\pi ^{\frac{d + 1}{2}}} \, \frac{1}{d - 2} \, \frac{1}{u^{d - 1}}, \\
        F_1^{\mathcal{D}}(u) &\underset{u\to\infty}{\sim} 
        \frac{1}{u} F_0^{\mathcal{D}}(u).
    \end{aligned}
\end{equation}
Therefore, \(\Pi^{\mathcal{D}}_{d - 1}(X_1, X_2; W_1, W_2)\) solves the equation of motion and correctly matches the Dirichlet boundary condition for any \(d \neq 2\). In EAdS\(_{2 + 1}\) the expression in \eqref{eq:Dirichlet-propagator-boundary-limit} is singular and this case needs to be studied separately. This is a manifestation of the incompatibility of this bc with AdS isometries for $d=2$ \cite{Marolf:2006nd, Faulkner:2012gt}.


\subsection{The Neumann propagator}\label{sec:NeuProp}
To derive the Neumann propagator, we first of all observe that 
\begin{equation}
    \boxed{
    \begin{aligned}
        \label{eq:homogeneous-solution}
        \int_{\lambda = \pm i\left(\frac{d}{2} - 1\right)}^{\substack{\circlearrowright \\ \circlearrowleft}}  &\dd\lambda \,\frac{1}{\lambda^2 + \left(\frac{d}{2} - 1\right)^2} \Omega_{\lambda}^{(1)}(X_1, X_2; W_1, W_2) \\
        & +\xi \int_{\lambda = \pm i\frac{d}{2}}^{\substack{\circlearrowright \\ \circlearrowleft}} \dd\lambda\,\frac{1}{\left(\lambda^2 + \frac{d^2}{4}\right)^2} \left(W_1 \cdot \nabla_1\right)\left(W_2 \cdot \nabla_2\right) \Omega_{\lambda}^{(0)}(X_1, X_2),
    \end{aligned}
    }
\end{equation}
is a homogeneous solution to the equation of motion \eqref{eq:equation-of-motion-embedding}.
This is verified explicitly in appendix \ref{sec:propagators-details}. 
The notation \(\int_{\lambda = \pm i \alpha}^{\substack{\circlearrowright \\ \circlearrowleft}}\) means that the integration contour computes (\(2 \pi i\) times) the anticlockwise residue around \(\lambda = - i \alpha\) and the clockwise residue around \(\lambda = + i \alpha\).

Moreover, in a tensor basis, the expression in \eqref{eq:homogeneous-solution} will be written as
\begin{equation}
    \eqref{eq:homogeneous-solution} = F_0^{\text{Hom}}(u) (W_1\cdot W_2) + F_1^{\text{Hom}}(u) (X_1\cdot W_2)(X_2\cdot W_1),
\end{equation}
and in appendix \ref{sec:propagators-details} we find that the boundary limit of \(F_0^{\text{Hom}}(u)\) is 
\begin{equation}
\label{eq:homogeneous-solution-boundary-limit}
\begin{aligned}
        F_0^{\text{Hom}}(u) &\underset{u\to\infty}{\sim}
        -\frac{\Gamma \left(\frac{d + 1}{2}\right)}
        {2\pi ^{\frac{d + 1}{2}}(d-2)}   \frac{1}{u^{d - 1}} -  \frac{1}{4 \pi^{\frac{d}{2}}\Gamma\left(2-\frac{d}{2}\right)}
        \left[\left(1 - \frac{d-2}{d}\xi\right)\log\left(\frac{u}{2}\right) + C \right]\frac{1}{u} , \\
        C := & -\frac{d-1}{d-2}+\left(1-\frac{d-2}{d}\xi\right)\left(\log(4)+\gamma_E + \psi^{(0)} \left(-\frac{d}{2}\right)-\frac{2}{d}\right)~.
    \end{aligned}
\end{equation}
Here $\gamma_E$ denotes Euler's constant and $\psi^{(0)}$ the digamma function. If this homogeneous solution is summed to the Dirichlet propagator, the first term cancels the \(\frac{1}{u^{d - 1}}\) fall-off series of the Dirichlet propagator, and the second term adds instead a \(\frac{1}{u} \, \log u \) decay. In Yennie gauge \(\xi = \frac{d}{d - 2}\) the logarithmic behaviours cancels leaving a $\frac{1}{u}$ asymptotics as in \cite{Ciccone:2024guw}. Moreover in this gauge the constant \(C\) simplifies to
\begin{equation}
C = -\frac{d - 1}{d-2}.
\end{equation} 
Similarly, the limit of \(F_1^{\text{Hom}}(u)\) is derived to be
\begin{align}
    \begin{split}
        F_1^{\text{Hom}}(u) \underset{u\to\infty}{\sim} \,
        &-\frac{\Gamma \left(\frac{d + 1}{2}\right)}
        {2\pi ^{\frac{d + 1}{2}} (d-2)} \,  \frac{1}{u^{d}} \\ 
        &- \frac{1}{4 \pi^{\frac{d}{2}}\Gamma\left(2-\frac{d}{2}\right)}
        \left[\left(1 - \frac{d-2}{d}\xi\right)\log\left(\frac{u}{2}\right) + C  + \frac{d - 2}{d}\xi\right]\frac{1}{u^2}~, 
    \end{split}
\end{align}
where \(C\) is the same as in \eqref{eq:homogeneous-solution-boundary-limit}. Therefore the sum of the Dirichlet propagator and the homogeneous solution \eqref{eq:homogeneous-solution} will solve the equation of motion and correctly match the Neumann boundary condition. Moreover, the coincident point limit will still be the same, since the homogeneous part is sub-leading at \(u\to 0\).

We can thus write the the Neumann propagator as
\begin{equation}
    \label{eq:Neumann-propagator}
    \boxed{
    \begin{aligned}
        \Pi^{\mathcal{N}}_{1}(X_1, X_2; W_1, W_2) &=
        \int_{\R \oplus \substack{\circlearrowright \\ \circlearrowleft}} \dd \lambda \, 
        \frac{1}{\lambda^2 + \left(\frac{d}{2} - 1\right)^2} \, \Omega^{(1)}_{\lambda}(X_1, X_2; W_1, W_2) \\
        &\quad+\int_{\R \oplus \substack{\circlearrowright \\ \circlearrowleft}}\dd \lambda \, 
        \frac{\xi}{\left(\lambda^2 + \frac{d^2}{4}\right)^2} \,  \left(W_1 \cdot \nabla_1\right)\left(W_2 \cdot \nabla_2\right) \Omega^{(0)}_{\lambda} (X_1, X_2)~,
    \end{aligned}
    }
\end{equation}
where the symbol below the integral represents the sum of the contour for the Dirichlet propagator and for the homogeneous solution.


\subsection{Boundary limit of the field strength 2--point function}
Combining the above results together with the equations derived in Appendix \ref{app:field-strength-2p-function}, we can  compute the bulk-to-boundary limit of the two-point function of the {\it electric} and {\it magnetic} fields for the gauge field with either boundary conditions. This will elucidate the physical meaning of the two boundary conditions for the gauge fields that we discussed above. 

Notice that we are working in Euclidean AdS, so by electric and magnetic fields we mean the corresponding components of the field strength tensor choosing as ``time'' the radial direction. We adopt Poincar\'e coordinates $z\geq 0$ and $\vec{x} \in \mathbb{R}^d$. The boundary is at \(z = 0\), and the AdS metric is
\begin{equation}\label{eq:Poincare}
ds^2 = \frac{1}{z^2}\left(\dd \vec{x}^2 + \dd z^2\right)~.    
\end{equation}
The radial direction is then parametrized by coordinate $z$. We use $i,j,k,\dots$ from $1$ to $d$ as indices in the $\mathbb{R}^d$ directions parallel to the boundary. With these conventions the 2-point functions of the electric and magnetic fields are
\begin{equation}
    \begin{split}
    \langle E_j(X_1)E_k(X_2) \rangle &= \langle F_{zj}(X_1)F_{zk}(X_2) \rangle ~,\\
    \langle B_{i_1\ldots i_{d - 2}}(X_1)B_{j_1\ldots j_{d - 2}}(X_2) \rangle 
    &= \varepsilon_{i_1\ldots i_{d - 2}}^{\phantom{i_1\ldots i_{d - 2}} ab}\,\varepsilon_{j_1\ldots j_{d - 2}}^{\phantom{j_1\ldots j_{d - 2}}cd} \,\langle F_{ab}(X_1)F_{cd}(X_2) \rangle~.
    \end{split}
\end{equation}
Here $\varepsilon_{i_1\ldots i_d}$ denotes the Levi-Civita tensor of the boundary $\mathbb{R}^d$ with flat metric $\dd \vec{x}^2$.
In this way we find that, when pushing \(X_1\) to the boundary, the field strength 2--point functions decay with the powers indicated in table \ref{table:F}.
\begin{table}[!ht]
    \centering
    \begin{tabular}{c || c | c}
        &  Dirichlet & Neumann \\
        \hline
        \hline
        & & \\
        \(\langle E_j(X_1)E_k(X_2) \rangle\) & \(\mathcal{O}\left(z_1^{d - 3}\right)\) & \(\mathcal{O}\left(z_1\right)\) \\
        & & \\
        \(\langle B_{i_1\ldots i_{d - 2}}(X_1)B_{j_1\ldots j_{d - 2}}(X_2) \rangle\) & \(\mathcal{O}\left(z_1^{d - 2}\right)\) & \(\mathcal{O}\left(z_1^0\right)\) \\
    \end{tabular}   
    \caption{Asymptotic behaviors near AdS boundary of electric and magnetic two-point functions, for the two choices of photon quantization.}
    \label{table:F}
\end{table}
In particular, we find that with Dirichlet boundary conditions 
\begin{equation}
     \langle E_j(X_1)E_k(X_2) \rangle \underset{z_1\to 0} {\gg} \langle B_j(X_1)B_k(X_2) \rangle,
\end{equation}
and so we identify the physical interpretation of this boundary condition as suppressing the magnetic mode at the boundary and keeping the electric one. This is consistent with the boundary limit of the gauge field 
\begin{equation}
	\begin{split}
		A_i \underset{z\to 0}{\sim} z^{d - 2}\,j_i \left(\vec{x}\right), \quad\quad
        A_z \underset{z\to 0}{\sim} \mathcal{O}(z^{d - 1}) \,,
	\end{split}
\end{equation}
as discussed in Appendix \ref{sec:propagators-details}.

For Neumann boundary conditions the 2--point functions have the opposite behavior:
\begin{equation}
     \langle E_j(X_1)E_k(X_2) \rangle \underset{z_1 \to 0}{\ll}  \langle B_j(X_1)B_k(X_2) \rangle \,.
\end{equation}
The magnetic mode is thus dominant in this case.

%% file: Dirichlet-tree-level.tex
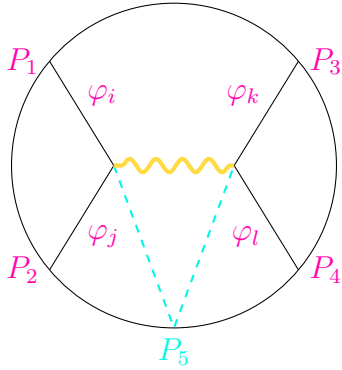
\begin{figure}[htbp]
    \centering
    \begin{tikzpicture}[scale = 0.8]
        \draw[black] (0,0) circle (2.7);
        \draw[black]  (140:2.7) -- (-1, 0) -- (220:2.7);
        \draw[black]  (40:2.7) -- (1, 0) -- (-40:2.7);
        \draw[turquoise, thick, dashed]  (-1, 0) -- (270:2.7) -- (1, 0);
        \draw[goldenyellow, ultra thick, decorate,decoration={snake, pre length=0.1cm}]  (-1, 0) -- (1, 0);
        \node[fuchsia, left] at (140:2.7) {\(P_1\)};
        \node[fuchsia, right] at (144:2) {\(\varphi_{i}\)};
        \node[fuchsia, left] at (220:2.7) {\(P_2\)};
        \node[fuchsia, right] at (216:2) {\(\varphi_{j}\)};
        \node[fuchsia, right] at (40:2.7) {\(P_3\)};
        \node[fuchsia, left] at (36:2) {\(\varphi_{k}\)};
        \node[fuchsia, right] at (-40:2.7) {\(P_4\)};
        \node[fuchsia, left] at (-36:2) {\(\varphi_{l}\)};
        \node[turquoise, below] at (270:2.7) {\(P_5\)};
    \end{tikzpicture}
    \caption{Witten diagram representing the  photon exchange. This diagram is gauge invariant for the exchange of a gauge field with Dirichlet boundary condition, but it is not for a Neumann exchange.}
    \label{fig:witten-diagram-current-photon-current}
\end{figure}

In this section we will calculate the exchange diagram of the photon with Dirichlet boundary conditions. This calculation has appeared previously with various different techniques in \cite{DHoker:1998bqu, DHoker:1999mqo, Fitzpatrick:2011ia, Fitzpatrick:2011hh}, here we follow more closely \cite{costa2014spinning, ankur2023scalar}, with the only modification that the interaction vertex is slightly more general, as we allow different boundary conditions for the two real matter fields. To be concrete, let us compute the contribution from the diagram in figure \ref{fig:witten-diagram-current-photon-current} to the connected four-point function at order \(e^2\): 
\begin{equation}
\langle\varphi_i(P_1)\varphi_j(P_2)\varphi_k(P_3)\varphi_l(P_4)\rangle_{\text{connected}} := 
\mathcal{A}_{\mathcal{D}}^{ijkl},
\end{equation}
which shall be antisymmetric under the exchange \(\varphi_i(P_1) \leftrightarrow \varphi_j(P_2)\) or \(\varphi_k(P_3) \leftrightarrow \varphi_l(P_4)\), and each field \(\varphi_i\) is understood to have conformal dimension \(\Delta_i\).  Here the indices $i,j,k,l=1,2$ run over the real components $\varphi_{1,2}$ of the fields, confusion with the $\mathbb{R}^d$ indices in Poincar\'e coordinates should not arise. We will work in a generic gauge and check explicitly how the gauge dependence disappears. This exercise is useful at least for the following reason: as we will see in section \ref{sec:neumann-photon-exchange} an equivalent diagram is not gauge invariant in the case of Neumann boundary conditions for the photon. To parametrize the interaction vertex it is convenient to define the following vertex structure 
\begin{equation}
    \label{eq:current-mixing-boundary-conditions}
    \begin{aligned}
        T_{ij}^A\left(P_1, P_2, X_m\right) = \Pi^{(0)}_{\frac{d}{2} + i\nu_i}\left(P_1, X_m\right) \nabla^A_m \Pi^{(0)}_{\frac{d}{2} + i\nu_j}\left(P_2, X_m\right) 
        - \begin{pmatrix}
            i \leftrightarrow j\\
            P_1 \leftrightarrow P_2
        \end{pmatrix},
    \end{aligned}
\end{equation}
where \(\nabla^A\) is the (geometric) covariant derivative with respect to embedding space coordinates.

In what follows we will calculate separately the contributions from the transverse and longitudinal components of the photon propagator.
Moreover, we will focus on the s-channel contribution to \(\mathcal{A}_{\mathcal{D}}^{ijkl}\): physically this can be motivated by promoting the the scalar field to also carry a flavour index. Then the s-channel exchange is the leading contribution in the large number of flavours \(N_f\) limit, once we project on the singlet sector of the flavour symmetry \(SU(N_f)\).


\subsection{Transverse piece}
\label{subsec:current-photon-current-diagram-transverse}

The transverse part of the photon exchange between two matter currents is
\begin{equation}
    \label{eq:TransverseContributionDefinition}
    \begin{aligned}
        \mathcal{A}^{ijkl, \perp}_{\mathcal{D}} = \frac{(-ie)^2 }{\left(\frac{d - 1}{2}\right)^2}
        \int_{AdS} &\dd^{d+1} X_1\int_{AdS} \dd^{d+1} X_2 \, K_{1,A} 
        \, T_{ij}^A\left(P_1, P_2, X_1\right)
        \, \\
        &\quad \Pi^{(1), \perp}_{d - 1} (X_1, X_2; W_1, K_2) \, W_{2,B}
        \, T_{kl}^B\left(P_3, P_4, X_2\right) \,,
    \end{aligned}
\end{equation}
where \(K_{i, A}\) is the projector operator \eqref{eq:K-index-projector-definition}.
\(\Pi^{(1), \perp}_{d - 1} (X_1, X_2; W_1, W_2)\) indicates the transverse part of the photon propagator with Dirichlet boundary conditions, and expressions are intended in the embedding-space formalism, with \(W_1, W_2\) the embedding polarization vectors.

For a Dirichlet photon, we have seen in \eqref{eq:Dirichlet-propagator} that the transverse part of the propagator is given by
\[
    W_1^A W_2^B \, \Pi^{(1), \perp}_{d-1, AB} (X_1, X_2) =
    \int_{\R} 
    \dd \lambda \, \frac{1}{\lambda^2 + \left(\frac{d}{2} - 1\right)^2}\, \Omega^{(1)}_{\lambda} (X_1, X_2; W_1, W_2) \,.
\]
Applying the split representation of the AdS harmonic function, we find:

\begin{equation}
    \label{eq:TransverseContributionSplitted}
    \begin{aligned}
        \mathcal{A}^{ijkl, \perp}_{\mathcal{D}} &= 
        \frac{1}{\pi \left(\frac{d}{2} - 1\right)}
        \frac{(-ie)^2 }{\left(\frac{d - 1}{2}\right)^2}
        \int_{\R}
        \dd \lambda \, 
        \frac{\lambda^2 \sqrt{
        \mathcal{C}_{\lambda}^{(1)}
        \mathcal{C}_{-\lambda}^{(1)}
        }}
        {\lambda^2 + \left(\frac{d}{2} - 1\right)^2} \int_{\partial AdS} \dd^{d} P_5 \,\\
        & \qquad \int_{AdS} \dd^{d+1} X_1 \,
        K_{1,A}
        \, T_{ij}^A\left(P_1, P_2, X_1\right)
         \, \Pi^{(1)}_{\frac{d}{2} + i\lambda}(X_1, P_5; W_1, D_Z) \\
        & \qquad \int_{AdS} \dd^{d+1} X_2 \, 
        \Pi^{(1)}_{\frac{d}{2} - i\lambda} (P_5, X_2; Z, K_2) \, W_{2,B}
        \, T_{kl}^B\left(P_3, P_4, X_2\right) \,.
    \end{aligned}
\end{equation}

Substituting the structures \(T_{ij}^C\left(P_n, P_m, X\right)\) as defined in \eqref{eq:current-mixing-boundary-conditions} in \eqref{eq:TransverseContributionSplitted} we can immediately recognize the two bulk integrals to be two three-point functions, since \((5.2)\) and \((5.7)\) of \cite{costa2014spinning} tells us that
\begin{equation}
    \label{eq:3-point-function}
    \begin{aligned}
        &\frac{1}{J!\left(\frac{d - 1}{2}\right)_J}\int_{AdS} \dd^{d+1} X_1 \, 
        \Pi_{\frac{d}{2} + i\nu_i}^{(0)}(P_1, X_1)\left(K_{1,A}\nabla_1^A\right)^J\Pi_{\frac{d}{2} + i\nu_j}^{(0)}\left(P_2, X_1\right) \, \Pi^{(J)}_{\frac{d}{2} + i\lambda}(X_1, P_5; W_1, D_Z) \\
        &= 
        \underbrace{\frac{\pi^\frac{d}{2}
        \Gamma\left(\frac{d}{4} + \frac{i\nu_i + i\nu_j + i\lambda + J}{2}\right)
        \Gamma\left(\frac{d}{4} + \frac{i\nu_i + i\nu_j - i\lambda + J}{2}\right)
        \Gamma\left(\frac{d}{4} + \frac{i\nu_i - i\nu_j + i\lambda + J}{2}\right)
        \Gamma\left(\frac{d}{4} + \frac{i\nu_j - i\nu_i + i\lambda + J}{2}\right)}
        {2^{1 - J}\Gamma\left(\frac{d}{2} + i\nu_i\right)\Gamma\left(\frac{d}{2} + i\nu_j\right)\Gamma\left(\frac{d}{2} + i\lambda + J\right)}}_{b_{\text{Bulk}}\left(\nu_i,\,  \nu_j,\, \lambda, \, J\right)}\\
        &\hspace{1.4cm}\,\sqrt{\mathcal{C}_{\nu_i}^{(0)}\mathcal{C}_{\nu_j}^{(0)}\mathcal{C}_{\lambda}^{(J)}}\,\langle \mathcal{O}_{\nu_i}(P_1) \mathcal{O}_{\nu_j}(P_2) \mathcal{O}_{\lambda}^{(J)}(P_5; D_Z)\rangle_1 \,,
    \end{aligned}
\end{equation}
    where we chose the normalization for bulk-to-boundary propagators to be consistent with the one adopted in \cite{di2022analyticity} (but different from the convention adopted in \cite{costa2014spinning}):
\begin{equation}
    \label{eq:bulk-to-boundary-propagator-normalization}
    \begin{aligned}
    \Pi_{\frac{d}{2} + i\nu}^{(J)}(X, P; W, Z) &= \sqrt{\mathcal{C}_{\nu}^{(J)}} \,
    \frac{((-2P\cdot X)(W\cdot Z) + 2(W\cdot P)(Z\cdot X))^J}
    {\left(-2P \cdot X\right)^{\frac{d}{2} + i\nu + J}} \\
    \mathcal{C}_{\nu}^{(J)} &= 
    \frac{\left(J + \frac{d}{2} + i\nu - 1\right)}{\left(\frac{d}{2} + i\nu - 1\right)}
    \frac{\Gamma\left(\frac{d}{2} + i\nu\right)}{2\pi^{\frac{d}{2}}\Gamma(1 + i\nu)},  
    \end{aligned}
\end{equation}
and we denote via \(\langle \mathcal{O}_i\mathcal{O}_j\mathcal{O}_k\rangle_1\) the unit-normalized structure of  three-point functions:
\[
\langle \mathcal{O}_{\nu_i}(P_1)\mathcal{O}_{\nu_j}(P_2)\mathcal{O}_{\lambda}^{(J)}(P_5, Z) \rangle_1 = 
\frac{\left[(Z\cdot P_1)P_{25} - (Z\cdot P_2)P_{15}\right]^J}
{P_{12}^{\frac{d}{4} + \frac{i\nu_i + i\nu_j - i\lambda + J}{2}}
P_{25}^{\frac{d}{4} + \frac{i\nu_j - i\nu_i + i\lambda + J}{2}}
P_{15}^{\frac{d}{4} + \frac{i\nu_i - i\nu_j + i\lambda + J}{2}}} \,.
\]
Now we have that the transverse amplitude becomes:
\begin{equation}
    \begin{aligned}
        \mathcal{A}^{ijkl, \perp}_{\mathcal{D}} = \frac{\left(-ie\right)^2}{\pi}
        &\frac{\sqrt{\mathcal{C}_{\nu_1}^{(0)}\mathcal{C}_{\nu_2}^{(0)}\mathcal{C}_{\nu_3}^{(0)}\mathcal{C}_{\nu_4}^{(0)}}}{\left(\frac{d}{2} - 1\right)}
        \int_{\R}
        \dd \lambda \, \frac{\lambda^2 \mathcal{C}_{\lambda}^{(1)}\mathcal{C}_{-\lambda}^{(1)}}{\lambda^2 + \left(\frac{d}{2} - 1\right)^2} 
        b_{\text{Bulk}}(\nu_i, \nu_j, \lambda, 1)b_{\text{Bulk}}(\nu_k, \nu_l, -\lambda, 1) \\
        \, \int_{\partial AdS} \dd^{d} P_5
        &\left[\left(
            \langle \mathcal{O}_{\nu_i}(P_1) \mathcal{O}_{\nu_j}(P_2) \mathcal{O}_{\lambda}^{(1)}(P_5; D_Z)\rangle_1 
            - \langle \mathcal{O}_{\nu_j}(P_2) \mathcal{O}_{\nu_i}(P_1) \mathcal{O}_{\lambda}^{(1)}(P_5; D_Z)\rangle_1 
            \right)\right. \,\\
        &\left. \left(
            \langle \mathcal{O}_{\nu_k}(P_3) \mathcal{O}_{\nu_l}(P_4) \mathcal{O}_{-\lambda}^{(1)}(P_5; Z)\rangle_1 
            - \langle \mathcal{O}_{\nu_l}(P_4) \mathcal{O}_{\nu_k}(P_3) \mathcal{O}_{-\lambda}^{(1)}(P_5; Z)\rangle_1 \right)\right].
    \end{aligned}
\end{equation}
Via the symmetry property of three-point functions with spin-1 operators
\[
    \langle \mathcal{O}_{\nu_i}(P_i) \mathcal{O}_{\nu_j}(P_j) \mathcal{O}_{\lambda}^{(1)}(P_5; D_Z)\rangle_1 = (-1)\langle \mathcal{O}_{\nu_j}(P_j) \mathcal{O}_{\nu_i}(P_i) \mathcal{O}_{\lambda}^{(1)}(P_5; D_Z)\rangle_1,
\]
we obtain:
\begin{equation}
    \label{eq:TransverseContribution3points}
    \begin{aligned}
        \mathcal{A}^{ijkl, \perp}_{\mathcal{D}} &= \frac{4(-ie)^2}{\pi}
        \frac{\sqrt{\mathcal{C}_{\nu_i}^{(0)}\mathcal{C}_{\nu_j}^{(0)}\mathcal{C}_{\nu_k}^{(0)}\mathcal{C}_{\nu_l}^{(0)}}}{\left(\frac{d}{2} - 1\right)}
        \int_{\R}
        \dd \lambda \, \frac{\lambda^2\mathcal{C}_{\lambda}^{(1)}\mathcal{C}_{-\lambda}^{(1)}}{\lambda^2 + \left(\frac{d}{2} - 1\right)^2} 
        b_{\text{Bulk}}(\nu_i, \nu_j, \lambda, 1)b_{\text{Bulk}}(\nu_k, \nu_l, -\lambda, 1) \\
        & \qquad\int_{\partial AdS} \dd^{d} P_5 
        \Big\langle \mathcal{O}_{\nu_i}(P_1) \mathcal{O}_{\nu_j}(P_2) \mathcal{O}_{\lambda}^{(1)}(P_5; D_Z)\Big\rangle _1 
        \, \Big\langle \mathcal{O}_{\nu_k}(P_3) \mathcal{O}_{\nu_l}(P_4) \mathcal{O}_{-\lambda}^{(1)}(P_5; Z)\Big\rangle _1.
    \end{aligned}
\end{equation}

Eventually, we can apply formula \((3.1)\) of \cite{simmons2018spacetime}:
\begin{equation}
    \label{eq:3p-function-to-partial-waves}
    \begin{aligned}
       & \Psi_{\lambda, J}^{\{\Delta_i\}} (\{P_i\})
        :=
        \frac{1}{P_{12}^{\frac{\Delta_i + \Delta_j}{2}} \, P_{34}^{\frac{\Delta_k + \Delta_l}{2}}} 
        \left(\frac{P_{24}}{P_{14}}\right)^{\frac{\Delta_{ij}}{2}}
        \left(\frac{P_{14}}{P_{13}}\right)^{\frac{\Delta_{kl}}{2}}\, \mathcal{F}_{\lambda, J}^{\{\Delta_i\}}(u,v) \\
     & = \frac{1}{\left(\frac{d}{2} - 1\right)_J}\int_{\partial AdS} \dd^d P_{5} \,
        \langle \mathcal{O}_{\nu_i}(P_1)\mathcal{O}_{\nu_j}(P_2)\mathcal{O}_{\frac{d}{2} + i\lambda}^{(J)}(P_5, D_Z)\rangle_1
        \langle \mathcal{O}_{\frac{d}{2} - i\lambda}^{(J)}(P_5, Z)\mathcal{O}_{\nu_k}(P_3)\mathcal{O}_{\nu_l}(P_4)\rangle_1 \mspace{-17mu}
    \end{aligned}
\end{equation}
to recover the expansion in conformal partial waves  
\begin{equation}
    \label{eq:TransverseContributionPartialWaves}
    \begin{aligned}
        \mathcal{A}^{ijkl, \perp}_{\mathcal{D}} &= 
        \frac{(-ie)^2}{\pi}
        \sqrt{\prod_{a \in \{ijkl\}} \mathcal{C}_{\nu_a}^{(0)}}
        \, \left(\frac{1}{P_{12}}\right)^{\frac{d}{2} + i\frac{\nu_i + \nu_j}{2}}
        \left(\frac{1}{P_{34}}\right)^{\frac{d}{2} + i\frac{\nu_k + \nu_l}{2}}
        \left(\frac{P_{24}}{P_{14}}\right)^{i\frac{\nu_i - \nu_j}{2}}
        \,
        \left(\frac{P_{14}}{P_{13}}\right)^{i\frac{\nu_k - \nu_l}{2}} \hspace{-0.9em}\\
        & \quad\int_{\R} \dd\lambda \, 
        \frac{4 \lambda^2}{\lambda^2 + \left(\frac{d}{2} - 1\right)^2} \mathcal{C}_{\lambda}^{(1)}\mathcal{C}_{-\lambda}^{(1)}\,
        b_{\text{Bulk}}(\nu_i, \nu_j, \lambda, 1)b_{\text{Bulk}}(\nu_k, \nu_l, -\lambda, 1)
        \, \mathcal{F}_{\lambda}^{(1), \{\Delta_i\}}(u, v) \,, \hspace{-0.9em}
    \end{aligned}
\end{equation}
where, as usual
\[
    P_{ij} = -2P_i\cdot P_j \,, \quad\quad u = \frac{P_{12}P_{34}}{P_{13}P_{24}} \,, \quad\quad v = \frac{P_{14}P_{23}}{P_{13}P_{24}} \,,
\]
and we chose to normalize conformal partial waves as in \cite{di2022analyticity, simmons2018spacetime} (which is different from the convention adopted in \cite{costa2014spinning}), so that

\begin{equation}
    \label{eq:conformal-partial-waves-2-conformal-blocks}
    \mathcal{F}_{\lambda, J}^{\{\Delta_i\}}(u,v) = 
    \mathcal{K}_{d-\Delta_{\lambda}}^{(J), \{\Delta_k, \Delta_l\}}
    \hat{\mathcal{G}}_{\Delta_{\lambda}}^{(J)}(u,v) + 
    \mathcal{K}_{\Delta_{\lambda}}^{(J), \{\Delta_i, \Delta_j\}}\hat{\mathcal{G}}_{d - \Delta_{\lambda}}^{(J)}(u,v) \,,
\end{equation}
with 
\begin{equation}
   \mathcal{K}_{\Delta_{\lambda}}^{(J), \{\Delta_1, \Delta_2\}} = \frac{\pi^{\frac{d}{2}}}{(-2)^J}
    \frac{
        \Gamma\left(\Delta_{\lambda} - \frac{d}{2}\right)
        \Gamma\left(\Delta_{\lambda} + J - 1\right)
        \Gamma\left(\frac{d - \Delta_{\lambda} + \Delta_1 - \Delta_2 + J}{2} \right)
        \Gamma\left(\frac{d - \Delta_{\lambda} + \Delta_2 - \Delta_1 + J}{2} \right)
        }
        {
            \Gamma\left(\Delta_{\lambda} - 1\right)
            \Gamma\left(d - \Delta_{\lambda} + J\right)
            \Gamma\left(\frac{\Delta_{\lambda} + \Delta_1 - \Delta_2 + J}{2} \right)
            \Gamma\left(\frac{\Delta_{\lambda} + \Delta_2 - \Delta_1 + J}{2} \right)
        } \,.
\end{equation}
This normalization is particularly convenient because both \(\mathcal{K}_{d-\Delta_{\lambda}, \{\Delta_i\}}^{(J)}\) and \(\hat{\mathcal{G}}_{\Delta_{\lambda}}^{(J)}(u,v)\) only depend on \(\Delta_{ij}\equiv\Delta_i - \Delta_j\) and \(\Delta_{kl}\equiv\Delta_k - \Delta_l\) from the external legs, so the full conformal partial wave \(\mathcal{F}_{\lambda, J}^{\{\Delta_i\}}(u,v)\) itself only depends on \(\Delta_{jk}\) and \(\Delta_{kl}\). 

For \(\nu_i = \nu_j = \nu_k = \nu_l = \nu\), \eqref{eq:TransverseContributionPartialWaves} reduces to
    \begin{equation}
        \label{eq:Dirichlet-current-photon-current-transverse}
        \begin{aligned}
        \mathcal{A}^{ijkl, \perp}_{\mathcal{D}} &= 
        (-ie)^2 \left(\frac{1}{P_{12}}\, \frac{1}{P_{34}}\right)^{\frac{d}{2} + i\nu} 
        \, \int_{\R}
        \dd \lambda \, \rho_{\mathcal{D}}^{J = 1}(\lambda; \{\tfrac{d}{2} + i\nu\})  \, \mathcal{F}_{\lambda, 1}^{\{\frac{d}{2} + i\nu\}}(u,v) \,, 
        \end{aligned}
    \end{equation}
    where $\{\tfrac{d}{2} + i\nu\}$ denotes the four equal external scaling dimensions $\Delta_i=\frac{d}{2} + i\nu$ and
    \begin{equation}
    \label{eq:dirichlet-spectral-density-tree-level}
    \begin{split}
        \rho_{\mathcal{D}}^{J = 1}(\lambda; \{\tfrac{d}{2} + i\nu\}) &= \Pi^{(1), \perp}_{d - 1}(\lambda) \, \mathcal{Q}^{J = 1}(\lambda, \{\tfrac{d}{2} + i\nu\}) \,, \\
        \Pi^{(1), \perp}_{d - 1} &= \frac{1}{\lambda^2 + \left(\tfrac{d}{2} - 1\right)^2} \,, \\
        \mathcal{Q}^{J = 1}(\lambda, \{\tfrac{d}{2} + i\nu\}) &=
        \frac{\lambda \sinh(\pi \lambda)}{4\pi^{d + 2}} 
        \frac{1}{\lambda^2 + \left(\frac{d}{2} - 1\right)^2}
        \frac{\Gamma\left(\frac{d + 2 \pm 2i\lambda}{4}\right)^2 \Gamma\left(\frac{d + 2 + 4i\nu \pm 2i\lambda}{4}\right)^2}
    {\Gamma (1+i \nu )^2 \Gamma \left(\frac{d}{2}+i \nu \right)^2} \,.
    \end{split}
    \end{equation}
    The spectral density \(\rho_{\mathcal{D}}^{J = 1}(\lambda; \{\tfrac{d}{2} + i\nu\})\) is the product of the factor \(\Pi^{(1), \perp}_{d - 1}(\lambda)\) coming from the spectral density of the photon propagator, and a kinematic factor \(\mathcal{Q}^{J = 1}\). It decays as \(\ee^{-\pi\lambda}\) for large \(\lambda\). 
To find the OPE decomposition, we write the conformal partial wave in terms of conformal blocks using \eqref{eq:conformal-partial-waves-2-conformal-blocks}. The density becomes
\begin{equation}
    \label{eq:Dirichelet-tree-level-OPE}
    \begin{split}
    & \rho_{\mathcal{D}}^{J = 1}(\lambda; \{\tfrac{d}{2} + i\nu\}) \mathcal{K}_{d-\Delta_{\lambda}}^{(J = 1)}  \\
    & \qquad =-\frac{\lambda \sinh (\pi  \lambda )}{2^{5 +\frac{d}{2}+i \lambda}\pi ^{\frac{d + 3}{2}}}
    \frac{ \Gamma \left(\frac{d - 2 + 2 i \lambda}{4}\right)^2}{\Gamma (1+i \nu )^2 \Gamma \left(\frac{d}{2}+i \nu \right)^2}
    \frac{\Gamma (-i \lambda )\Gamma \left(\frac{d + 2 +2 i \lambda}{4}\right) \Gamma \left(\frac{d + 2 +4 i \nu \pm 2 i \lambda}{4}\right)^2}{\left(\frac{d}{2} - 1 - i\lambda\right) \Gamma \left(\frac{d+ 4 + 2 i \lambda}{4}\right)} \,.
    \end{split}
\end{equation}
Note that, at the location of the massless photon pole from the propagator $\Pi^{(1), \perp}_{d - 1}$, i.e. \(\lambda = - i \left(\frac{d}{2} - 1\right)\), there is also a pole of the kinematical factor $\mathcal{Q}$, which however is canceled by a zero of $\mathcal{K}$ and does not contribute to the expansion in conformal blocks. We then end up with a single pole as expected
\begin{equation}
    \rho_{\mathcal{D}}^{J = 1}(\lambda, \{\tfrac{d}{2} + i\nu\})\mathcal{K}_{d-\Delta_{\lambda}}^{(J = 1)} \underset{\lambda \to - i \left(\frac{d}{2} - 1\right)}{\sim}
    -\frac{\Gamma\left(\frac{d}{2}-1\right)^2}{2^{d+5}\pi^{\frac{d + 1}{2}}\Gamma \left(\frac{d+1}{2}\right)}
    \, \frac{i (d-2)}{\lambda +i \left(\frac{d}{2} - 1\right)}
    + \mathcal{O}\left(1\right),
\end{equation}
corresponding to the exchange of an operator with \(\Delta_{\lambda} = d - 1\), the conserved current.

For symmetry-breaking boundary conditions, i.e. \(\Delta_i = \Delta_k = \frac{d}{2}+ i\nu\) and \(\Delta_j = \Delta_l = \frac{d}{2}-i\nu\), shortened as $\{\frac{d}{2} \pm i\nu\}$, the 4-point function evaluates to
  \begin{equation}
        \label{eq:Dirichlet-current-photon-current-transverse-opposite}
        \begin{aligned}
        \mathcal{A}^{ijkl, \perp}_{\mathcal{D}} &= 
        (-ie)^2 \left(\frac{1}{P_{12}}\, \frac{1}{P_{34}}\right)^{\frac{d}{2}} 
        \, \left(\frac{P_{24}}{P_{14}}\, \frac{P_{14}}{P_{13}} \right)^{i\nu} 
        \, \int_{\R}
        \dd \lambda \, \rho_{\mathcal{D}}^{J = 1}(\lambda; \{\tfrac{d}{2} \pm i\nu\})  \, \mathcal{F}_{\lambda, 1}^{\{\frac{d}{2} \pm i\nu\}}(u,v) \,, 
        \end{aligned}
    \end{equation}
    where
     \begin{equation}
    \label{eq:dirichlet-spectral-density-tree-level-opposite}
    \begin{split}
        \rho_{\mathcal{D}}^{J = 1}(\lambda; \{\tfrac{d}{2} \pm i\nu\}) &= \Pi^{(1), \perp}_{d - 1}(\lambda) \, \mathcal{Q}^{J = 1}(\lambda, \{\tfrac{d}{2} \pm i\nu\}) \,, \\
        \Pi^{(1), \perp}_{d - 1} &= \frac{1}{\lambda^2 + \left(\frac{d}{2} - 1\right)^2} \,, \\
        \mathcal{Q}^{J = 1}(\lambda, \{\tfrac{d}{2} \pm i\nu\}) &=
        \frac{\lambda \sinh(\pi \lambda)}{4\pi^{d + 3}}
        \frac{1}{\lambda^2 + \left(\frac{d}{2} - 1\right)^2}
        \frac{\sinh(\pi\nu)}{\nu}\frac{\Gamma\left(\frac{d + 2 \pm 2i\lambda}{4}\right)^2 \Gamma\left(\frac{d + 2 \pm 4i\nu \pm 2i\lambda}{4}\right)}{\Gamma \left(\frac{d}{2}\pm i \nu \right)}~.
    \end{split}
    \end{equation}
Like before, the density whose poles determine the expansion in conformal blocks is obtained adding the coefficient $\mathcal{K}$. Similarly to the case of symmetry preserving bc, also with the symmmetry breaking bc there is a cancellation between a pole of $\mathcal{Q}$ and a zero of $\mathcal{K}$, and we are left with a single pole due to the photon propagator at the location of the conserved current 
\begin{align}
    \label{eq:Dirichelet-tree-level-OPE-opposite}
    \begin{split}
    & \rho_{\mathcal{D}}^{J = 1}(\lambda; \{\tfrac{d}{2} \pm i\nu\}) \mathcal{K}_{d-\Delta_{\lambda}}^{(J = 1)}  \\
    & \mspace{30mu} =-\frac{\lambda \sinh (\pi  \lambda )}{2^{5}\pi ^{\frac{d}{2} + 3}}
    \frac{\sinh\left(\pi\nu\right)}{\nu}
    \frac{\Gamma \left(\frac{d - 2 + 2 i \lambda}{4}\right)^2}{ \Gamma \left(\frac{d}{2}\pm i \nu \right)}
    \frac{\Gamma (-i \lambda )\Gamma \left(\frac{d + 2 - 2 i \lambda}{4}\right)^2 
    \Gamma\left(\frac{d + 2 \pm 4 i \nu + 2 i \lambda}{4}\right)^2}{\left(\frac{d}{2} - 1 - i\lambda\right)  \Gamma \left(\frac{d}{2} +  i \lambda + 1\right)} \\
    & \underset{\lambda \to - i \left(\frac{d}{2} - 1\right)}{\sim}
    -\frac{\Gamma\left(\frac{d}{2}-1\right) \sinh (\pi  \nu ) \Gamma \left(\frac{d}{2}\pm i \nu \right)}
    {2^5 \pi^{\frac{d + 1}{2}}\nu  \Gamma (d)} 
    \, \frac{i}{\lambda +i \left(\frac{d}{2} - 1\right)}
    + \mathcal{O}\left(1\right).
\end{split}    
\end{align}

\subsection{Longitudinal piece}
\label{subsec:longitudinal-dirichlet-photon-exchange}
We will now show here that the longitudinal (gauge-dependent) contibution vanishes. First of all, we can show that the longitudinal piece always reduces to a boundary term. Its contribution to the Witten diagram represented in figure \ref{fig:witten-diagram-current-photon-current} is
\begin{equation}
    \label{eq:LongitudinalContributionDefinition}
    \begin{split}
        \mathcal{A}^{ijkl, \parallel}_{\mathcal{D}} = (-ie)^2 \int_{AdS} \dd^{d+1} X_1 \, \dd^{d+1} X_2 \,
        T_{ij}^A\left(P_1, P_2, X_1\right)
        \, \Pi^{(1), \parallel}_{d - 1, AB} (X_1, X_2)
        \, T_{kl}^B\left(P_3, P_4, X_2\right),
    \end{split}
\end{equation}
where, from \eqref{eq:Dirichlet-propagator}, we know
\[
    \Pi^{(1), \parallel}_{d - 1, AB} (X_1, X_2) = \int_{-\infty}^{+\infty}\dd \lambda \, \frac{\xi}{\left(\lambda^2 + \frac{d^2}{4}\right)^2}\, \nabla_{1,A}\nabla_{2,B} \, \Omega^{(0)}_{\lambda} (X_1, X_2),
\]
and \(T_{ij}^A\) is the  vertex structure, as defined in \eqref{eq:current-mixing-boundary-conditions}.

Again a convenient way to perform the integral in \eqref{eq:LongitudinalContributionDefinition} is to write the AdS harmonic function in terms of scalar bulk--to--boundary propagators with the split representation:
\begin{equation}
    \label{eq:dirchlet-longitudinal-4-point}
    \begin{aligned}
        \mathcal{A}^{ijkl, \parallel}_{\mathcal{D}}  = (-ie)^2\frac{\xi}{\pi} &\int_{-\infty}^{+\infty} \dd\lambda \, \frac{\lambda^2 \sqrt{\mathcal{C}_{\lambda}^{(0)}\,\mathcal{C}_{-\lambda}^{(0)}}}{\left(\lambda^2 + \frac{d^2}{4}\right)^2} 
        \int_{\partial AdS} \dd^d P_5 \\
        & \quad \int_{AdS}\dd^{d + 1} X_1 \, T_{ij}^A\left(P_1, P_2, X_1\right)
        \nabla_{1,A}\Pi^{(0)}_{\frac{d}{2} + i\lambda}(X_1, P_5) \\
        \, 
        & \quad \int_{AdS}\dd^{d + 1} X_2 \,
        T_{kl}^B\left(P_3, P_4, X_2\right) 
        \nabla_{2,B}\Pi^{(0)}_{\frac{d}{2} - i\lambda}(X_2, P_5) \,,
    \end{aligned}
\end{equation}
where the constants \(\mathcal{C}_{\pm \lambda}^{(0)}\) are 
\begin{equation}
    \label{eq:bulk-to-boundary-normalization}
    \mathcal{C}_{\lambda}^{(J)} = 
    \frac{\left(J + \frac{d}{2} + i\lambda - 1\right)}{\left(\frac{d}{2} + i\lambda - 1\right)}
    \frac{\Gamma\left(\frac{d}{2} + i\lambda\right)}{2\pi^{\frac{d}{2}}\Gamma(1 + i\lambda)} \,. 
\end{equation}
 The conservation of the matter current implies \(\nabla \cdot T_{ij} = 0\),\footnote{We are adopting the prescription of having the four matter fields inserted exactly at the boundary points \(P_i\), while interactions are switched on in the bulk up to a cut-off surface placed at some small distance \(z \rightarrow 0\) from the boundary so that we don't have to worry of potential contact terms in \(\nabla \cdot T_{ij} = 0\).} 
so integration by parts leaves only a boundary term 
\begin{equation}
    \label{eq:LongitudinalContributionBoundary}
    \begin{split}
        \mathcal{A}^{ijkl, \parallel}_{\mathcal{D}}  = (-ie)^2\frac{\xi}{\pi} 
        &\int_{-\infty}^{+\infty} \dd\lambda \,
        \frac{\lambda^2 \sqrt{\mathcal{C}_{\lambda}^{(0)}\mathcal{C}_{-\lambda}^{(0)}}}{\left(\lambda^2 + \frac{d^2}{4}\right)^2} \int_{\partial AdS} \dd^d P_5 
        \, \int\frac{\dd^d \vec{x}_{\tilde{1}}}{z_1^{d + 1}} \, \int\frac{\dd^d \vec{x}_{\tilde{2}}}{z_2^{d + 1}} \\
        & \qquad T_{ij}^z\left(P_1, P_2, X_1\right) 
        \Pi^{(0)}_{\frac{d}{2} + i\lambda}(X_1, P_5)
        \, T_{kl}^z\left(P_3, P_4, X_2\right)
         \Pi^{(0)}_{\frac{d}{2} - i\lambda}(X_2, P_5)~,
    \end{split}
\end{equation}
where, using Poincar\'e coordinates \eqref{eq:Poincare}, we plug \(X_a = \left(z_a, \vec{x}_{\tilde{a}}\right)\), $a=1,2$, with \(\vec{x}_{\tilde{a}}\) points on the regulating cut-off surface at distance \(z_1,z_2 \rightarrow 0\) from the AdS boundary, and the \(z\) component of \(T_{ij}^z\) stands for the component orthogonal to AdS boundary. The bulk--to--boundary propagator in Poincaré coordinates is (with the normalization of \cite{di2022analyticity})
\begin{equation}
\Pi^{(0)}_{\frac{d}{2} + i\nu}(X_1, P_2) = \sqrt{\mathcal{C}_{\nu}^{(0)}} \left(\frac{z_1}{z_1^2 + \vec{x}_{12}^2}\right)^{\frac{d}{2}+ i\nu} 
\,,\quad 
\mathcal{C}_{\nu}^{(0)} = \frac{\Gamma\left(\frac{d}{2}+ i\nu\right)}{2\pi^{\frac{d}{2}}\Gamma(1 + i\nu)}~. 
\end{equation}
Now we can take the boundary limits $z_a\to 0$. They need to be taken carefully, accounting for all the contact terms.
For that, it is useful to remind ourselves of the following identity 
(see (2.39) of \cite{witten1998anti} and appendix \ref{app:boundary-limit-bulk-boundary-propagators}). If \(\vec{x}\) is a \(d\)-dimensional vector, then:
\begin{equation}
    \lim_{z\rightarrow 0} \frac{z^{d + 2\alpha}}{(z^2 + \vec{x}^2)^{d + \alpha}} = \pi^{\frac{d}{2}} \frac{\Gamma\left(\frac{d}{2} + \alpha\right)}{\Gamma\left(d + \alpha\right)} \delta^{d}(\vec{x})~.
\end{equation}
We denote the coordinate of the boundary point \(P_a\) as \(\vec{x}_a\), and use \(\vec{x}_{i\tilde{j}} = \vec{x}_{i} - \vec{x}_{\tilde{j}}\). Setting \(z_1 = z_2 \equiv z \rightarrow 0\), the boundary asymptotics that we are interested in are (see appendix \ref{app:boundary-limit-bulk-boundary-propagators})
\begin{align}
\begin{split}
   \hspace{-0.2cm} \Pi^{(0)}_{\frac{d}{2} + i\nu}(X_1, P_1) \underset{z\to 0}{\sim} \sqrt{\mathcal{C}_{\nu}^{(0)}}&
    \left[\tilde{\kappa}_{\nu}\delta^d\left(\vec{x}_{1\tilde{1}}\right) z^{\frac{d}{2}-i\nu} + \left(\frac{z}{\vec{x}_{1\tilde{1}}^2}\right)^{\frac{d}{2}+i\nu} 
    \right]
    + \mathcal{O}\left(z^{\frac{d}{2} + i\nu + 2}\right)~, \label{eq:boundary-limit-bulk-boundary-propagator}
\end{split}    
\end{align}
\begin{align}
\begin{split}    
    \hspace{-0.5cm} \nabla^z\Pi^{(0)}_{\frac{d}{2} + i\nu}(X_1, P_1) \underset{z\to 0}{\sim} z^2\sqrt{\mathcal{C}_{\nu}^{(0)}} \left(\frac{d}{2} + i\nu\right) &\left[
            \tilde{\kappa}_{\nu} \frac{\frac{d}{2} - i \nu}{\frac{d}{2} + i \nu}\delta^d\left(\vec{x}_{1\tilde{1}}\right) z^{\frac{d}{2} - i\nu - 1}  \right. \\
            &+\left.
            \frac{z^{\frac{d}{2} + i\nu - 1}}{\left(\vec{x}_{1\tilde{1}}^2\right)^{\frac{d}{2} + i\nu}} 
            + \mathcal{O}\left(z^{\frac{d}{2} + i\nu + 1}\right)
            \right]~, 
\end{split}  \label{eq:boundary-limit-nabla-bulk-boundary-propagator} 
\end{align}
\begin{align}
\begin{split}
   \hspace{-5cm}  T_{12}^z\left(P_1, P_2, X_1\right)
        & \underset{z\to 0}{\sim} \,z^{d + 1}
        \left(
            \frac{\delta^d(\vec{x}_{1\tilde{1}})}{\left(\vec{x}_{2\tilde{1}}^2\right)^{\frac{d}{2} + i\nu}}
            - \frac{\delta^d(\vec{x}_{2\tilde{1}})}{\left(\vec{x}_{1\tilde{1}}^2\right)^{\frac{d}{2} + i\nu}} + 
            \mathcal{O}\left(z^2\right) + \mathcal{O}\left(z^{2(1 + i\nu)}\right)
            \right)\\
        \label{eq:boundary-limit-vertex}
\end{split}
\end{align}
where \(\tilde{\kappa}_{\nu}\) is computed in appendix \ref{app:boundary-limit-bulk-boundary-propagators} to be \(\tilde{\kappa}_{\nu} = \pi^{\frac{d}{2}}\frac{\Gamma\left(i\nu\right)}{\Gamma\left(\frac{d}{2} + i\nu\right)}\). 
Note the factor of $z^2$ in equation \eqref{eq:boundary-limit-nabla-bulk-boundary-propagator}, due to the inverse of the AdS metric needed to raise the index of the covariant derivative. We plug these expansions in \eqref{eq:LongitudinalContributionBoundary} and retrieve the expansion of \(\mathcal{A}^{\parallel}\) in powers of \(z\), reported in equations \eqref{eq:equal-matter-insertions-parallel-current-z-expansion} for the choice of equal boundary conditions of the matter field. Such expansion contains terms that do not contain \(\lambda\) in the exponent of \(z\), and will vanish by counting powers of $z$, while terms proportional to \(z^{d \pm 2 i \lambda}\) will vanish once the \(\lambda\) integration is performed. We provide the details of this calculations in appendix \ref{app:parallel-current-boundary-expansion}; we find that in \(d > 0\), and with matter fields that satisfy the unitarity bound of AdS, \(\mathcal{A}^{\parallel}(z) \longrightarrow 0\) for \(z \rightarrow 0\), proving the gauge invariance of the amplitude. Finally, we notice that this argument works for more generic gauge fixing families: we can promote \(\xi\) to be a function of \(\lambda\) as long as it decays at infinity fast enough so to close the integration contour.

%% file: Neumann-tree-level.tex
In this section we will study the \(4\)-point function given by the exchange in the \(s\)-channel of a photon with {\it alternate} quantization, or as we referred to, with Neumann boundary condition. With the Neumann boundary condition, there exists the $U(1)$ gauge symmetry also on the boundary of AdS. Therefore, unlike the case with the Dirichlet boundary condition, we need to impose the $U(1)$-preserving boundary condition for the charged scalar $\Phi$ (i.e.~identical boundary conditions for both real scalars $\varphi_{1,2}$). This in particular implies that the complex scalar fields $\Phi$ and $\Phi^{\dagger}$ have definite scaling dimensions.

We will start by showing that the longitudinal part of the current-photon-current exchange diagram is non-vanishing, so the four-point function depicted in figure \ref{fig:witten-diagram-current-photon-current} is not gauge invariant. We will then define the Wilson line prescription, show that this defines a gauge invariant observable, that also respects conformal symmetry, and finally compute it. It will be more convenient to use complex fields \(\Phi, \Phi^{\dagger}\) for the computation of the $4$-point functions. It is of course straightforward to rewrite them in terms of the real scalar fields; for instance the s-channel projection of connected 4-point functions in complex-fields basis can be written in terms of real-fields amplitudes as
\begin{equation}
    \langle \Phi(P_1)\Phi^{\dagger}(P_2)\Phi^{\dagger}(P_3)\Phi(P_4) \rangle 
    = \frac{1}{4}\epsilon_{ij}\epsilon_{kl}\mathcal{A}_{\mathcal{N}}^{ijkl}.
\end{equation}

\subsection{Necessity of dressing}
\label{subsec:longitudinal-neumann-photon-exchange}


\paragraph{Non-gauge invariance.}The propagator for the Neumann gauge field \eqref{eq:Neumann-propagator} in  \(d\neq 2\) can be expressed as
\begin{equation}\label{eq:difND}
    \begin{aligned}
        \Pi^{\mathcal{N}} _1 (X_1, X_2; W_1, W_2) &= \Pi^{\mathcal{D}}_{d - 1} (X_1, X_2; W_1, W_2) \\
        & \quad\ +\int_{\lambda = \pm i\left(\frac{d}{2} - 1\right)}^{\substack{\circlearrowright \\ \circlearrowleft}} \dd \lambda \,\frac{1}{\lambda^2 + \left(\frac{d}{2} - 1\right)^2} \Omega_{\lambda}^{(1)}(X_1, X_2; W_1, W_2) \\
        & \quad\ +\xi \int_{\lambda = \pm i\frac{d}{2}}^{\substack{\circlearrowright \\ \circlearrowleft}} \dd\lambda \,\frac{1}{\left(\lambda^2 + \frac{d^2}{4}\right)^2} \left(W_1 \cdot \nabla_1\right)\left(W_2 \cdot \nabla_2\right) \Omega_{\lambda}^{(0)}(X_1, X_2),
    \end{aligned}
\end{equation}
where the residues of the additional transverse part are taken in \(\pm i\left(\frac{d}{2} - 1\right)\), and the ones of the longitudinal part in \(\pm i\frac{d}{2}\), with the depicted contour prescriptions.

Using this expression for the propagator, it is straightforward to compute the difference between the longitudinal contributions to the \(4\)-point function with Neumann and Dirichlet boundary conditions, \(\mathcal{A}^{\parallel}_{\mathcal{N}}\) and \(\mathcal{A}^{\parallel}_{\mathcal{D}}\). As is clear from \eqref{eq:difND}, the only difference is the integration contour of the spectral parameter in the gauge propagator, which accounts for the addition of the correct homogeneous solution. Therefore \(\mathcal{A}^{\parallel}_{\mathcal{N}}\) differs from \(\mathcal{A}^{\parallel}_{\mathcal{D}}\) as computed in \eqref{eq:dirchlet-longitudinal-4-point} by the term
\begin{align}
    \label{eq:homogeneous-longitudinal-4-point}
    \begin{split}
        \mathcal{A}^{ijkl, \parallel}_{\mathcal{N}} - \mathcal{A}^{ijkl, \parallel}_{\mathcal{D}} = (-ie)^2\frac{\xi}{\pi} 
        & \int_{\lambda = \pm i\frac{d}{2}}^{\substack{\circlearrowright \\ \circlearrowleft}} \dd\lambda \,
        \frac{\lambda^2}{\left(\lambda^2 + \frac{d^2}{4}\right)^2} \sqrt{\mathcal{C}_{\lambda}^{(0)}\mathcal{C}_{-\lambda}^{(0)}}\int_{\partial AdS} \dd^d P_5  \\
        & \quad \int_{AdS}\dd^{d + 1} X_1 \, T_{ij}^A\left(P_1, P_2, X_1\right) \nabla_{1,A}\Pi^{(0)}_{\frac{d}{2} + i\lambda}(X_1, P_5)\\
        & \quad \int_{AdS}\dd^{d + 1} X_2 \,T_{kl}^B\left(P_3, P_4, X_2\right) \nabla_{2,B}\Pi^{(0)}_{\frac{d}{2} - i\lambda}(X_2, P_5)~.
    \end{split}
\end{align}
Analogously to what was done in subsection \ref{subsec:longitudinal-dirichlet-photon-exchange}, we can integrate by part (again using \(\nabla \cdot T_{ij} = 0 \)), perform the boundary limit of the above expression, and find an expansion in powers of \(z\). 

The integrand in \eqref{eq:homogeneous-longitudinal-4-point}, is precisely the same integrand as in \eqref{eq:dirchlet-longitudinal-4-point}, so the power expansions in \(z\) {\it before} carrying out the integration over \(\lambda\) is again \eqref{eq:equal-matter-insertions-parallel-current-z-expansion}.
As in section~\ref{subsec:longitudinal-dirichlet-photon-exchange}, the first three lines in \eqref{eq:equal-matter-insertions-parallel-current-z-expansion} do not depend on \(\lambda\), and always vanish because the power of $z$ is positive if we are dealing with unitary AdS theories. 

In order to treat the \(\lambda\)-dependent powers, we need to perform the integration over $\lambda$.  
The details of the computation are similar to the ones in appendix \ref{app:parallel-current-boundary-expansion}, although the integration contour is composed of the two small circles around \(\lambda = \pm i\frac{d}{2}\) instead of the real line. The residue around \(\lambda = -i\frac{d}{2}\) goes to zero as in appendix \ref{app:parallel-current-boundary-expansion}, but the residue in the upper half-plane gives a non-vanishing contribution. In in \(d = 3\) we find
\begin{align}
    \label{eq:longitudinal-contribution-neumann-current-photon-current}
    \begin{split}
        & \hspace{-0.5cm}\mathcal{A}^{ijkl, \parallel}_{\mathcal{N}} 
        = (-ie)^2\frac{\xi}{\pi} \,  
        \frac{ (2i\nu)^2\pi^{d} \left(\mathcal{C}_{\nu}^{(0)}\right)^2 \Gamma\left(i\nu\right)^2}{\Gamma\left(\frac{d}{2} + i\nu\right)^2} (2\pi i) \underset{\lambda = i \frac{d}{2}}{\text{Res}}\left[ 
        \frac{\lambda^2}{\left(\lambda^2 + \frac{d^2}{4}\right)^2}
         \frac{\mathcal{C}_{-\lambda}^{(0)}  \mathcal{C}_{\lambda}^{(0)}}{|\vec{x}_{12}|^{2 \left(\frac{d}{2}+i \nu \right)} |\vec{x}_{34}|^{2 \left(\frac{d}{2} + i \nu \right)}} \right.\\
        & \ \left.\left(z^{d + 2 i\lambda} 
        \frac{\pi^{\frac{d}{2}}\Gamma\left(-i\lambda\right)}{\Gamma\left(\frac{d}{2} - i\lambda\right)}
        \left(\frac{1}{|\vec{x}_{13}|^{2\left(\frac{d}{2} + i\lambda\right)}} - \frac{1}{|\vec{x}_{14}|^{2\left(\frac{d}{2} + i\lambda\right)}} - \frac{1}{|\vec{x}_{23}|^{2\left(\frac{d}{2} + i\lambda\right)}} + \frac{1}{|\vec{x}_{24}|^{2\left(\frac{d}{2} + i\lambda\right)}}\right) \right.\right. \\
        &\left.\left. \quad{}+
            z^{d - 2 i\lambda} 
            \frac{\pi^{\frac{d}{2}}\Gamma\left(i\lambda\right)}{\Gamma\left(\frac{d}{2} + i\lambda\right)}
            \left(\frac{1}{|\vec{x}_{13}|^{2\left(\frac{d}{2} - i\lambda\right)}} - \frac{1}{|\vec{x}_{14}|^{2\left(\frac{d}{2} - i\lambda\right)}} - \frac{1}{|\vec{x}_{23}|^{2\left(\frac{d}{2} - i\lambda\right)}} + \frac{1}{|\vec{x}_{24}|^{2\left(\frac{d}{2} - i\lambda\right)}}\right) 
        \right)
        \right] \\
        & \underset{d=3}{=}- (-i e)^2 
         \frac{\xi}{6\pi^2}  \frac{1}{|\vec{x}_{12}|^{2 \left(\frac{3}{2}+i \nu \right)} |\vec{x}_{34}|^{2 \left(\frac{3}{2} + i \nu \right)}}   \left[\left(2 \log (z) - \gamma_E - \psi ^{(0)}\left(-\tfrac{1}{2}\right)\right)  \log \left(\frac{|\vec{x}_{13}| |\vec{x}_{24}|}{|\vec{x}_{14}| |\vec{x}_{23}|}\right) \right.\\
        & \mspace{150mu}\left.\phantom{\frac{|\vec{x}_{13}|}{|\vec{x}_{14}|}}- \left(\log ^2 |\vec{x}_{13}| - \log ^2 |\vec{x}_{14}| - \log ^2|\vec{x}_{23}| + \log ^2 |\vec{x}_{24}|\right)  \right] + \mathcal{O} \left(z^{2\alpha}\right). 
    \end{split}
\end{align}
The subleading term has exponent \(\alpha = (1 + i\nu)\) if \(-1 < i\nu \le 0\), and \(\alpha = 1\) otherwise.

The expression in \eqref{eq:longitudinal-contribution-neumann-current-photon-current} shows that the longitudinal piece is non-vanishing. It contains a logarithmically divergent piece proportional to $\log z$ and a finite piece, both of which are nontrivial functions of cross ratios. This clearly signals the non-gauge-invariance of the 4-point function\footnote{
When the scalars are sitting exactly on the AdS unitarity bound \(i\nu = - 1\), there are some additional contributions, which give the same \(\vec{x}\) and \(z\) dependence as in \eqref{eq:longitudinal-contribution-neumann-current-photon-current}, but with a divergent numerical coefficient in front: we don't treat this case further here, and postpone some comments about it to the end of section~\ref{sec:1-loop-self-energy}. }.

\begin{figure}[t]
    \centering
    \begin{tikzpicture}
        \begin{scope}[xshift = -3.3cm]
            \draw[black] (0,0) circle (2.7);
            \draw[black]  (140:2.7) -- (220:2.7);
            \draw[green, very thick]  (140:2.7) arc (40:-40:2.7);
            \draw[black]  (40:2.7) -- (-40:2.7);
            \draw[green, very thick]  (40:2.7) arc (140:220:2.7);
            \draw[turquoise, thick, dashed]  (-1.4, 0) -- (270:2.7) -- (1.4, 0);
            \draw[goldenyellow, ultra thick, decorate,decoration={coil,aspect=0, pre length=0.1cm}]  (-1.4, 0) -- (1.4, 0);
            \node[fuchsia, left] at (140:2.7) {\(P_1\)};
            \node[fuchsia, left] at (220:2.7) {\(P_2\)};
            \node[fuchsia, right] at (40:2.7) {\(P_3\)};
            \node[fuchsia, right] at (-40:2.7) {\(P_4\)};
            \node[turquoise, below] at (270:2.7) {\(P_5\)};
        \end{scope}
        \begin{scope}[xshift = 3.3cm]
            \draw[black] (0,0) circle (2.7);
            \draw[black] (140:2.7) -- (220:2.7);
            \draw[green, very thick]  (140:2.7) arc (40:-40:2.7);
            \draw[black]  (40:2.7) -- (1, 0) -- (-40:2.7);
            \draw[turquoise, thick, dashed]  (-1.4, 0) -- (270:2.7) -- (1, 0);
            \draw[goldenyellow, ultra thick, decorate,decoration={coil,aspect=0, pre length=0.07cm}]  (-1.4, 0) -- (1, 0);
            \node[fuchsia, left] at (140:2.7) {\(P_1\)};
            \node[fuchsia, left] at (220:2.7) {\(P_2\)};
            \node[fuchsia, right] at (40:2.7) {\(P_3\)};
            \node[fuchsia, right] at (-40:2.7) {\(P_4\)};
            \node[turquoise, below] at (270:2.7) {\(P_5\)};
        \end{scope}
    \end{tikzpicture}
    \caption{Witten diagrams representing the geodesic-photon-geodesic exchange on the left and the geodesic-photon-current exchange on the right; the geodesics are depicted in green, and the exchanged photon is intended with Neumann boundary conditions.}
    \label{fig:geodesic-witten-diagrams}
\end{figure}
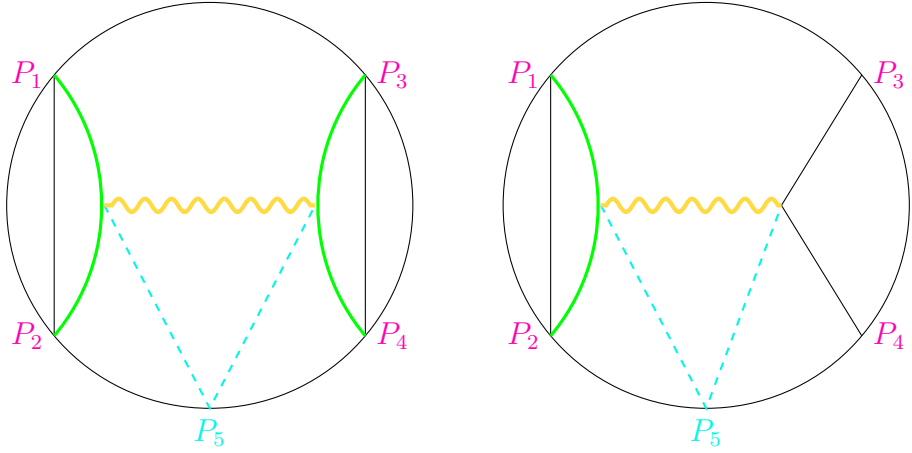
\paragraph{Geodesic Wilson line dressing.} The reason for the aforementioned non-gauge invariance is intuitively obvious; with the Neumann boundary condition, the gauge fields are dynamical even at the boundary of AdS, and thus the local correlation functions of charged operators are not well-defined observables. One typical way to remedy this problem is to attach charged operators to Wilson lines so that the entire object becomes gauge invariant. We choose to define the gauge-invariant observables as follows:
\[
\langle \Phi(P_1) e^{-ie\int_{P_1}^{P_2}dX^{A}_1 A_{A}(X_1)}   \Phi^{\dagger}(P_2) \Phi^{\dagger}(P_3) e^{ie\int_{P_3}^{P_4}dX^{B}_2 A_{B}(X_2)}   \Phi(P_4) \rangle~.
\]

For the purpose of defining a gauge-invariant operator, Wilson lines of any shape will do. However, using such generic Wilson lines would make the correlation functions highly non-local and also break the conformal structure completely. Instead, in this paper we focus on a special class of Wilson lines---Wilson lines along the geodesics of AdS. The advantage of this choice is twofold. First, since they are defined along geodesics, they don't break any additional symmetry of the AdS isometry group i.e.~the conformal symmetry at the boundary, as compared to the insertion of two points. In fact, integrations over such geodesic lines were studied in holography \cite{hijano2016witten, Chen:2017yia} and were shown to be dual to the exchange of conformal blocks in the boundary theory. We will make use of this connection in the computation below. Second, since the Wilson lines are extended in the bulk rather than on the boundary, we expect that local correlation functions in Euclidean kinematics are better-behaved\footnote{What we discuss here should be true at least for correlation functions with neutral operators. For charged operators, additional singularities may arise from the intersection of Wilson lines in the bulk.}: the Wilson lines on the boundary would introduce a defect (see e.g.~\cite{Gabai:2025hwf}), along which local correlation functions may have singularities. In contrast, the Wilson lines in the bulk should be considered as appropriately smeared insertions of gauge fields\footnote{It would be interesting to make the connection between the Wilson line in the bulk and the smeared insertion at the boundary more precise.} and the correlation functions are expected to be less singular. 

In the rest of this section, we use this geodesic Wilson line dressing and compute the four-point function. The dressing generates two additional kinds of diagrams, represented in figure \ref{fig:geodesic-witten-diagrams}. Unless otherwise specified, we will take the prescription of sending Wilson lines along geodesics connecting \(P_i\) to \(P_j\). We will see in subsection \ref{subsec:wilson-line-prescription-longitudinal} that the dressing perfectly compensates the longitudinal contribution computed above, {\it regardless of the curve} we send the lines along. We then compute the transverse piece using the geodesic Wilson line in subsection \ref{subsec:wilson-line-prescription-transverse}.

\subsection{Longitudinal piece}
\label{subsec:wilson-line-prescription-longitudinal}
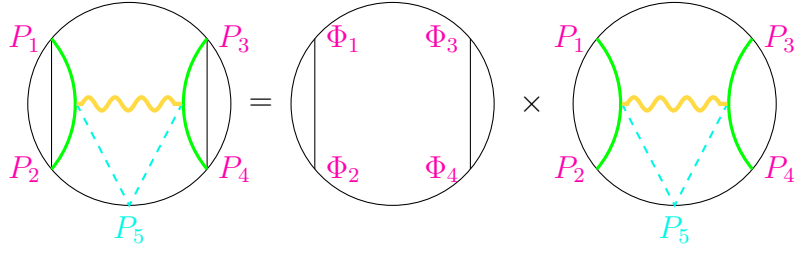
\begin{figure}[!ht]
    \centering
    \begin{tikzpicture}[scale = 0.5]
        \begin{scope}[xshift = -3.5cm]
            \draw[black] (0,0) circle (2.7);
            \draw[black]  (140:2.7) -- (220:2.7);
            \draw[green, very thick]  (140:2.7) arc (40:-40:2.7);
            \draw[black]  (40:2.7) -- (-40:2.7);
            \draw[green, very thick]  (40:2.7) arc (140:220:2.7);
            \draw[turquoise, thick, dashed]  (-1.4, 0) -- (270:2.7) -- (1.4, 0);
            \draw[goldenyellow, ultra thick, decorate,decoration={coil,aspect=0, pre length=0.07cm}]  (-1.4, 0) -- (1.4, 0);
            \node[fuchsia, left] at (140:2.7) {\(P_1\)};
            \node[fuchsia, left] at (220:2.7) {\(P_2\)};
            \node[fuchsia, right] at (40:2.7) {\(P_3\)};
            \node[fuchsia, right] at (-40:2.7) {\(P_4\)};
            \node[turquoise, below] at (270:2.7) {\(P_5\)};
        \end{scope}
        \node[black] at (0, 0) {\(=\)};
        \begin{scope}[xshift = 3.5cm]
            \draw[black] (0,0) circle (2.7);
            \draw[black] (140:2.7) -- (220:2.7);
            \draw[black]  (40:2.7) -- (-40:2.7);
            \node[fuchsia, right] at (140:2.7) {\(\Phi_1\)};
            \node[fuchsia, right] at (220:2.7) {\(\Phi_2\)};
            \node[fuchsia, left] at (40:2.7) {\(\Phi_3\)};
            \node[fuchsia, left] at (-40:2.7) {\(\Phi_4\)};
        \end{scope}
        \node[black] at (7.2, 0) {\(\times\)};
        \begin{scope}[xshift = 11cm]
            \draw[black] (0,0) circle (2.7);
            \draw[green, very thick]  (140:2.7) arc (40:-40:2.7);
            \draw[green, very thick]  (40:2.7) arc (140:220:2.7);
            \draw[turquoise, thick, dashed]  (-1.4, 0) -- (270:2.7) -- (1.4, 0);
            \draw[goldenyellow, ultra thick, decorate,decoration={coil,aspect=0, pre length=0.07cm}]  (-1.4, 0) -- (1.4, 0);
            \node[fuchsia, left] at (140:2.7) {\(P_1\)};
            \node[fuchsia, left] at (220:2.7) {\(P_2\)};
            \node[fuchsia, right] at (40:2.7) {\(P_3\)};
            \node[fuchsia, right] at (-40:2.7) {\(P_4\)};
            \node[turquoise, below] at (270:2.7) {\(P_5\)};
        \end{scope}
    \end{tikzpicture}
    \caption{The geodesic-photon-geodesic diagram factorizes into the product of CFT four-point function of the matter insertions, and a photon exchange integrated along geodesics.}
    \label{fig:factorization-geodesic-photon-geodesic}
\end{figure}

Below we compute longitudinal pieces of additional diagrams that arise from the geodesic dressing. 
\paragraph{Geodesic-photon-geodesic exchange.}
Let us first consider the photon exchange between two Wilson lines. Its contribution factorizes as in figure \ref{fig:factorization-geodesic-photon-geodesic}.

The first factor on the right hand side can be computed by taking the limit of a product of bulk two-point functions\footnote{We can think of a boundary-to-boundary propagator as the limit of a bulk-to-boundary propagator, and the correct limiting procedure with our choice of normalization in \eqref{eq:bulk-to-boundary-normalization} is
\[
    \langle \varphi_1(X_1) \varphi_2 (X_2) \ldots \varphi_k(\vec{x}_k) \ldots \varphi_n (X_n) \rangle 
    = \lim_{z \rightarrow 0} \frac{z^{-\left(\frac{d}{2} + i\nu_k\right)}}{\sqrt{\mathcal{C}_{\nu_k}^{(0)}}} 
    \langle \varphi_1(X_1) \varphi_2 (X_2) \ldots \varphi_k(z, \vec{x}_k) \ldots \varphi_n (X_n) \rangle.
\]
A quick way to verify this is computing the bulk-to-boundary propagator with the normalization chosen in \cite{di2022analyticity}, as the limit of the corresponding bulk-to-bulk propagator.
} as shown in figure \ref{fig:CFT-4-point-function-normalization}.
\begin{figure}[!h]
    \centering
    \begin{tikzpicture}[scale = 0.5]
        \begin{scope}[xshift = -3.1cm]
            \draw[black] (0,0) circle (2.7);
            \draw[black] (140:2.7) -- (220:2.7);
            \draw[black]  (40:2.7) -- (-40:2.7);
            \node[fuchsia, left] at (140:2.7) {\(\Phi(P_1)\)};
            \node[fuchsia, left] at (220:2.7) {\(\Phi^{\dagger}(P_2)\)};
            \node[fuchsia, right] at (40:2.7) {\(\Phi(P_3)\)};
            \node[fuchsia, right] at (-40:2.7) {\(\Phi^{\dagger}(P_4)\)};
        \end{scope}
        \node[black,right] at (0, 0) {\(= \left(\frac{1}{P_{12}}  \frac{1}{P_{34}}\right)^{\frac{d}{2} + i\nu}\).};
    \end{tikzpicture}
    \caption{The four-point function of the matter fields must be unit normalized to be consistent with the normalization of the bulk-to-boundary propagator adopted in the rest of the document.}
    \label{fig:CFT-4-point-function-normalization}
\end{figure}
Applying the split representation to the second factor, we obtain the following expression for the full amplitude
\begin{align}
    \label{eq:geodesic-photon-geodesic-amplitude}
    \begin{split}
        \mathcal{A}_{gg}^{\parallel}& = (-ie)^2\xi  \left(\frac{1}{P_{12}} \, \frac{1}{P_{34}}\right)^{\frac{d}{2} + i\nu} 
        \int_{\R \oplus \substack{\circlearrowright \\ \circlearrowleft}}\dd \lambda \,\frac{1}{\left(\lambda^2  + \frac{d^2}{4}\right)^2} \\
        & \mspace{150mu}\int_{P_1}^{P_2}\dd X^{A}_1\int_{P_3}^{P_4}\dd X^{B}_2 \,
        \nabla_{1, A} \nabla_{2, B} \Omega_{\lambda}^{(0)} (X_1, X_2) \\
       &  = (-ie)^2\frac{\xi}{\pi}
        \left(\frac{1}{P_{12}} \, \frac{1}{P_{34}}\right)^{\frac{d}{2} + i\nu} 
        \int_{\R \oplus \substack{\circlearrowright \\ \circlearrowleft}}\dd \lambda \, \frac{\lambda^2 \sqrt{\mathcal{C}_{-\lambda}^{(0)}\mathcal{C}_{\lambda}^{(0)}}}{\left(\lambda^2  + \frac{d^2}{4}\right)^2} \int_{\R^d} \dd^d P_5 \\
        & \mspace{150mu} \int_{P_1}^{P_2}\dd X^{A}_1 \, \nabla_{1, A}\Pi^{(0)}_{\frac{d}{2} + i\lambda}(X_1, P_5)
        \int_{P_3}^{P_4}\dd X^{B}_2 \, \nabla_{1, B}\Pi^{(0)}_{\frac{d}{2} - i\lambda}(X_2, P_5) \,.
    \end{split}
\end{align}
Note that the integrals $\dd X_1^{A}$ and $\dd X_2^{B}$ are along the contours of the Wilson lines.
Since the integrands for both of these integrals are total derivatives, we can simplify them to
\begin{equation}
    \label{eq:geodesic-photon-geodesic-amplitude-integrated-parts}
    \begin{aligned}
        \mathcal{A}_{gg}^{\parallel}  = (-ie)^2\frac{\xi}{\pi} &\left(\frac{1}{P_{12}} \frac{1}{P_{34}}\right)^{\frac{d}{2} + i\nu} \, \int_{\R \oplus \substack{\circlearrowright \\ \circlearrowleft}}\dd \lambda  \,
    \frac{\lambda^2}{\left(\lambda^2  + \frac{d^2}{4}\right)^2} \sqrt{\mathcal{C}_{-\lambda}^{(0)}\mathcal{C}_{\lambda}^{(0)}}\int_{\R^d} \dd^d P_5 \\
    & \qquad\mspace{-50mu}\left(\Pi^{(0)}_{\frac{d}{2} + i\lambda}(P_2, P_5) - \Pi^{(0)}_{\frac{d}{2} + i\lambda}(P_1, P_5)\right)
    \left(\Pi^{(0)}_{\frac{d}{2} - i\lambda}(P_4, P_5) - \Pi^{(0)}_{\frac{d}{2} - i\lambda}(P_3, P_5)\right)~.
    \end{aligned}
\end{equation}
This is clearly independent of the curves the Wilson lines have been sent along.

To evaluate \eqref{eq:geodesic-photon-geodesic-amplitude-integrated-parts}, we proceed in a similar way as in section \ref{subsec:longitudinal-dirichlet-photon-exchange}; we take insertion points \(P_1, P_2, P_3\) and \(P_4\) to be on a regulating surface at distance \(z\) from the boundary, and expand for \(z \rightarrow 0\) using formulas \eqref{eq:boundary-limit-bulk-boundary-propagator}, \eqref{eq:boundary-limit-nabla-bulk-boundary-propagator} and \eqref{eq:boundary-limit-vertex}. We can see that the leading terms of this diagram, that do not vanish by power-counting, are proportional to \(z^{d \pm 2i\lambda}\), and so will give a logarithmic contribution of the same order as in \eqref{eq:longitudinal-contribution-neumann-current-photon-current} once the \(\lambda\)-residue in the upper-half plane is computed. At this point, we choose not to further expand the equation \eqref{eq:geodesic-photon-geodesic-amplitude-integrated-parts}, since this form will prove more useful in the following.

\paragraph{Geodesic-photon-current exchange.}
The second diagram to analyze is the one on the right side of figure \ref{fig:geodesic-witten-diagrams}. It is given by the following amplitude:
\begin{equation}
    \begin{aligned}
        \mathcal{A}^{\parallel}_{gc}  = (-ie)^2\xi &\left(\frac{1}{P_{12}}\right)^{\frac{d}{2} + i\nu} 
        \int_{\R \oplus \substack{\circlearrowright \\ \circlearrowleft}}\dd \lambda \,\frac{1}{\left(\lambda^2  + \frac{d^2}{4}\right)^2}
        \\
        & \quad\int_{P_1}^{P_2}\dd X^{A}_1\int_{AdS}\dd^{d + 1} X_2 \,
        \nabla_{1, A} \nabla_{2, B} \Omega_{\lambda}^{(0)} (X_1, X_2) T_{34}^{B}(P_3, P_4, X_2)~. \\
    \end{aligned}
\end{equation}
As usual, we can proceed with the split representation and separate the \(X_1\) and \(X_2\) integrals:
\begin{equation}
    \begin{aligned}
        \mathcal{A}^{\parallel}_{gc} &= (-ie)^2\frac{\xi}{\pi}  \left(\frac{1}{P_{12}}\right)^{\frac{d}{2} + i\nu} 
        \int_{\R \oplus \substack{\circlearrowright \\ \circlearrowleft}}\dd \lambda \, \frac{\lambda^2}{\left(\lambda^2  + \frac{d^2}{4}\right)^2} \sqrt{\mathcal{C}_{-\lambda}^{(0)}\mathcal{C}_{\lambda}^{(0)}}
        \int_{\partial AdS}\dd^dP_5\\
        & \qquad\int_{P_1}^{P_2}\dd X^{A}_1 \, \nabla_{1, A} \Pi^{(0)}_{\frac{d}{2} + i \lambda}(X_1, P_5)\int_{AdS}\dd^{d + 1} X_2 \,
        \nabla_{2, B} \Pi^{(0)}_{\frac{d}{2} - i \lambda}(P_5, X_2) T_{34}^{B}(P_3, P_4, X_2)~. 
    \end{aligned}
\end{equation}
Again, we can directly perform the \(x_1\) integral since it is a total derivative:
\[
    \int_{\bar{P}_1}^{\bar{P}_2}\dd X^{A}_1 \, \nabla_{1, A} \Pi^{(0)}_{\frac{d}{2} + i \lambda}(X_1, P_5) = \Pi^{(0)}_{\frac{d}{2} + i \lambda}(\bar{P}_2, P_5) - \Pi^{(0)}_{\frac{d}{2} + i \lambda}(\bar{P}_1, P_5)~.
\]
where the points \(\bar{P}_1\rightarrow P_1\) and \(\bar{P}_2 \rightarrow P_2\) are again taken on a regulating surface very close to the boundary, outside of which the interactions are switched off. In Poincar\'e coordinates $\bar{P}_{1,2}=(z,\vec{x}_{1,2})$ where $z\to 0$ is the location of the regulating surface.

Similarly, the \(X_2\) integral reduces to a boundary integral since \(\nabla \cdot T_{34} = 0\), so we obtain
\begin{align}
    \begin{split}
    \int_{AdS}\dd^{d + 1} X_2 \,
    \nabla_{2, \beta} \Pi^{(0)}_{\frac{d}{2} - i \lambda}(P_5, X_2) & T_{34}^{\beta}(P_3, P_4, X_2)  \\
    & \quad =\int_{\partial AdS}\frac{\dd^d \vec{x}_{\tilde{2}}}{z^{1 + d}} \,
    \Pi^{(0)}_{\frac{d}{2} - i \lambda}(P_5, X_2) T_{34}^{z}(P_3, P_4, X_2)~,
    \end{split}
\end{align}
where in the right-hand side we plug $X_2 = (z, \vec{x}_{\tilde{2}})$ and integrate on the regulating surface with $z\to 0$. Expanding in $z$ we get a contribution of the same order as \eqref{eq:longitudinal-contribution-neumann-current-photon-current}, but it is better to keep the expression in this form and combine it with the other diagrams of the four-point function of dressed scalars.

\paragraph{Full longitudinal piece.}
We are now ready to compute the full longitudinal contribution to the dressed current exchange mediated by a Neumann photon, accounting for the Wilson lines contributions. This is: 

\begin{equation}
    \label{eq:full-longitudinal-neumann-photon-exchange}
    \begin{aligned}
        \mathcal{A}^{\parallel}_{cc} +  \mathcal{A}^{\parallel}_{gg} + \mathcal{A}^{\parallel}_{gc} + \mathcal{A}^{\parallel}_{cg}  = \underset{z\to 0}{ \mathrm{lim}}\,
        & (-ie)^2 \frac{\xi}{\pi}
        \int_{\R \oplus \substack{\circlearrowright \\ \circlearrowleft}}\dd \lambda\, \frac{\lambda^2}{\left(\lambda^2  + \frac{d^2}{4}\right)^2} \sqrt{\mathcal{C}_{-\lambda}^{(0)}\mathcal{C}_{\lambda}^{(0)}}
        \int_{\partial AdS}\dd^d P_5\\
        & \left(
            \int_{\partial AdS}\frac{\dd^d \vec{x}_{\tilde{1}}}{z^{d + 1}} \;
            \Pi^{(0)}_{\frac{d}{2} + i \lambda}(P_5, X_1) T_{12}^{z}(P_1, P_2, X_1)\Big\vert_{X_1 = (z,\vec{x}_{\tilde{1}})}
        \right. \\
            & \qquad + \left.
            \left(\frac{1}{P_{12}}\right)^{\frac{d}{2} + i\nu} 
            \int_{\bar{P}_1}^{\bar{P}_2}\dd X^{A}_1 \, \nabla_{1, A} \Pi^{(0)}_{\frac{d}{2} + i \lambda}(X_1, P_5)
        \right) \\
        & \left(
            \int_{\partial AdS}\frac{\dd^d \vec{x}_{\tilde{2}}}{z^{d + 1}} \;
            \Pi^{(0)}_{\frac{d}{2} - i \lambda}(P_5, X_2) T_{34}^{z}(P_3, P_4, X_2)\Big\vert_{X_2 = (z,\vec{x}_{\tilde{2}})}
        \right. \\
            & \qquad + \left.
            \left(\frac{1}{P_{34}}\right)^{\frac{d}{2} + i\nu} 
            \int_{\bar{P}_3}^{\bar{P}_4}\dd X^{A}_2 \, \nabla_{2,A} \Pi^{(0)}_{\frac{d}{2} - i \lambda}(X_2, P_5)
        \right) \,.
    \end{aligned}
\end{equation}
Let us study the contribution from each term in the square bracket and expand in the limit \(z \rightarrow 0\). Remembering \eqref{eq:boundary-limit-vertex}, we can write the integrand as
\begin{align}
    \label{eq:longitudinal-neumann-photon-exchange-one-side}
    \begin{split}
            &\int_{\partial AdS}\frac{\dd^d \vec{x}_{\tilde{1}}}{z^{d + 1}} \,
            \Pi^{(0)}_{\frac{d}{2} + i \lambda}(P_5, X_1) 
            T_{12}^{z}(P_1, P_2, X_1)\Big\vert_{X_1=(z,\vec{x}_{\tilde{1}})}
            \\
            & \qquad 
            + \left(\frac{1}{P_{12}}\right)^{\frac{d}{2} + i\nu} 
            \int_{\bar{P}_1}^{\bar{P}_2}\dd X^{A}_1 \, \nabla_{1, A} \Pi^{(0)}_{\frac{d}{2} + i \lambda}(X_1, P_5) \\
        &\underset{z\to 0}{\sim}
        \int_{\partial AdS}\dd^d \vec{x}_{\tilde{1}} \,
        \Pi^{(0)}_{\frac{d}{2} + i \lambda}(P_5, X_1)\Big\vert_{X_1=(z,\vec{x}_{\tilde{1}})} 
          \left(
            \frac{\delta^d(\vec{x}_{1\tilde{1}})}{\left(\vec{x}_{2\tilde{1}}^2\right)^{\frac{d}{2} + i\nu}}
            - \frac{\delta^d(\vec{x}_{2\tilde{1}})}{\left(\vec{x}_{1\tilde{1}}^2\right)^{\frac{d}{2} + i\nu}}
            + \mathcal{O}\left(z^2\right) + \mathcal{O}\left(z^{2(1 + i\nu)}\right)
        \right) \\
        & \qquad + \left(\frac{1}{P_{12}}\right)^{\frac{d}{2} + i\nu} 
        \left(
            \Pi^{(0)}_{\frac{d}{2} + i \lambda} \right.
            \left. (\bar{P}_2, P_5) 
            - \Pi^{(0)}_{\frac{d}{2} + i \lambda}(\bar{P}_1, P_5)
        \right) \\
        & = \int_{\partial AdS}\dd^d \vec{x}_{\tilde{1}} \,
        \Pi^{(0)}_{\frac{d}{2} + i \lambda}(P_5, X_1)\Big\vert_{X_1=(z,\vec{x}_{\tilde{1}})} 
        \left( 
        \mathcal{O}\left(z^2\right) + \mathcal{O}\left(z^{2(1 + i\nu)}\right)
        \right)~.
    \end{split}
\end{align}
Going from the second to the third expression in the above equation, the contribution from the Wilson line (third row in the second expression) cancels the leading term in the expansion of the vertex \(T^z_{ij}\), namely the term with \(\delta^d(\vec{x}_{ik})\), leaving behind subleading contributions with the powers of $z$ that we indicated. We see that when $i\nu < 0$ the leading power in the boundary expansion is $z^{2(1+i\nu)}$, while for $i\nu \geq 0$ the leading power is $z^2$. The result for the other terms in the integrand of \eqref{eq:full-longitudinal-neumann-photon-exchange} can be obtained by substituting $\lambda \to -\lambda$, $P_1 \to P_3$ and $P_2 \to P_4$. 

Putting everything together we find that the integrand in \eqref{eq:full-longitudinal-neumann-photon-exchange} is of the form \begin{equation}
    \label{eq:full-longitudinal-neumann-photon-exchange-integrand-expansion}
    z^{4\alpha}\,\Pi^{(0)}_{\frac{d}{2} + i \lambda}(P_5, X_1)\Pi^{(0)}_{\frac{d}{2} - i \lambda}(P_5, X_2)\Big\vert_{X_a = (z,\vec{x}_{\tilde{a}})} \propto z^{4\alpha}\left(z^d + \ldots + z^{d \pm 2i\lambda}
+ \ldots \right),
\end{equation}
with $\alpha = \mathrm{Min}(1, 1+i\nu)$. Unitarity imposes that $i
\nu \geq -1$ and therefore \(\alpha > 0\) (if no scalar is sitting exactly on the unitarity bound). We can thus see immediately that all the powers of \(z\) which do not depend on \(\lambda\) vanish in the limit of \(z\rightarrow 0\). 

To see the fate of the $\lambda$-dependent terms, we have to perform the integration. First, the integration on the real line  can be performed by closing the contour on the upper or lower half-plane respectively for powers with \(z^{\mp 2i\lambda}\). Since there are no poles with \(\mathrm{Im}\left(\lambda\right) < \frac{d}{2}\) in the integrand, one can see that it vanishes in the limit $z\to 0$. Second, the residues at \(\lambda = \pm i\frac{d}{2}\) also vanish in the limit thanks to the positive power \(z^{\alpha}\), that sits in front of the integrand as in \eqref{eq:full-longitudinal-neumann-photon-exchange-integrand-expansion}. Together this shows the vanishing of the longitudinal piece of the 4-point function as expected from the gauge invariance.

Note that this argument can be generalized to gauge fixings for which \(\xi\) is promoted to a generic function of \(\xi(\lambda)\) which decays fast enough at infinity, establishing the gauge invariance for more general gauge fixing conditions. Our observables are also  gauge invariant when the dressed operators are placed in the bulk, even though we did not check this explicitly.
\subsection{Transverse piece}\label{subsec:wilson-line-prescription-transverse}

Next we compute the transverse part of the 4-point function. Analogous computations with the Dirichlet boundary condition have been performed extensively in the literature, and were reviewed in section \ref{sec:dirichlet-photon-exchange}; here we compute the Neumann photon exchange, including the correction from the Wilson line dressing.
We will compute separately the contributions given by currents and geodesic insertions, and finally add them all together at the end. 

\paragraph{Current-photon-current diagram.}
This is the diagram where the two currents exchange a (Neumann) photon, as shown in figure~\ref{fig:witten-diagram-current-photon-current}. 
The computation of this diagram is nearly identical to that for the Dirichlet case reviewed in subsection \ref{subsec:current-photon-current-diagram-transverse}, with the sole exception of replacing the Dirichlet gauge propagator with the Neumann gauge propagator. As it was detailed in section \ref{sec:propagators}, in spectral representation this amounts to changing the integration contour for the spectral parameter.

Everything else goes through the same way, so the amplitude we need to compute is:
\begin{align}
    \label{eq:Neumann-TransverseContribution3points}
    \begin{split}
        \mathcal{A}^{\perp}_{\mathcal{N}} & = \frac{4(-ie)^2}{\pi}\frac{\sqrt{\mathcal{C}_{\nu_1}^{(0)}\mathcal{C}_{\nu_2}^{(0)}\mathcal{C}_{\nu_3}^{(0)}\mathcal{C}_{\nu_4}^{(0)}}}{\left(\frac{d}{2} - 1\right)}\\
        & \quad\int_{\R \oplus \substack{\circlearrowright \\ \circlearrowleft}}
        \dd \lambda \, \frac{\lambda^2\mathcal{C}_{\lambda}^{(1)}\mathcal{C}_{-\lambda}^{(1)}}{\lambda^2 + \left(\frac{d}{2} - 1\right)^2} 
        b_{\text{Bulk}}(\nu_1, \nu_2, \lambda, 1)b_{\text{Bulk}}(\nu_3, \nu_4, -\lambda, 1) \\
        & \quad\int_{\partial AdS} \dd^{d} P_5 \,
        \Big\langle \mathcal{O}_{\nu_1}(P_1) \mathcal{O}_{\nu_2}(P_2) \mathcal{O}_{\lambda}^{(1)}(P_5; D_Z)\Big\rangle_1 
        \Big\langle \mathcal{O}_{\nu_3}(P_3) \mathcal{O}_{\nu_4}(P_4) \mathcal{O}_{-\lambda}^{(1)}(P_5; Z)\Big\rangle_1 \,,
    \end{split}
\end{align}
which in terms of conformal partial waves (again normalized as in \cite{simmons2018spacetime}) becomes\footnote{Here we added a subscript ``bare" to the spectral density in order to indicate that the result does not include contributions from the Wilson line dressing.}
    \begin{equation}
        \label{eq:Neumann-current-photon-current-transverse}
        \begin{aligned}
        \mathcal{A}^{\perp}_{\mathcal{N}} &= 
        (ie)^2 \left(\frac{1}{P_{12}}\, \frac{1}{P_{34}}\right)^{\frac{d}{2} + i\nu} 
        \, \int_{\R \oplus \substack{\circlearrowright \\ \circlearrowleft}}
        \dd \lambda \, \rho_{\mathcal{N},  {\rm bare}}^{J = 1}(\lambda)  \, \mathcal{F}_{\lambda, 1}^{\{\frac{d}{2} + i\nu \equiv \Delta_{\nu}\}}(u,v) \\
        &= (ie)^2 \left(\frac{1}{P_{12}}\, \frac{1}{P_{34}}\right)^{\frac{d}{2} + i\nu} \,
        \int_{\R \oplus \substack{\circlearrowright \\ \circlearrowleft}}
        \dd \lambda \,
        \frac{\lambda \sinh(\pi \lambda)}{64\pi^{d + 2}}
        \frac{\Gamma\left(\frac{d - 2 \pm 2i\lambda}{4}\right)^2 \Gamma\left(\frac{d + 2 + 4i\nu \pm 2i\lambda}{4}\right)^2}
        {\Gamma (1+i \nu )^2 \Gamma \left(\frac{d}{2}+i \nu \right)^2} 
        \mathcal{F}_{\lambda, 1}^{\{\Delta_{\nu}\}}(u,v) \,,
        \end{aligned}
    \end{equation}
    Note that the integrand is identical to the Dirichlet photon exchange studied in \eqref{eq:Dirichlet-current-photon-current-transverse}, and it displays a pole-singularity for \(\lambda = \pm i \left(\frac{d}{2} - 1\right)\) due to the presence of \(\Gamma\left(\frac{d - 2 \pm 2i\lambda}{4}\right)^2\), while it decays as \(\ee^{-\pi\lambda}\) for large \(\lambda\). 

\paragraph{Geodesic-partial amplitude.}
Having computed the standard exchange diagram, the remaining task is to compute diagrams involving geodesic Wilson lines. Using the split representation, the computation of such diagrams reduces to evaluating the diagram in figure \ref{fig:geodesic-partial-amplitude}, which we call {\it the geodesic partial amplitude} and denote by \(\varphi^{(1)}_{\frac{d}{2} + i \lambda} (P_1, P_2; P_5)\)  following the notation in \cite{hijano2016witten}. We will be performing a computation very similar to the one done in \cite{Chen:2017yia}, although its authors end up computing a rather different quantity from the one we are interested in.

\begin{figure}
    \centering
    \begin{tikzpicture}[scale = 0.8]
        \node[black, left] at (-3, 0) {\(\varphi^{(1)}_{\frac{d}{2} + i \lambda} (P_1, P_2; P_5) = \)};
        \begin{scope}
            \draw[black] (0,0) circle (2.7);
            \draw[green, very thick]  (140:2.7) arc (40:-40:2.7);
            \draw[turquoise, ultra thick, decorate,decoration={coil,aspect=0, pre length=0.07cm}]  (-1.4, 0) -- (2.7, 0);
            \node[turquoise, above] at (0.5, 0) {\(\left(\textcolor{black}{\frac{d}{2} + i \lambda, J = 1}\right)\)};
            \node[fuchsia, left] at (140:2.7) {\(P_1\)};
            \node[fuchsia, left] at (220:2.7) {\(P_2\)};
            \node[turquoise, right] at (0:2.7) {\(P_5\)};
        \end{scope}
    \end{tikzpicture}
    \caption{Adopting the split representation of spin-1 harmonic function, it becomes necessary to compute the spin-1 propagator from a boundary point \(P_5\) to a point integrated over the geodesic connecting \(P_1\) and \(P_2\).}
    \label{fig:geodesic-partial-amplitude}
\end{figure}

Notice that the spin-\(1\) partial amplitude \(\varphi^{(1)}_{\frac{d}{2} + i \lambda} (P_1, P_2; P_5)\) carries one free boundary index, due to the spin \(1\) bulk-to-boundary propagator. When lifted to embedding space, we contract such index with a generic polarization vector \(Z\) on the boundary,
\begin{equation}
    \begin{aligned}
        \varphi^{(1)}_{\frac{d}{2} + i \lambda} (P_1, P_2; P_5; Z) 
        &= \int_{P_1}^{P_2} \dd X^{A} \, \Pi^{(1)}_{\frac{d}{2} + i \lambda, AB}(X, P_5) Z^B\\
        &= \int_{-\infty}^{+\infty} \dd \sigma \, \Pi^{(1)}_{\frac{d}{2} + i \lambda}\left(X(\sigma), P_5; \frac{dX}{d\sigma}(\sigma), Z\right)
    \end{aligned}
\end{equation}
The line integral between \(P_1\) and \(P_2\) is performed along the geodesic connecting them; such geodesic can be parameterized as
\begin{equation}
    X(\sigma) = \frac{P_1 \, e^{-\sigma} + P_2 \, e^{+\sigma}}{\sqrt{-2 P_1 \cdot P_2}} \qquad \text{with} \qquad \sigma \in \R.
\end{equation}

Using the explicit expression for the bulk-to-boundary spin-1 propagator given in (3.39) of \cite{costa2014spinning}, we obtain the following expression for $\varphi^{(1)}_{\frac{d}{2}+i\lambda}$, with the  normalization in \eqref{eq:bulk-to-boundary-propagator-normalization}:
\begin{align}
    \label{eq:phi-geodesic-integral}
    \begin{split}
        & \varphi^{(1)}_{\frac{d}{2} + i \lambda} (P_1, P_2; P_5; Z) \\
        & \ =  \sqrt{\mathcal{C}_{\lambda}^{(1)}}
        \int_{-\infty}^{+\infty} \dd \sigma \, 
        \frac{\left[\left(-2 X(\sigma)\cdot P_5\right)\left(\frac{dX(\sigma)}{d\sigma}\cdot Z\right) 
        - \left(X(\sigma)\cdot Z\right)\left(-2P_5\cdot\frac{dX(\sigma)}{d\sigma} \right)\right]}{\left(-2 X(\sigma) \cdot P_5\right)^{\frac{d}{2} + i \lambda + 1}} \\
        & \ = 2 \sqrt{\mathcal{C}_{\lambda}^{(1)}}
        \frac{\left[\left(P_2\cdot Z\right)P_{15} - \left(P_1\cdot Z\right)P_{25}\right]}
        {(-2 P_1 \cdot P_2)^{\frac{1 - \Delta_{\lambda}}{2}}}
        \int_{-\infty}^{+\infty} \dd \sigma \, 
        \frac{1}{\left[-2 P_1 \cdot P_5 \, e^{-\sigma} - 2P_2 \cdot P_5 \, e^{\sigma} \right]^{1 + \Delta_{\lambda}}} \,, 
    \end{split}
\end{align}
The integral in \eqref{eq:phi-geodesic-integral} can be computed via a simple substitution:
\begin{equation}
    \begin{aligned}
        \int_{-\infty}^{+\infty} \dd \sigma \, \frac{1}{\left[A \, e^{-\sigma} + B \, e^{\sigma} \right]^{1 + \Delta_{\lambda}}} 
        &= \frac{1}{\left(A B\right)^{\frac{1 + \Delta_{\lambda}}{2}}} 
        \int_{-\infty}^{+\infty} \dd \sigma \, \frac{1}
        {\left[\sqrt{\frac{A}{B}}e^{-\sigma} 
        + \sqrt{\frac{B}{A}} e^{\sigma} \right]
        ^{1 + \Delta_{\lambda}}} \\
        &= \frac{1}{\left(A B\right)^{\frac{1 + \Delta_{\lambda}}{2}}} \int_{0}^{+\infty} \frac{\dd x}{x} \frac{1}{\left[x + \frac{1}{x}\right]^{1 + \Delta_{\lambda}}} \\
        &= \frac{1}{\left(A B\right)^{\frac{1 + \Delta_{\lambda}}{2}}} 
       \frac{1}{2} \frac{\Gamma\left(\frac{\Delta_{\lambda} + 1}{2}\right)^2}{\Gamma\left(\Delta_{\lambda} + 1\right)}\,.
    \end{aligned}
\end{equation}
So, finally, we have
\begin{equation}
    \label{eq:phi-geodesic-final}
    \begin{aligned}
    \varphi^{(1)}_{\frac{d}{2} + i \lambda} (P_1, P_2; P_5; Z)
    &= \underbrace{-\frac{\Gamma\left(\frac{\Delta_{\lambda} + 1}{2}\right)^2}{\Gamma\left(\Delta_{\lambda} + 1\right)}}_{b_g(\lambda)} \sqrt{\mathcal{C}_{\lambda}^{(1)}} \\
    &\hspace{1cm} 
    \frac{\left[\left(P_1\cdot Z\right)P_{25} - \left(P_2\cdot Z\right)P_{15}\right]}
    {(-2 P_1 \cdot P_2)^{\frac{1 - \Delta_{\lambda}}{2}} 
    (-2 P_1 \cdot P_5)^{\frac{1 + \Delta_{\lambda}}{2}}
    (-2 P_2 \cdot P_5)^{\frac{1 + \Delta_{\lambda}}{2}}} \\
    \vspace{0.1cm}\\
    &= \sqrt{\mathcal{C}_{\lambda}^{(1)}} b_g(\lambda) \cdot \langle \mathcal{O}_{\Delta = 0}^{J = 0}(P_1) \mathcal{O}_{\Delta = 0}^{J = 0}(P_2) \mathcal{O}_{\Delta_{\lambda}}^{J = 1}(P_5; Z)\rangle_1.
    \end{aligned}
\end{equation}
In the last line of \eqref{eq:phi-geodesic-final}, we rewrote it in terms of the kinematic structure of the three-point function of two scalar operators and a spin-1 operator. 
Indeed, such a rewriting was expected, since \(\varphi^{(1)}_{\frac{d}{2} + i \lambda} (P_1, P_2; P_5; Z)\) is a conformally covariant object, and under scaling it behaves as
\[
\begin{cases}
    \varphi^{(1)}_{\frac{d}{2} + i \lambda} (\alpha P_1, P_2; P_5; Z) = \varphi^{(1)}_{\frac{d}{2} + i \lambda} (P_1, \alpha P_2; P_5; Z) = \varphi^{(1)}_{\frac{d}{2} + i \lambda} (P_1, P_2; P_5; Z) &\implies  \Delta_1 = \Delta_2 = 0,\\
    \varphi^{(1)}_{\frac{d}{2} + i \lambda} (P_1, P_2; \alpha P_5; Z) = \alpha^{-\Delta_{\lambda} - 1} \varphi^{(1)}_{\frac{d}{2} + i \lambda} (P_1, P_2; P_5; Z) &\implies \Delta_3 = \Delta_{\lambda}.
\end{cases}    
\]
So \(\varphi^{(1)}_{\frac{d}{2} + i \lambda} (P_1, P_2; P_5; Z)\) needs to be of the form of a three-point function of two scalar operators with \(\Delta = 0\) and a spin-1 operator with \(\Delta = \Delta_{\lambda}\), which is completely fixed by conformal invariance.


\paragraph{Geodesic-photon-geodesic exchange.}
Now we are finally equipped with all the tools necessary to compute the diagrams with geodesic Wilson lines. We start with analyzing the geodesic-photon-geodesic exchange.

As for the longitudinal part, the matter insertions factorize out, as depicted in figures \ref{fig:factorization-geodesic-photon-geodesic} and 
\ref{fig:CFT-4-point-function-normalization}. 
Decomposing the spin-1 propagator using the split representation, we obtain
\begin{equation}
    \begin{aligned}
        \mathcal{A}_{gg}^{\perp} &= (-ie)^2 \left(\frac{1}{P_{12}}  \frac{1}{P_{34}}\right)^{\frac{d}{2} + i\nu} 
        \frac{1}{\pi \left(\frac{d}{2} - 1\right)} 
        \int_{\R \oplus \substack{\circlearrowright \\ \circlearrowleft}}\dd \lambda \, \frac{\lambda^2}{\lambda^2 + \left(\frac{d}{2} - 1\right)^2}  \sqrt{\mathcal{C}_{\lambda}^{(1)}\mathcal{C}_{-\lambda}^{(1)}}\\
         & \qquad \int_{\partial AdS} \dd^d P_5 \, 
        \varphi^{(1)}_{\frac{d}{2} + i \lambda} (P_1, P_2; P_5; D_Z) 
        \varphi^{(1)}_{\frac{d}{2} - i \lambda} (P_3, P_4; P_5; Z)\\
        & = (-ie)^2 \left(\frac{1}{P_{12}}  \frac{1}{P_{34}}\right)^{\frac{d}{2} + i\nu} 
        \frac{1}{\pi \left(\frac{d}{2} - 1\right)} 
        \int_{\R \oplus \substack{\circlearrowright \\ \circlearrowleft}}\dd \lambda \, \frac{\lambda^2}{\lambda^2 + \left(\frac{d}{2} - 1\right)^2}
        \mathcal{C}_{\lambda}^{(1)}\mathcal{C}_{- \lambda}^{(1)}
        b_g(\lambda) b_g(-\lambda)  \\
        & \qquad\!\int_{\partial AdS} \dd^d P_5 \, \langle \mathcal{O}_{\Delta = 0}^{J = 0}(P_1) \mathcal{O}_{\Delta = 0}^{J = 0}(P_2) \mathcal{O}_{\Delta_{\lambda}}^{J = 1}(P_5; D_Z)\rangle_1 
        \langle \mathcal{O}_{\Delta = 0}^{J = 0}(P_3) \mathcal{O}_{\Delta = 0}^{J = 0}(P_4) \mathcal{O}_{\Delta_{-\lambda}}^{J = 1}(P_5; Z)\rangle_1 \,. \mspace{-55mu}   \\
    \end{aligned}
\end{equation}
As is well known, the boundary convolution of two three-point structures gives a conformal partial wave of the spin of the exchanged operator (see \((3.1)\) of \cite{simmons2018spacetime} and \eqref{eq:3p-function-to-partial-waves} for the normalization constant), so we have
\begin{equation}
    \label{eq:geodesic-photon-geodesic-transverse-final}
    \begin{aligned}
        \mathcal{A}_{gg}^{\perp} = \frac{(-ie)^2}{\pi} \left(\frac{1}{P_{12}} \, \frac{1}{P_{34}}\right)^{\frac{d}{2} + i\nu} \, 
        \int_{\R \oplus \substack{\circlearrowright \\ \circlearrowleft}}\dd \lambda \, \frac{\lambda^2}{\lambda^2 + \left(\frac{d}{2} - 1\right)^2}
        \mathcal{C}_{\lambda}^{(1)}\mathcal{C}_{- \lambda}^{(1)}
        b_g(\lambda) b_g(-\lambda) \,
        \mathcal{F}_{\lambda, 1}^{\{\Delta_i = 0\}}(u, v) \,.
    \end{aligned}
\end{equation}
As we remarked at the end of section \ref{subsec:current-photon-current-diagram-transverse}, \(\mathcal{F}_{\lambda, 0}^{\{\Delta_i\}}(u, v)\) only depends on \(\Delta_1 - \Delta_2\) and \(\Delta_3 - \Delta_4\), so \(\mathcal{F}_{\lambda, 1}^{\{\Delta_i = 0\}}(u, v) = \mathcal{F}_{\lambda, 1}^{\{\Delta_{\nu}\}}(u, v)\), the same partial waves appearing in \eqref{eq:Neumann-current-photon-current-transverse}, the current exchange with equal external legs.

\paragraph{Geodesic-photon-current exchange.}
We next study the geodesic-photon-current exchange, see right diagram in figure \ref{fig:geodesic-witten-diagrams}, that is
\begin{equation}
    \begin{aligned}
        \mathcal{A}_{gc}^{\perp} &= (-ie)^2 \left(\frac{1}{P_{12}}\right)^{\frac{d}{2} + i\nu} 
        \frac{1}{\pi \left(\frac{d}{2} - 1\right)} \,
        \int_{\R \oplus \substack{\circlearrowright \\ \circlearrowleft}}\dd \lambda \, \frac{\lambda^2}{\lambda^2 + \left(\frac{d}{2} - 1\right)^2} \sqrt{\mathcal{C}_{\lambda}^{(1)}\mathcal{C}_{-\lambda}^{(1)}}\\
        & \qquad \int_{\partial AdS} \dd^d P_5 \, 
        \varphi^{(1)}_{\frac{d}{2} + i \lambda} (P_1, P_2; P_5; D_Z) \,\\
        & \qquad \int_{AdS} \dd^{d + 1} X_2 \, \frac{1}{\left(\frac{d - 1}{2}\right)}\Pi^{(1)}_{\frac{d}{2} - i \lambda}(X_2, P_5; Z, K_2) W_{2,B} T_{34}^B(P_3, P_4; X_2) \,.
    \end{aligned}
\end{equation}
Plugging in the explicit expressions of the building blocks, as derived in equations \eqref{eq:phi-geodesic-final} and \eqref{eq:3-point-function}, we get
\begin{equation}
    \label{eq:geodesic-photon-current-exchange-3point-functions}
    \begin{aligned}
        \mathcal{A}_{gc}^{\perp} = 2(ie)^2 \sqrt{\mathcal{C}_{\nu_3}^{(0)}\mathcal{C}_{\nu_4}^{(0)}} \left(\frac{1}{P_{12}}\right)^{\frac{d}{2} + i\nu} 
        \frac{1}{\pi \left(\frac{d}{2} - 1\right)} \,
        \int_{\R \oplus \substack{\circlearrowright \\ \circlearrowleft}}\dd \lambda \, \frac{\lambda^2 \, \mathcal{C}_{\lambda}^{(1)}\mathcal{C}_{-\lambda}^{(1)}}{\lambda^2 + \left(\frac{d}{2} - 1\right)^2} b_g(\lambda)b_{\text{Bulk}}(\nu_3, \nu_4, -\lambda, 1)\\
        \, \int_{\partial AdS} \dd^d P_5 \, 
        \Big\langle \mathcal{O}_{\Delta = 0}^{J = 0}(P_1) \mathcal{O}_{\Delta = 0}^{J = 0}(P_2) \mathcal{O}_{\Delta_{\lambda}}^{J = 1}(P_5; D_Z)\Big\rangle _1
        \Big\langle \mathcal{O}_{\Delta_{-\lambda}}^{J = 1}(P_5; Z) \mathcal{O}_{\Delta = \frac{d}{2} + i \nu_3}^{J = 0}(P_3) \mathcal{O}_{\Delta = \frac{d}{2} + i \nu_4}^{J = 0}(P_4) \Big\rangle _1\\
    \end{aligned}
\end{equation}
Since we are considering matter insertions with the same conformal dimension \(\nu_3 = \nu_4 = \nu\), we can use again equation (6.3) of \cite{costa2014spinning} to simplify the expression \eqref{eq:geodesic-photon-current-exchange-3point-functions} as
\begin{equation}
    \label{eq:geodesic-photon-current-transverse}
    \begin{aligned}
        \mathcal{A}_{gc}^{\perp} &= 2 \frac{(-ie)^2}{\pi} \mathcal{C}_{\nu}^{(0)} \left(\frac{1}{P_{12}}\, \frac{1}{P_{34}}\right)^{\frac{d}{2} + i\nu}  \, \\
        \vspace{3.7cm}
        & \qquad\int_{\R \oplus \substack{\circlearrowright \\ \circlearrowleft}}\dd \lambda \, \frac{\lambda^2}{\lambda^2 + \left(\frac{d}{2} - 1\right)^2}\mathcal{C}_{\lambda}^{(1)}\mathcal{C}_{-\lambda}^{(1)} \, b_g(\lambda)b_{\text{Bulk}}(\nu, \nu, -\lambda, 1) \, 
        \mathcal{F}_{\lambda, 1}^{\{\Delta_{1,2} = 0, \Delta_{3,4} = \Delta_{\nu}\}}(u,v) \,.
    \end{aligned}
\end{equation}
Once again, the only external-leg dependence of \(\mathcal{F}_{\lambda, 1}^{\{\Delta_{1,2} = \, 0, \Delta_{3,4} = \Delta_{\nu}\}}(u,v)\) is \(\Delta_1 - \Delta_2\) and \(\Delta_3 - \Delta_4\), so \(\mathcal{F}_{\lambda, 1}^{\{\Delta_{1,2} = \, 0, \Delta_{3,4} = \Delta_{\nu}\}}(u,v)\) is exactly the same function as \(\mathcal{F}_{\lambda, 1}^{\{\Delta_i =  \Delta_{\nu}\}}(u,v)\).

\paragraph{The full transverse Neumann amplitude.}
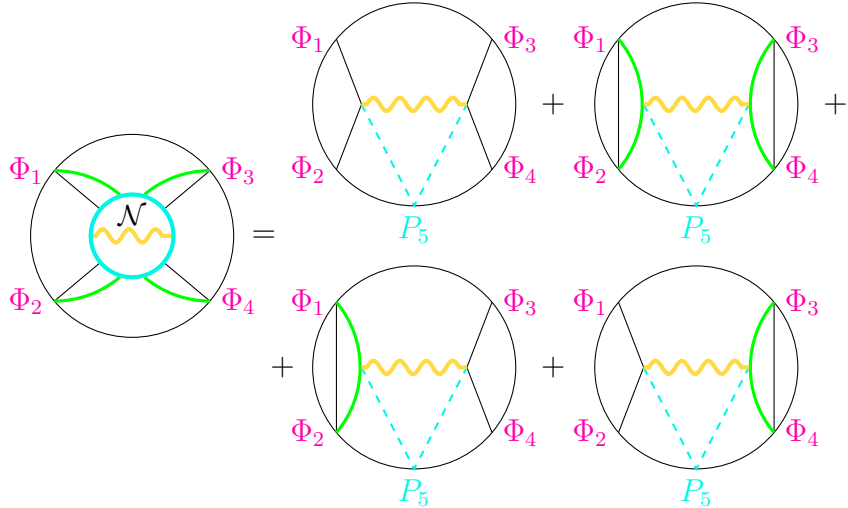
\begin{figure}[!h]
    \centering
    \begin{tikzpicture}[scale = 0.5]
        \begin{scope}[xshift = -3.5cm]
            \draw[black] (0,0) circle (2.7);
            \draw[black]  (140:2.7) -- (140:1.1);
            \draw[green, very thick]  (140:2.7) arc (90:50:2.7);
            \draw[black]  (220:1.1) -- (220:2.7);
            \draw[green, very thick]  (220:2.7) arc (-90:-50:2.7);
            \draw[black]  (40:2.7) -- (40:1.1);
            \draw[green, very thick]  (40:2.7) arc (90:130:2.7);
            \draw[black]  (-40:2.7) -- (-40:1.1);
            \draw[green, very thick]  (-40:2.7) arc (-90:-130:2.7);
            \draw[goldenyellow, ultra thick, decorate,decoration={coil,aspect=0, pre length=0.07cm}]  (-1.1, 0) -- (1.1, 0);
            \node[black, above] at (0,0) {\small \(\mathcal{N}\)};
            \draw[turquoise, ultra thick] (0,0) circle (1.1);
            \node[fuchsia, left] at (140:2.7) {\(\Phi_1\)};
            \node[fuchsia, left] at (220:2.7) {\(\Phi_2\)};
            \node[fuchsia, right] at (40:2.7) {\(\Phi_3\)};
            \node[fuchsia, right] at (-40:2.7) {\(\Phi_4\)};
        \end{scope}
        \node[black] at (0, 0) {\(=\)};
        \begin{scope}[xshift = 4cm, yshift=3.5cm]
            \draw[black] (0,0) circle (2.7);
            \draw[black]  (140:2.7) -- (-1.4, 0);
            \draw[black]  (220:2.7) -- (-1.4, 0);
            \draw[black]  (40:2.7) -- (1.4, 0);
            \draw[black]  (-40:2.7) -- (1.4, 0);
            \draw[turquoise, thick, dashed]  (-1.4, 0) -- (270:2.7) -- (1.4, 0);
            \draw[goldenyellow, ultra thick, decorate,decoration={coil,aspect=0, pre length=0.07cm}]  (-1.4, 0) -- (1.4, 0);
            \node[fuchsia, left] at (140:2.7) {\(\Phi_1\)};
            \node[fuchsia, left] at (220:2.7) {\(\Phi_2\)};
            \node[fuchsia, right] at (40:2.7) {\(\Phi_3\)};
            \node[fuchsia, right] at (-40:2.7) {\(\Phi_4\)};
            \node[turquoise, below] at (270:2.7) {\(P_5\)};
        \end{scope}
        \node[black] at (7.7, 3.5) {\(+\)};
        \begin{scope}[xshift = 11.5cm, yshift=3.5cm]
            \draw[black] (0,0) circle (2.7);
            \draw[black]  (140:2.7) -- (220:2.7);
            \draw[green, very thick]  (140:2.7) arc (40:-40:2.7);
            \draw[black]  (40:2.7) -- (-40:2.7);
            \draw[green, very thick]  (40:2.7) arc (140:220:2.7);
            \draw[turquoise, thick, dashed]  (-1.4, 0) -- (270:2.7) -- (1.4, 0);
            \draw[goldenyellow, ultra thick, decorate,decoration={coil,aspect=0, pre length=0.07cm}]  (-1.4, 0) -- (1.4, 0);
            \node[fuchsia, left] at (140:2.7) {\(\Phi_1\)};
            \node[fuchsia, left] at (220:2.7) {\(\Phi_2\)};
            \node[fuchsia, right] at (40:2.7) {\(\Phi_3\)};
            \node[fuchsia, right] at (-40:2.7) {\(\Phi_4\)};
            \node[turquoise, below] at (270:2.7) {\(P_5\)};
        \end{scope}
        \node[black] at (15.2, 3.5) {\(+\)};
        \node[black] at (0.5, -3.5) {\(+\)};
        \begin{scope}[xshift = 4cm, yshift=-3.5cm]
            \draw[black] (0,0) circle (2.7);
            \draw[black]  (140:2.7) -- (220:2.7);
            \draw[green, very thick]  (140:2.7) arc (40:-40:2.7);
            \draw[black]  (40:2.7) -- (1.4, 0);
            \draw[black]  (-40:2.7) -- (1.4, 0);
            \draw[turquoise, thick, dashed]  (-1.4, 0) -- (270:2.7) -- (1.4, 0);
            \draw[goldenyellow, ultra thick, decorate,decoration={coil,aspect=0, pre length=0.07cm}]  (-1.4, 0) -- (1.4, 0);
            \node[fuchsia, left] at (140:2.7) {\(\Phi_1\)};
            \node[fuchsia, left] at (220:2.7) {\(\Phi_2\)};
            \node[fuchsia, right] at (40:2.7) {\(\Phi_3\)};
            \node[fuchsia, right] at (-40:2.7) {\(\Phi_4\)};
            \node[turquoise, below] at (270:2.7) {\(P_5\)};
        \end{scope}
        \node[black] at (7.7, -3.5) {\(+\)};
        \begin{scope}[xshift = 11.5cm, yshift=-3.5cm]
            \draw[black] (0,0) circle (2.7);
            \draw[black]  (140:2.7) -- (-1.4, 0);
            \draw[black]  (220:2.7) -- (-1.4, 0);
            \draw[black]  (40:2.7) -- (-40:2.7);
            \draw[green, very thick]  (40:2.7) arc (140:220:2.7);
            \draw[turquoise, thick, dashed]  (-1.4, 0) -- (270:2.7) -- (1.4, 0);
            \draw[goldenyellow, ultra thick, decorate,decoration={coil,aspect=0, pre length=0.07cm}]  (-1.4, 0) -- (1.4, 0);
            \node[fuchsia, left] at (140:2.7) {\(\Phi_1\)};
            \node[fuchsia, left] at (220:2.7) {\(\Phi_2\)};
            \node[fuchsia, right] at (40:2.7) {\(\Phi_3\)};
            \node[fuchsia, right] at (-40:2.7) {\(\Phi_4\)};
            \node[turquoise, below] at (270:2.7) {\(P_5\)};
        \end{scope}
    \end{tikzpicture}
    \caption{The \(4-\)point interaction with the exchange of a Neumann photon consists of \(4\) contributions: current-current, geodesic-geodesic, current-geodesic and geodesic-current exchanges, mediated by a photon with Neumann boundary conditions.}
    \label{fig:4-point-neumann-exchnge-decomposition}
\end{figure}
We are finally able to compute the full amplitude for the Neumann photon exchange in scalar QED in AdS, with \(4\) equal matter insertions at the boundary. 
At the end of section \ref{subsec:wilson-line-prescription-longitudinal} we proved that the amplitude is gauge invariant. Hence its longitudinal part is exactly zero and it is given purely by the transverse piece.

Putting together equations \eqref{eq:Neumann-current-photon-current-transverse}, \eqref{eq:phi-geodesic-final} and \eqref{eq:geodesic-photon-current-transverse} 
\begin{equation}
    \label{eq:Neumann-4-point-amplitude}
    \begin{aligned}
        \mathcal{A}_{\mathcal{N}} &= \mathcal{A}_{cc}^{\perp}+ \mathcal{A}_{gg}^{\perp} + \mathcal{A}_{gc}^{\perp} + \mathcal{A}_{cg}^{\perp} \\
        &= \frac{(-ie)^2}{\pi} \left(\frac{1}{P_{12}}\, \frac{1}{P_{34}}\right)^{\frac{d}{2} + i\nu}  \,
        \int_{\R \oplus \substack{\circlearrowright \\ \circlearrowleft}}\dd \lambda \, \frac{\lambda^2}{\lambda^2 + \left(\frac{d}{2} - 1\right)^2} \mathcal{C}_{\lambda}^{(1)}\mathcal{C}_{-\lambda}^{(1)}  \, \mathcal{F}_{\lambda, 1}^{\{\Delta_i =  \Delta_{\nu}\}}(u,v) \\
        \vspace{0.7cm}
        & \mspace{120mu}\Big[
            4\left(\mathcal{C}_{\nu}^{(0)}\right)^2b_{\text{Bulk}}(\nu, \nu, \lambda, 1)b_{\text{Bulk}}(\nu, \nu, -\lambda, 1) + b_g(\lambda)b_g(-\lambda) 
            \,  \\
            & \mspace{150mu} + 2 \mathcal{C}_{\nu}^{(0)}b_g(\lambda)b_{\text{Bulk}}(\nu, \nu, -\lambda, 1) 
            + 2\mathcal{C}_{\nu}^{(0)}b_{\text{Bulk}}(\nu, \nu, \lambda, 1) b_g(-\lambda) 
        \Big] \\
            & = (ie)^2 \left(\frac{1}{P_{12}}\, \frac{1}{P_{34}}\right)^{\frac{d}{2} + i\nu} \, \int_{\R \oplus \substack{\circlearrowright \\ \circlearrowleft}}\dd \lambda \, \rho_{\mathcal{N}}^{J = 1}(\lambda; \{\tfrac{d}{2} + i\nu\})  \, \mathcal{F}_{\lambda, 1}^{\{\Delta_{\nu}\}}(u,v)\,,
    \end{aligned}
\end{equation}
where the full spectral density \(\rho_{\mathcal{N}}^{J = 1}(\lambda; \{\tfrac{d}{2} + i\nu\})\) is given by
\begin{equation}
    \label{eq:rho-Neumann-AdS}
   \begin{split}
        \rho_{\mathcal{N}}^{J = 1}(\lambda; \{\tfrac{d}{2} + i\nu\}) &= \Pi^{(1), \perp}_{1}(\lambda) \, \mathcal{Q}^{J = 1}_{\text{Dressed}}(\lambda, \{\tfrac{d}{2} + i\nu\}), \\
        \Pi^{(1), \perp}_{1} &= \frac{1}{\lambda^2 + \left(\tfrac{d}{2} - 1\right)^2}, \\
        \mathcal{Q}^{J = 1}_{\text{Dressed}}(\lambda, \{\tfrac{d}{2} + i\nu\}) &=
        \frac{\lambda \sinh(\pi \lambda)}{4\pi^{d + 2}} 
        \frac{1}{\lambda^2 + \left(\frac{d}{2} - 1\right)^2}
        \frac{\Gamma\left(\frac{d + 2 \pm 2i\lambda}{4}\right)^2}
        {\Gamma (1+i \nu )^2 \Gamma \left(\frac{d}{2}+i \nu \right)^2} \,\\
        & \qquad\left[
            \Gamma\left(\frac{d + 2 + 4i\nu \pm 2i\lambda}{4}\right)
            - \Gamma (1+i \nu ) \Gamma \left(\frac{d}{2}+i \nu \right)
        \right]^2. \\
    \end{split}
    \end{equation}
From this expression, one can also obtain the conformal block expansion by expressing the conformal partial wave in terms of conformal blocks, through equation \eqref{eq:conformal-partial-waves-2-conformal-blocks}, specifying the coefficient to spin \(J = 1\):
\begin{equation}
    \mathcal{K}_{d - \Delta}^{(1)} = 
    -\frac{\pi^{\frac{d}{2}}\Gamma\left(\frac{d}{2} - \Delta\right)
    \Gamma\left(d - \Delta\right)\Gamma\left(\frac{\Delta + 1}{2}\right)^2}{2 \Gamma\left(d - \Delta - 1\right)\Gamma\left(1 + \Delta\right)
    \Gamma\left(\frac{1 + d - \Delta}{2}\right)^2}.
\end{equation}
In particular, we have
\begin{equation}
    \label{eq:density-blocks-Neumann-AdS}
    \begin{aligned}
        \rho_{\mathcal{N}}^{J = 1}(\lambda; \{\tfrac{d}{2} + i\nu\})\mathcal{K}_{d - \Delta_{\lambda}}^{(1)} = 
        -&\frac{\lambda \sinh(\pi \lambda)}{2^{5 + \frac{d}{2} + i\lambda}\pi^{\frac{d + 3}{2}}}
        \,
        \frac{\Gamma\left(\frac{d - 2 + 2i\lambda}{4}\right)^2 
        }
        {\Gamma (1+i \nu )^2 \Gamma \left(\frac{d}{2}+i \nu \right)^2} \, 
        \frac{
            \Gamma\left(-i\lambda\right)
            \Gamma\left(\frac{d + 2 + 2i\lambda}{4}\right)}
        {\left(\frac{d}{2} - 1 - i\lambda\right)
        \Gamma\left(\frac{d + 4 + 2i\lambda}{4}\right)}\\
        & \qquad\left[
            \Gamma\left(\frac{d + 2 + 4i\nu \pm 2i\lambda}{4}\right)
            - \Gamma (1+i \nu ) \Gamma \left(\frac{d}{2}+i \nu \right)
        \right]^2 \,.
    \end{aligned}
\end{equation}
To read off the conformal block expansion, we close the $\lambda$ contour in the lower (upper) half plane for the piece proportional to $\hat{G}^{(1)}_{\Delta_{\lambda}}(u,v)$ ($\hat{G}^{(1)}_{d-\Delta_{\lambda}}(u,v)$). Since the two pieces produce identical answers, we focus on the piece proportional to $\hat{G}^{(1)}_{\Delta_{\lambda}}(u,v)$. Let us first make a few technical comments:
\begin{itemize}
    \item First, the factor \(\Gamma\left(\frac{d - 2 + 2i\lambda}{4}\right)^2\Gamma\left(\frac{d + 2 + 2i\lambda}{4}\right) / (\frac{d}{2}-1-i \lambda)\) has a double pole at $\lambda = i (\frac{d}{2}-1)$, a simple pole at $\lambda=-i(\frac{d}{2}-1)$, and third-order poles at \(\lambda = i \left(\frac{d}{2} - 1 + 2k\right)\), \(\forall k \in \mathbb{N}_{> 0}\). For $d> 2$, the third-order poles do not contribute since they are in the upper half plane. The simple pole at $\lambda =-i(\frac{d}{2}-1)$ is in the lower half plane, but because of the contour choice for the Neumann photon exchange $\R \oplus \substack{\circlearrowright \\ \circlearrowleft}$, it does not contribute either when closing the contour. In contrast, the double pole $\lambda= i (\frac{d}{2}-1)$ can in principle contribute, despite being on the upper half plane, and has an important physical consequence. We will discuss it separately below. 
    \item Second, since \(\Gamma\left(-i\lambda\right)\sinh(\pi \lambda)\) is regular everywhere, it does not give rise to extra pole in the lower-half plane. 
    \item Third, the term \(\Gamma\left(\frac{d + 2 + 4i\nu \pm 2i\lambda}{4}\right)\) inside the square brackets comes from a standard exchange Witten diagram and gives rise to double-trace operators of external scalars. This can also be also verified by comparing the result with \eqref{eq:Dirichlet-current-photon-current-transverse}. On the other hand, the other term inside the brackets, factor \(- \Gamma (1+i \nu ) \Gamma \left(\frac{d}{2}+i \nu \right)\), comes from the geodesic diagrams. 
    \item Fourth, in the large \(\lambda\) limit the contribution from geodesic diagrams grows polynomially in $\lambda$, as is the case with free fields in AdS or equivalently mean field theory, (see (3.118) of \cite{karateev2019harmonic}, which uses the same normalization as us for conformal blocks and partial waves). Note that this asymptotic behavior is exponentially larger than that of the standard exchange Witten diagram. 
    \end{itemize}

Let us now turn to the fate of the double pole at $\lambda=i(\frac{d}{2}-1)$ mentioned above. The first thing to notice is that, if it were not for the Wilson line dressing, it would give rise to nontrivial contributions in the conformal block expansion as can be seen from below
\begin{equation}
        \begin{aligned}
        \rho_{\mathcal{N} , {\rm bare}}^{J = 1}&(\lambda; \{\tfrac{d}{2} + i\nu\}) = \Pi^{(1), \perp}_{1}(\lambda) \, \mathcal{Q}^{J = 1}(\lambda, \{\tfrac{d}{2} + i\nu\}), \\
        \rho_{\mathcal{N} , {\rm bare}}^{J = 1}&(\lambda; \{\tfrac{d}{2} + i\nu\})\,\mathcal{K}_{d-\Delta_{\lambda}}^{(J = 1)}
        \underset{\lambda \to  i \left(\frac{d}{2} - 1\right)}{\sim}-\frac{1}
        {16 \pi^{\frac{d}{4}} \Gamma\left(2-\frac{d}{2}\right)\left(\lambda -i\left(\frac{d}{2} -1\right)\right)^2} \,  \\
        & \mspace{50mu}+\frac{i   \left( \psi ^{(0)}\left(\frac{d}{2}+i \nu \right)-\psi ^{(0)}(1+i \nu)+\psi ^{(0)}\left(1-\frac{d}{2}\right)+\gamma_E +\frac{d-3}{d-2}\right)}
        {16 \pi^{\frac{d}{4}} \Gamma\left(2-\frac{d}{2}\right)\left(\lambda -i\left(\frac{d}{2} -1\right)\right)}
        + \mathcal{O}\left(1\right)\,.
        \end{aligned}
    \end{equation}
However, such contributions are incompatible with unitarity since they correspond to an exchange of a gauge field $a_{j}$, which has spin 1 and $\Delta =1$, and violates the unitarity bound for $d> 2$. Fortunately, the Wilson line contributions exactly cancel them since the term in the square bracket \(\left[\Gamma\left(\frac{d + 2 + 4i\nu \pm 2i\lambda}{4}\right)- \Gamma (1+i \nu ) \Gamma \left(\frac{d}{2}+i \nu \right)
    \right]^2\) has a double zero precisely at $\lambda =i(\frac{d}{2}-1)$. The fact that the Wilson line dressing which cancels the longitudinal part of the amplitude, also cancels this contribution is a very nice sanity-check of our prescription\footnote{It would be interesting to check if this property also holds regardless of the path Wilson lines are sent along.}. 

In fact, something stronger is happening here. Our computation shows that, not only the gauge field $a_j$ but also the field strength $f_{ij}=\partial_i a_j -\partial_j a_i$ is absent in the conformal block expansion. The field strength has $\Delta=2$ and is above the unitarity bound in $d=3,4$, thus arguments of gauge invariance and unitarity only do not explain the decoupling of $f_{ij}$. As we will see later, this decoupling of $f_{ij}$ is a consequence of the conformal symmetry and higher-form symmetry in AdS, and it therefore continues to hold even at the non-perturbative level. The same symmetry ensures that the dimension of $f_{ij}$ is protected. See section \ref{subsec:decoupling_of_F_from_symmetry} for details.

%% file: 1-loop-self-energy.tex
 In this section we compute the one loop self-energy of the photon field in AdS. As before, we consider the coupling of the photon field with two real scalars, of equal mass and charge, but possibly different boundary conditions. We find that in the case of different boundary conditions a mass for the photon is induced radiatively, and we compute it explicitly at the one-loop level.

We interpret this fact as a particular type of {\it Higgs mechanism} in AdS, occurring in the presence of boundary conditions that are not invariant under the action of the gauge transformation on the matter fields, triggering a SSB  in AdS. This type of Higgs mechanism has been studied in \cite{Rattazzi:2009ux} in the case of massless fermionic matter, and more recently in \cite{Sadekov:2023ivd} for conformally-coupled scalar matter, and in \cite{Sleight:2025dmt} for more general gauge theories, including scalar QED with generic masses, Yang-Mills and gravity. As far as we know, the formula for the mass of the photon at one-loop, with generic mass of the matter field in scalar QED, has not appeared before. In the next section we will also confirm this formula with a complementary approach based on multiplet recombination, of wider applicability.  
 
In the entirety of this section we choose the gauge field to have {\it Dirichlet} boundary conditions; as we explain, only with this choice it makes sense to consider SSB boundary conditions for the matter fields. On the other hand, even though we do not stress it in this section, all the statements we make about the one-loop corrections with symmetry-preserving boundary condition can be also readily imported to the case of Neumann boundary condition, the main difference being only the choice of contour for the spectral integral, which is not affected by the one-loop correction (we refer the reader to section \ref{sec:NeuProp} for a discussion of the contour change, and to e.g.~\eqref{eq:Neumann-4-point-amplitude} to see how the contour change affects the dressed boundary four-point function).

\subsection{1PI two-point function}
\label{subsec:spectral-representation-self-energy}

We recall that the interaction Lagrangian, written in terms of the real and imaginary part of the complex scalar $\Phi = \frac{\varphi_1 + i \varphi_2}{\sqrt{2}}$, is
\begin{equation}
\mathcal{L}^{int} = e A_{\mu}(\varphi_2\nabla^{\mu}\varphi_1 - \varphi_1\nabla^{\mu}\varphi_2) + \frac{e^2}{2} A_\mu A^\mu (\varphi_1^2 +\varphi_2^2)~.
\end{equation}
 We denote the boundary conditions for the scalars as $(\nu_1,\nu_2)$, indicating that the operator at the boundary corresponding to $\varphi_i$ has scaling dimension $\frac{d}{2}+i\nu_i$.  We will initially keep $(\nu_1,\nu_2)$ generic. Only when $\nu_1 =\pm \nu_2 =\nu$ the bulk Lagrangian is gauge-invariant, therefore we later restrict to the two most relevant cases of the symmetry-preserving boundary condition $(\nu,\nu)$ and the SSB one $(\nu,-\nu)$. The contribution of the quartic interaction is crucial to keep the photon massless in the case of the symmetry-preserving boundary condition $(\nu,\nu)$.

As depicted in figure \ref{fig:1-PI-AdS}, the propagator at one loop is computed by resumming the 1PI one loop two-point function, which we denote with $B_{\nu_1,\nu_2}^{(1)}(X_1, X_2)$.\footnote{To clarify the definition of the 1PI one loop two-point, we mean that the one-loop correction to the propagator $\langle A(X_1) A(X_2) \rangle$ is 
\begin{equation}
\int_{Y_1 }\int_{Y_2}\langle A(X_1) A(Y_1) \rangle B_{\nu_1,\nu_2}^{(1)}(Y_1, Y_2) \langle A(Y_2) A(X_2) \rangle~,
\end{equation}
where we are omitting index contraction.
} Each term in the sum involves an increasing number of convolution integrals in position space: it is convenient to represent the function $B_{\nu_1,\nu_2}^{(1)}(X_1, X_2)$ using the spectral representation, that converts convolutions to products and gives a geometric series that can be easily resummed
\begin{equation}
    \label{eq:1PI-bubble-AdS}
\hspace{-0.2cm} B_{\nu_1,\nu_2}^{(1)}(X_1, X_2; W_1, W_2)
    = -e^2
        \int_{-\infty}^{+\infty} \dd \lambda 
            \underbrace{\left(\langle JJ \rangle_{\nu_1, \nu_2}(\lambda)
            + \mathcal{T}_{\nu_1} + \mathcal{T}_{\nu_2}
            \right)}_{B^{(1)}_{\nu_1, \nu_2}(\lambda)}   
        \Omega^{(1)}_{\lambda} (X_1, X_2; W_1, W_2)~.
\end{equation}
We denoted the spectral representation as $-e^2 B^{(1)}_{\nu_1, \nu_2}(\lambda)$, where $e^2$ collects the one-loop dependence on the coupling, and a sign has been collected for future convenience. We further decomposed the function $B^{(1)}_{\nu_1, \nu_2}(\lambda)$ in three terms: the first \(\langle JJ \rangle_{\nu_1, \nu_2}(\lambda)\) is the spectral density of the two-point function of the current at separated points, i.e. the bubble diagram containing two scalar propagators, which is the first diagram in the second line of fig.~\ref{fig:1-PI-AdS}. 
\begin{figure}
    \centering
    \begin{tikzpicture}
        \node[black] at (-5, 0) {\(\langle W_1^{\mu}A_{\mu}(X_1) W_2^{\nu}A_{\nu}(X_2) \rangle_{\text{Full}} = \)};
        \draw[goldenyellow, ultra thick, decorate,decoration={snake, amplitude= 2mm}]  (-2, 0) -- (-0.3, 0) node[black, right] {+};
        \begin{scope}[xshift = 0.3cm]
            \draw[goldenyellow, ultra thick, decorate,decoration={snake, amplitude= 2mm}]  (0, 0) -- (0.8, 0);
            \filldraw[fuchsia, ultra thick, fill = fuchsia!20] (1.2, 0) circle (0.4);
            \draw[goldenyellow, ultra thick, decorate,decoration={snake, amplitude= 2mm}]  (1.6, 0) -- (2.4, 0) node[black, right] {+};
        \end{scope}
        \begin{scope}[xshift = 3.3cm]
            \draw[goldenyellow, ultra thick, decorate,decoration={snake, amplitude= 2mm}]  (0, 0) -- (0.8, 0);
            \filldraw[fuchsia, ultra thick, fill = fuchsia!20] (1.2, 0) circle (0.4);
            \draw[goldenyellow, ultra thick, decorate,decoration={snake, amplitude= 2mm}]  (1.6, 0) -- (2.4, 0);
            \filldraw[fuchsia, ultra thick, fill = fuchsia!20] (2.8, 0) circle (0.4);
            \draw[goldenyellow, ultra thick, decorate,decoration={snake, amplitude= 2mm}]  (3.2, 0) -- (4, 0) node[black, right] {+ \ldots};
        \end{scope}
        \begin{scope}[yshift = -2.5 cm, xshift = -3.1cm]
            \node at (-2.3, 0) {\(B_{\nu_1,\nu_2}^{(1)}(X_1, X_2; W_1, W_2) = \)};
            \begin{scope}[yshift = 0 cm, xshift = 0.6cm, scale = 0.9]
                \draw[goldenyellow, ultra thick, decorate,decoration={snake, amplitude= 2mm}]  (-0.3, 0) -- (0.5, 0);
                \filldraw[fuchsia, ultra thick, fill = fuchsia!20] (1.2, 0) circle (0.7);
                \draw[goldenyellow, ultra thick, decorate,decoration={snake, amplitude= 2mm}]  (1.9, 0) -- (2.7, 0) ;
                \filldraw[turquoise] (-0.3, 0) circle (0.05) node[below, black] {\(X_1\)};
                \filldraw[turquoise] (2.7, 0) circle (0.05) node[below, black] {\(X_2\)};
            \end{scope}
            \node at (3.6, 0) {=};
            \begin{scope}[yshift = 0 cm, xshift = 4.5cm, scale = 1.2]
                \draw[goldenyellow, ultra thick, decorate,decoration={snake, amplitude= 2mm}]  (-0.3, 0) -- (0.5, 0);
                \draw[fuchsia, ultra thick] (1.2, 0) circle (0.7);
                \draw[goldenyellow, ultra thick, decorate,decoration={snake, amplitude= 2mm}]  (1.9, 0) -- (2.7, 0) ;
                \node[fuchsia, above] at (1.2, 0.7) {\(\varphi_1\)};
                \node[fuchsia, below] at (1.2, -0.7) {\(\varphi_2\)};
                \filldraw[turquoise] (0.5, 0) circle (0.05) node[right, black] {};
                \filldraw[turquoise] (1.9, 0) circle (0.05) node[left, black] {};
                \filldraw[turquoise] (-0.3, 0) circle (0.05) node[below, black] {\(X_1\)};
                \filldraw[turquoise] (2.7, 0) circle (0.05) node[below, black] {\(X_2\)};
            \end{scope}
            \begin{scope}[yshift = 0 cm, xshift = 9.3cm, scale = 1.4]
                    \node[black] at (-0.7, 0) {\(+\)};
                    \draw[goldenyellow, ultra thick, decorate,decoration={snake, amplitude= 2mm}]  (-0.3, 0) -- (0.5, 0);
                    \draw[fuchsia, ultra thick] (0.5, 0) 
                    to[out = 140, in = 180, looseness = 2] (0.5, 1)  
                    to[out = 0, in = 40, looseness = 2] (0.5, 0);
                    \draw[goldenyellow, ultra thick, decorate,decoration={snake, amplitude= 2mm}]  (0.5, 0) -- (1.3, 0);
                    \node[fuchsia, above] at (0.5, 1) {\(\varphi_1\)};
                    \filldraw[turquoise] (-0.3, 0) circle (0.05) node[below, black] {\(X_1\)};
                    \filldraw[turquoise] (1.3, 0) circle (0.05) node[below, black] {\(X_2\)};
                    \filldraw[turquoise] (0.5, 0) circle (0.05) node[below, black] {};
            \end{scope}
            \begin{scope}[yshift = -3cm, xshift = 9.3cm, scale = 1.4]
                \node[black] at (-0.7, 0) {\(+\)};
                \draw[goldenyellow, ultra thick, decorate,decoration={snake, amplitude= 2mm}]  (-0.3, 0) -- (0.5, 0);
                \draw[fuchsia, ultra thick] (0.5, 0) 
                to[out = 140, in = 180, looseness = 2] (0.5, 1)  
                to[out = 0, in = 40, looseness = 2] (0.5, 0);
                \draw[goldenyellow, ultra thick, decorate,decoration={snake, amplitude= 2mm}]  (0.5, 0) -- (1.3, 0);
                \node[fuchsia, above] at (0.5, 1) {\(\varphi_2\)};
                \filldraw[turquoise] (-0.3, 0) circle (0.05) node[below, black] {\(X_1\)};
                \filldraw[turquoise] (1.3, 0) circle (0.05) node[below, black] {\(X_2\)};
                \filldraw[turquoise] (0.5, 0) circle (0.05) node[below, black] {};
            \end{scope}
        \end{scope}
    \end{tikzpicture}
    \caption{The 1 loop correction to the Anti de Sitter photon propagator in scalar QED. As per definition of the 1PI bubble in equation \eqref{eq:1PI-bubble-AdS} the external photon lines are amputated, but are depicted in the picture to clarify the interaction vertices considered.}
    \label{fig:1-PI-AdS}
\end{figure}
The two additional terms $\mathcal{T}_{\nu_1}$ and $\mathcal{T}_{\nu_2}$ are the tadpole diagrams with just one real scalar propagator, namely the second and third diagrams. The tadpoles do not depend on the spectral parameter $\lambda$ and as a result they give contact contributions in position space. The function $B_{\nu_1,\nu_2}^{(1)}$ can be thought of as the two-point function of the conserved current composite operator in the free scalar theory, in which the ambiguity of adding contact terms has been fixed in such a way to ensure conservation also at coincident points. This is evident from the fact that we are only including the harmonic function $\Omega^{(1)}$, that is transverse even at coincident points.

Let us discuss in turn the bubble diagram and the tadpole. In \cite{ShenMasterThesis} the following expression was obtained for the bubble diagram with generic $\nu_1$ and $\nu_2$\footnote{Comparing this result for the spin 1 bubble for generic masses with the scalar bubble with generic masses (i.e. the two point function of $\varphi_1\varphi_2$ in the free theory) we observe that the former is obtained from the latter by shifting \(d \mapsto d + 2\), see e.g. formula (D.6) of \cite{di2022analyticity}. }  
\begin{align}
\begin{split}
    \label{eq:simple-bubble-AdS}
    & \langle J J \rangle_{\nu_1,\nu_2} (\lambda) = \sum_{n = 0}^{+\infty} \frac{8\pi a^{d + 2}_{\nu_1, \nu_2}(n)}{2 i \alpha_n^-} \left[\frac{1}{i(\lambda - \alpha_n)} + (\lambda\to-\lambda))\right]~,\\
    & \alpha_n = - i\left(\frac{d}{2} + i\nu_1 + i \nu_2 + 2n + 1\right)~, \\
    \!\!\!a^{d + 2}_{\nu_1, \nu_2}(n) & =
    \frac{\left(\frac{d + 2}{2}\right)_n 
    \left(1 + i\nu_1 + i\nu_2 + n\right)_n 
    \left(2n + 2 + d + i\nu_1 + i\nu_2\right)_{-\frac{d}{2}}}
    {2\pi^{\frac{d + 2}{2}}n! 
    \left(\frac{d}{2} + i\nu_1 + n + 1\right)_{-\frac{d}{2}}
    \left(\frac{d}{2} + i\nu_2 + n + 1\right)_{-\frac{d}{2}}
    \left(\frac{d}{2} + i\nu_1 + i\nu_2 + n + 1\right)_{n}}~.
\end{split}
\end{align}
This infinite sum is divergent for $d\geq 1$. This is a UV divergence, and indeed the fact that it arises in this range of $d$ can be checked with a simple power counting argument in flat space. In dimensional regularization, we perform the sum in the range $d<1$ and then consider the analytic continuation in the complex $d$ plane. The result of the sum for $d<1$, in the two relevant cases of $\nu_1=\nu_2=\nu$ and $\nu_1=-\nu_2=\nu$ can be written as
\begin{align}
\label{eq:resummed-simple-bubble}
\begin{split}
 \langle JJ \rangle_{\nu,\nu}(\lambda) & = -\frac{2\, \Gamma \left(\frac{d}{2}+1+2 i \nu\right) \Gamma \left(\frac{d}{2}+1+i \nu \right)^2}{\pi ^{d/2} \Gamma (d+2+2 i \nu )\Gamma (1+i \nu )^2 }\\
& \hspace{-2.5cm}\times\left(\frac{ \, _5F_4\left(
\begin{array}{c}
\frac{d}{2}+1,i \nu +\frac{1}{2},\frac{d}{2}+i \nu +1,\frac{d}{4}+i \nu -\frac{i \lambda }{2}+\frac{1}{2},\frac{d}{2}+2 i \nu +1 \\
i \nu +1,\frac{d}{2}+i \nu +\frac{3}{2},\frac{d}{4}+i \nu -\frac{i \lambda }{2}+\frac{3}{2},2 i \nu +1
\end{array}
;1\right)}{\frac{d}{2}+1 +2 i \nu  -i \lambda } + (\lambda\to-\lambda)\right)~, \\
\langle JJ \rangle_{\nu,-\nu}(\lambda) & = - \frac{ \Gamma \left(\frac{d}{2}+1+ i \nu \right)\Gamma \left(\frac{d}{2}+1 - i \nu \right)}{(4 \pi )^{\frac{d-1}{2}} (d+1) \Gamma \left(\frac{d+1}{2}\right) \Gamma (1 + i \nu)\Gamma (1 - i \nu)} \\
& \hspace{-0.5cm}\times \left(\frac{\, _5F_4\left(
\begin{array}{c}
\frac{1}{2},\frac{d}{2}+1,\frac{d}{4}-\frac{i \lambda }{2}+\frac{1}{2},\frac{d}{2}-i \nu +1,\frac{d}{2}+i \nu +1 \\
\frac{d}{2}+\frac{3}{2},\frac{d}{4}-\frac{i \lambda }{2}+\frac{3}{2},1-i \nu ,i \nu +1
\end{array};
1\right)}{\frac{d}{2}+1-i \lambda} + (\lambda\to -\lambda)\right)~.
\end{split}
\end{align}
In the appendix \ref{app:spin-1-simple-bubble} we discuss how to analytically continue these expressions in $d$ and expand them around $d=3$ to obtain the pole associated to the logaritmic UV divergence in AdS$_4$, and the finite part. The result is\footnote{As explained in the appendix \ref{app:spin-1-simple-bubble}, the expression that we obtain for the residue at the pole is analytic but too lengthy to be simplified analytically. The simplified expression in equation \eqref{eq:poleJJ} is obtained via numerical comparison for several values of $\lambda$ and $\nu$.}
\begin{equation}\label{eq:poleJJ}
\langle JJ \rangle_{\nu,\pm\nu} (\lambda) \underset{d\to 3}{\sim} \frac{1}{3-d}\left(\frac{\lambda^2 + \frac{1}{4}}{24\pi^2} - \frac{\nu^2 + \frac 14}{4\pi^2}\right) + \overline{\langle JJ \rangle}_{\nu,\pm\nu} (\lambda) + \mathcal{O}(3-d)~.
\end{equation}
With this method we determine the finite part $\overline{\langle JJ \rangle}_{\nu,\pm\nu} (\lambda)$ only numerically. 

Moving to the tadpole diagram, we observe that it is computed by the coincident point limit of the scalar propagator $\Pi^{(0)}_{\frac{d}{2}+i\nu}$. This limit diverges for $d\geq1$, but it gives a finite constant when $d<1$ (again, this agrees with flat space power counting of UV divergences, as it should). Upon analytic continuation in $d$, this constant gives the tadpole diagram in dimensional regularization. The result is
\begin{equation}
\label{eq:tadpole-dim-reg}
\mathcal{T}_\nu = \frac{\Gamma \left(\frac{1-d}{2}\right) \Gamma \left(\frac{d}{2}+i \nu \right)}{(4 \pi )^{\frac{d+1}{2}} \Gamma \left(-\frac{d}{2}+i \nu +1\right)}~.
\end{equation}
Expanding around $d\to 3$ we isolate the pole due to the logarithmic UV divergence in AdS$_4$ and a finite term $\overline{\mathcal{T}}_\nu$, obtaining
\begin{align}
\label{eq:tadpole-dim-reg}
\begin{split}
& \mathcal{T}_\nu  \underset{d\to 3}{\sim} \frac{\nu ^2+\frac{1}{4}}{8 \pi ^2} \frac{1}{3-d} + \overline{\mathcal{T}}_\nu+ \mathcal{O}(3-d)~,\\
 \overline{\mathcal{T}}_\nu =\frac{\nu ^2+\frac{1}{4}}{16 \pi ^2}&\left(\psi(i \nu -1)-\psi \left(i \nu +\tfrac{3}{2}\right)-2 \psi(2 i \nu -2) -\gamma_\text{E} +1+\log (16 \pi ) \right)~,
\end{split}
\end{align}
where $\psi(x)$ is the digamma function and $\gamma_{\text{E}}$ is Euler's gamma. 

Summing up the bubble and the tadpoles, we get
\begin{align}
\begin{split}\label{eq:Bbareexp}
& B^{(1)}_{\nu, \pm\nu}(\lambda)  \underset{d\to 3}{\sim} \frac{1}{3-d}\frac{\lambda^2 + \frac 14}{24\pi^2} + \overline{\langle JJ \rangle}_{\nu,\pm\nu} (\lambda) +\overline{\mathcal{T}}_\nu +\overline{\mathcal{T}}_{\pm\nu} +\mathcal{O}(3-d)~.
\end{split}
\end{align}
It is interesting to observe that, while separately $\mathcal{T}_\nu$ and $\langle JJ \rangle_{\nu,\pm \nu}$ are finite for $d<1$ and have a pole at $d=1$, their combination \(B^{(1)}_{\nu, \pm\nu}(\lambda)\) is finite in the wider regime \(d < 3\). This is consistent with the UV divergences of the current two-point function that are known from flat space.

\subsection{Charge renormalization}
\label{subsec:charge-renormalization}

Using the $R_\xi$ gauge fixing, and summing the geometric series gives the following expression for the one-loop propagator of the photon 
\begin{align}\label{eq:resummed-photon-propagator-AdS}
\begin{split}
         &\langle A(X_1,W_1) A(X_2,W_2) \rangle_{\text{one-loop}} \\
        & \hspace{2cm}= \int_{-\infty}^{+\infty}
        \dd \lambda\,\frac{1}{\lambda^2 +(\frac d2 -1)^2  + e^2
        B^{(1)}_{\nu_1, \nu_2}(\lambda)}
       \,
        \Omega^{(1)}_{\lambda} (X_1, X_2; W_1, W_2) \\
& \hspace{2.5cm}+\int_{-\infty}^{+\infty}\dd \lambda  \, \frac{\xi}{\left(\lambda^2 + \frac{d^2}{4}\right)^2} \,  \left(W_1 \cdot \nabla_1\right)\left(W_2 \cdot \nabla_2\right) \Omega^{(0)}_{\lambda} (X_1, X_2)~. 
\end{split}
\end{align}
Note that the longitudinal part of the propagator is left unchanged by the resummation, because it has vanishing convolution with the spin 1 harmonic function, since the latter is transverse and we can integrate by parts in the convolution integral. 

In \eqref{eq:Bbareexp}  we obtained that the divergent part in $B^{(1)}_{\nu,\pm\nu}(\lambda)$ is proportional to $
\lambda^2 +\frac 14$, i.e. the inverse tree-level propagator of the gauge field for $d=3$. This means that the divergence can be reabsorbed in the wavefunction renormalization of the gauge field, or equivalently in the ``charge renormalization'', i.e. the renormalization of the coupling $e^2$.\footnote{If we had additional divergences with coefficients that do not depend on $\lambda$, reabsorbing them would have required adding a mass counterterm for the gauge field. The fact that this does not happen is of course expected due to gauge invariance, but it is a check of the calculation. We stress that this happens both for $(\nu,\nu)$ and $(\nu,-\nu)$ boundary conditions, because the latter is only a spontaneous breaking of the symmetry, which cannot give rise to a photon-mass counterterm.} Defining the renormalized gauge field and the renormalized coupling
\begin{align}
\begin{split}
& A_\mu = \sqrt{Z_A} A_\mu^{\text{ren}}~,~e^2 =  \mu^{3-d}\, 
\frac{e^2_\text{ren}(\mu)}{Z_A} \\ \text{with}~& Z_A = 1+ e_{\text{ren}}^2 \underbrace{\left(\tfrac{1}{3-d}\delta z_A^{\text{div}} + \delta z_A^{\text{finite}}\right)}_{\delta z_A} +\mathcal{O}(e_\text{ren}^4)~,
\end{split}
\end{align}
the 1PI two-point function receives an additional contribution from the counterterm given by
\begin{align}
\begin{split}
B^{(1),\text{ren}}_{\nu, \pm\nu}(\lambda)& = \mu^{3-d} B^{(1)}_{\nu, \pm\nu}(\lambda) + \delta z_A\left(\lambda^2 + (\tfrac d2 -1)^2\right) +\mathcal{O}(e_\text{ren}^2) = \text{finite} \\
& \Rightarrow \delta z_A^{\text{div}} = -\frac{1}{24\pi^2} ~.
\end{split}
\end{align}
As required by the locality of UV divergences, this is the same renormalization coefficient that is found in flat space, and gives rise to the one-loop $\beta$ function in scalar QED 
\begin{equation}
\beta_{e_\text{ren}^2} = \frac{d e^2_{\text{ren}}}{d
\log\mu} \underset{d=3}{=} \frac{e_\text{ren}^4}{24\pi^2} +\mathcal{O}(e_\text{ren}^6)~.
\end{equation}

Substituting the expansion \eqref{eq:Bbareexp} 
we obtain the following finite result for the renormalized 1PI two-point function at one loop
\begin{align}
\label{eq:generic-finite-part-bubble-1PI}
\begin{split}
B^{(1),\text{ren}}_{\nu, \pm\nu}(\lambda)  
        = \overline{\langle JJ \rangle}_{\nu,\pm\nu} (\lambda) +\overline{\mathcal{T}}_\nu +\overline{\mathcal{T}}_{\pm\nu}+\frac{1}{48\pi^2}+ C \left(\lambda^2 + \frac{1}{4}\right)~,
    \end{split}
\end{align}
where $C= \delta z_A^{\text{finite}} - \delta z_A^{\text{div}}\log\mu$ is a constant that depends on the renormalization scheme, and it is the only part of the function that is scheme-dependent. Note that, in addition to the finite parts of the diagrams, there is a shift of $1/(48\pi^2)$ from the product of the dimreg pole \(\delta z_A\) and the term $(\tfrac{d}{2}-1)^2$ in the inverse tree-level propagator. This is purely an effect of the curvature, restoring the AdS length $L$ it is suppressed by $L^{-2}$ compared to the other terms in the flat-space limit. The resulting expression for the one-loop renormalized two-point function in $d=3$ is 
\begin{align}\label{eq:renormalized-photon-propagator-AdS}
\begin{split}
         &\langle A^{\text{ren}}(X_1,W_1) A^{\text{ren}}(X_2,W_2) \rangle_{\text{one-loop}} \\
        & \hspace{2cm}= \int_{-\infty}^{+\infty}
        \dd \lambda\,\frac{1}{\lambda^2 +\frac 14  + e^2_{\text{ren}}
        B^{(1),\text{ren}}_{\nu_1, \nu_2}(\lambda)}
       \,
        \Omega^{(1)}_{\lambda} (X_1, X_2; W_1, W_2) \\
& \hspace{2.5cm}+\int_{-\infty}^{+\infty}\dd \lambda  \, \frac{\xi}{\left(\lambda^2 + \frac{9}{4}\right)^2} \,  \left(W_1 \cdot \nabla_1\right)\left(W_2 \cdot \nabla_2\right) \Omega^{(0)}_{\lambda} (X_1, X_2)~. 
\end{split}
\end{align}
We now analyze the propagator \eqref{eq:renormalized-photon-propagator-AdS} separately in the cases of symmetry-preserving boundary conditions \((\nu, \nu)\), and symmetry-breaking boundary conditions \((\nu, - \nu)\).

\subsection{Symmetry-preserving boundary conditions}
\label{subsec:gauge-preserving-boundary-conditions}

The boundary condition \((\nu, \nu)\) is invariant under the $U(1)$ gauge symmetry in the bulk, or equivalently under the $U(1)$ global symmetry at the boundary. Therefore in this case we do not expect any Higgs mechanism. This means that the photon should not acquire a mass, i.e. the pole in the tree-level propagator, at $\lambda=\pm i (\tfrac{d}{2}-1)$, is expected to remain in the full one-loop resummed propagator \eqref{eq:resummed-photon-propagator-AdS}. The condition on the bubble function, for any $d$, is then
\begin{equation}\label{eq:nophotmass}
B^{(1)}_{\nu, \nu}\left(\pm i\left(\tfrac{d}{2} - 1\right)\right) = 0~.
\end{equation}
This condition was used in \cite{ankur2023scalar} to fix a constant that remained undetermined in the function $B^{(1)}_{\nu, \nu}(\lambda)$, when the latter was obtained from a bootstrap condition in the limit of a large number of flavors. From the more explicit expression \eqref{eq:1PI-bubble-AdS} we can explicitly check that this condition indeed holds true. The check is performed in dimensional regularization, by analytically continuing the formulas for \(\langle J J \rangle_{\nu, \nu}(\lambda)\) and the tadpole \(\mathcal{T}(\nu)\) and evaluating them for \(d < 1\), where each term is finite. Since $B^{(1)}_{\nu, \nu}\left(\pm i\left(\tfrac{d}{2} - 1\right)\right)$ vanishes identically as a function of $d$ in this range, by analytic continuation this must hold also for $d>1$. 

\begin{figure}
    \centering 
    \includegraphics[width=0.8\textwidth]{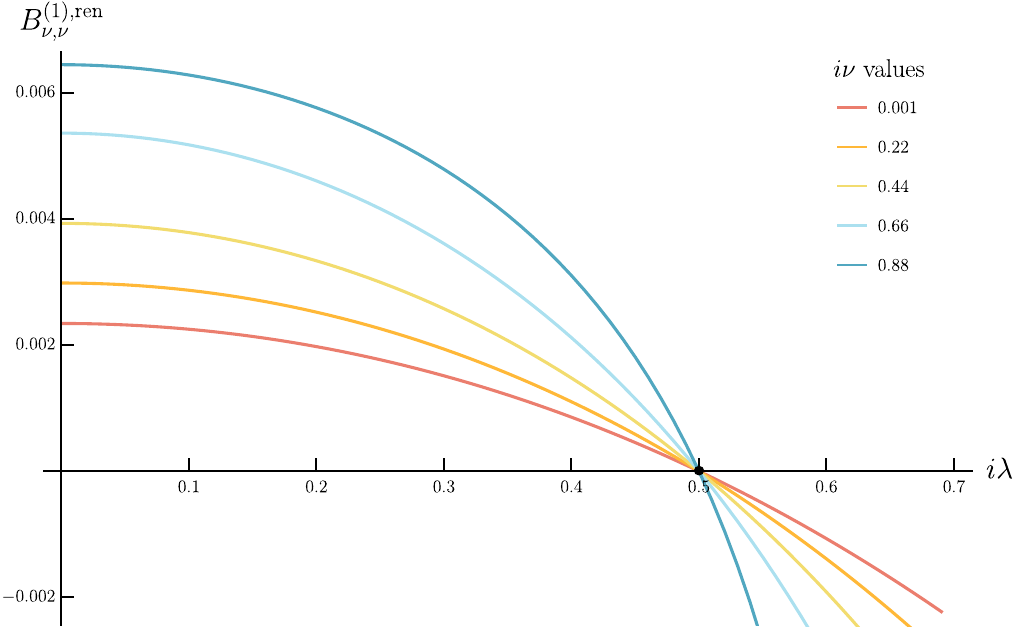}
    \caption{The renormalized photon self--energy for several values of \(\nu\), with gauge-preserving boundary conditions and \(d = 3\), see \eqref{eq:generic-finite-part-bubble-1PI}. This picture shows a zoom around \(\lambda = \frac{i}{2}\), where the self energy vanishes regardless of the value of \(\nu\), therefore preserving a vanishing photon mass. This property does not depend on choices of scheme. However for generic $\lambda$ we need to fix the scheme dependent constant \(C\), here we took \(C=0\).}
    \label{fig:lambda-plot-finite-part-equal-insertions-zoom-on-tree-level-zero}
\end{figure}

Since \eqref{eq:nophotmass} is valid for any $d$, taking the limit $d\to 3$ it implies that a similar statement holds for the renormalized function (note that the scheme-dependent part vanishes for this value of $\lambda$)
\begin{equation}\label{eq:zeronoHiggs}
B^{(1),\text{ren}}_{\nu, \nu}\left(\pm\tfrac{i}{2}\right)= 0~.
\end{equation}
We checked numerically the validity of this equation, see figure \ref{fig:lambda-plot-finite-part-equal-insertions-zoom-on-tree-level-zero}. 

\begin{figure}[ht]
\centering
    \includegraphics[width=0.8\textwidth]{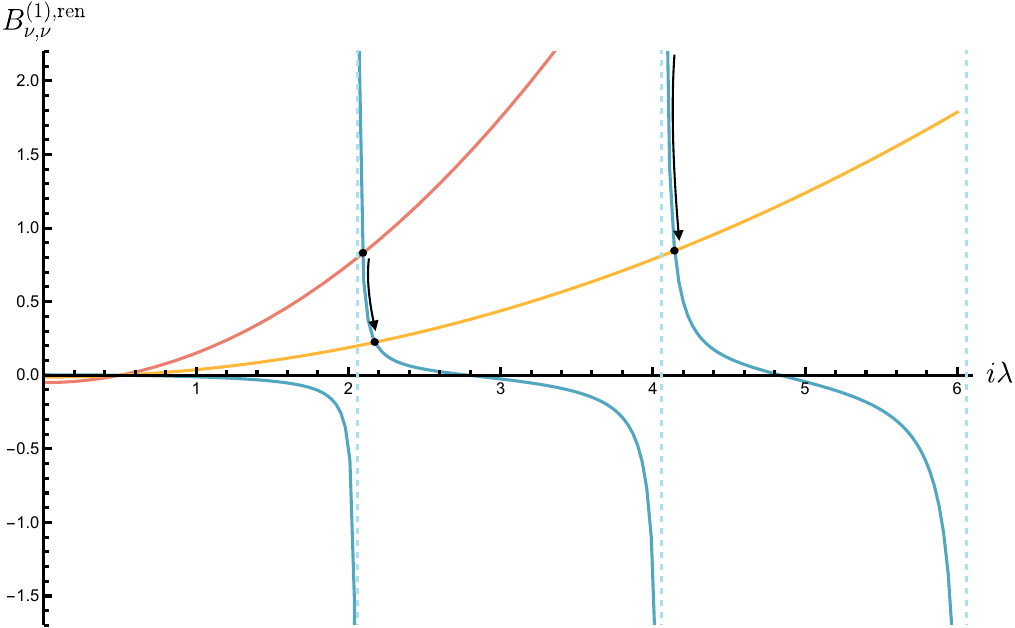}
    \caption{The blue curve is the photon self-energy in \eqref{eq:generic-finite-part-bubble-1PI} for $d=3$, with $(\nu,\nu)$ boundary condition and $\nu=0.22 i$. The scheme dependent constant \(C\) is set to zero. 
    The orange and yellow curves represent the tree level spectral densities, multiplied by the inverse charge, namely \(\frac{1}{e^2_{ren}}\left(\lambda^2 + \frac{1}{4}\right)\), for two different values of the renormalized charge: the yellow curves corresponds to the numerical value \(e_{ren}^2 = 20\), while the orange one to \(e_{ren}^2 = 5\). The intersections with the blue curve give the dimensions of the operators appearing in the boundary OPE of the photon due to the one-loop resummed propagator. Thanks to the condition \eqref{eq:nophotmass}, that ensure that the photon remains massless at one loop, among these operators there is a conserved current. The dashed vertical lines correspond to the spin 1 ``double-trace'' operators of the free complex scalar.}
    \label{fig:lambda-plot-finite-part-equal-insertions}
\end{figure}

The zeroes of the denominator in the one-loop resummed propagator \eqref{eq:generic-finite-part-bubble-1PI} give rise, upon closing the contour integral with an arc at infinity, to the spectrum of spin 1 operators appearing in the boundary OPE expansion of the bulk gauge field. Thanks to \eqref{eq:zeronoHiggs} one zero is ensured to stay at the dimension of a conserved current. We display the other zeros in figure \ref{fig:lambda-plot-finite-part-equal-insertions}, by showing the intersection between the one-loop bubble function and the tree-level propagator. Since this is only a one-loop perturbative result, the resulting dimensions are trustworthy only as long as they remain sufficiently close to the free theory values, given by the dashed vertical lines in the figure.

\subsection{Symmetry-breaking boundary conditions}
\label{subsec:gauge-breaking-boundary-conditions}

The boundary condition $(\nu, -\nu)$ breaks the $U(1)$ symmetry. Being localized at the boundary, this is a spontaneous symmetry breaking in AdS. As a result, a Higgs mechanism is expected to take place, generating a mass for the photon. We postpone a more general discussion of SSB and Higgs mechanism in AdS to the next section \ref{sec:SecHiggs}, however, we will use some of the results that are derived there.

The Higgs mechanism at play here is not the usual one due to the condensation of the scalar field itself, for which the mass is visible already by an analysis at tree-level. Instead in this case the photon mass arises at one-loop. Indeed, considering the matter sector in the limit $e\to 0$, we see that $\langle \Phi \rangle = 0$, but instead the bilinear charge 2 operator gets a condensate 
\begin{align}
\begin{split}
\langle \Phi^2 \rangle & = \frac 12(\langle \varphi_1^2 \rangle - \langle \varphi_2^2 \rangle )\\ & = \frac 12(\mathcal{T}_\nu - \mathcal{T}_{-\nu} )\\
& = -\frac{\sin (\pi  i \nu ) \Gamma \left(\frac{d}{2}+i \nu \right) \Gamma \left(\frac{d}{2}-i \nu \right)}{(4 \pi )^{\frac{d+1}{2}} \Gamma \left(\frac{d+1}{2}\right)} \\
& \underset{d=3}{=} -\frac{\left(4 \nu ^2+1\right) \tan (\pi  i \nu )}{64 \pi }~.
\end{split}
\end{align}
Note that for any $d\geq 2$ and $0\leq i \nu \leq 1$ this calculation gives a finite, scheme-independent answer. At leading order, the anomalous dimension of the boundary current, or equivalently the photon mass, is proportional to this condensate, with a calculable coefficient. This formula can be derived using a multiplet recombination argument, as we explain in section  \ref{sec:SecHiggs}, and the final result is
\begin{align}
\begin{split}\label{eq:phmassphisqd3}
M^2_\gamma & = - e^2 \frac{4i\nu}{d} \langle \Phi^2 \rangle \\
& = e^2 \frac{4 i \nu \sin (\pi  i \nu ) \Gamma \left(\frac{d}{2}+i \nu \right) \Gamma \left(\frac{d}{2}-i \nu \right)}{(4 \pi )^{\frac{d+1}{2}} d\,\Gamma \left(\frac{d+1}{2}\right)}+\mathcal{O}(e^4)~.
\end{split}
\end{align}

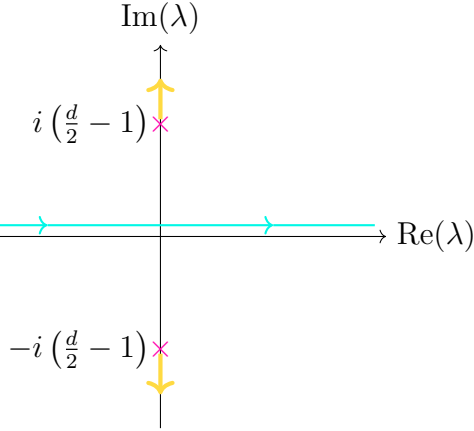
\begin{figure}
        \centering
        \begin{tikzpicture}[scale=1.5]
            \draw[black, ->] (-2,0) -- (2,0) node[right] {\(\text{Re}(\lambda)\)};
            \draw[black, ->] (0,-1.7) -- (0,1.7) node[above] {\(\text{Im}(\lambda)\)};
            \node[left] at (0,1) {\(i\left(\frac{d}{2} - 1\right)\)};
            \node[fuchsia] at (0,1) {\(\times\)};
            \node[left] at (0,-1) {\(-i\left(\frac{d}{2} - 1\right)\)};
            \node[fuchsia] at (0,-1) {\(\times\)};
            \draw[turquoise, thick, ->] (-2,0.1) -- (-1, 0.1);
            \draw[turquoise, thick, ->] (-1, 0.1) -- (1, 0.1);
            \draw[turquoise, thick] (1, 0.1) -- (1.9, 0.1);
            \draw[goldenyellow, ultra thick, ->] (0, 1.04) -- (0, 1.4);
            \draw[goldenyellow, ultra thick, ->] (0, -1.04) -- (0, -1.4);
        \end{tikzpicture}
        \caption{Integration contour for 1--loop 4--point functions with Dirichlet boundary conditions on the gauge field. At 1-loop the spontaneous symmetry breaking given by the boundary conditions of the scalar field shifts the location of the most dominant pole in the direction marked by the yellow arrows.}
    \label{fig:4-point-function-integration-contours-1-loop}
\end{figure}

This result can be compared with the explicit calculation of the one-loop propagator, since the mass-squared is simply the location of the pole, which shifts at one-loop due to the symmetry-breaking boundary condition, as illustrated in figure \ref{fig:4-point-function-integration-contours-1-loop}. Specifying to $d=3$, the one-loop shift of the pole is obtained from the value of the renormalized 1PI two-point function \eqref{eq:generic-finite-part-bubble-1PI} at the location of the massless pole $\lambda=\pm \frac i 2$. We find
\begin{equation}\label{eq:numMsquared}
B^{(1),\text{ren}}_{\nu, -\nu}(\pm \tfrac i 2)  
        = \overline{\langle JJ \rangle}_{\nu,-\nu} (\pm \tfrac i 2) +\overline{\mathcal{T}}_\nu +\overline{\mathcal{T}}_{-\nu}+\frac{1}{48\pi^2} = \frac{\left(4 \nu ^2+1\right)\, i\nu  \tan (\pi  i \nu )}{48 \pi }~,
\end{equation}
in agreement with \eqref{eq:phmassphisqd3}, which is a strong check of the result for the 1PI two-point function. Note that, evaluating the function at this point, the contribution of the scheme-dependent coefficient $C$ drops, and we get a physical answer. The finite term $\overline{\langle JJ \rangle}_{\nu,-\nu} (\pm \frac i 2)$ is evaluated numerically, following the regularization procedure detailed in the appendix \ref{app:spin-1-simple-bubble}. This numerical evaluation is sufficient to verify the validity of \eqref{eq:numMsquared}.

\begin{figure}[ht]
    \centering
    \includegraphics[width =0.8\textwidth]{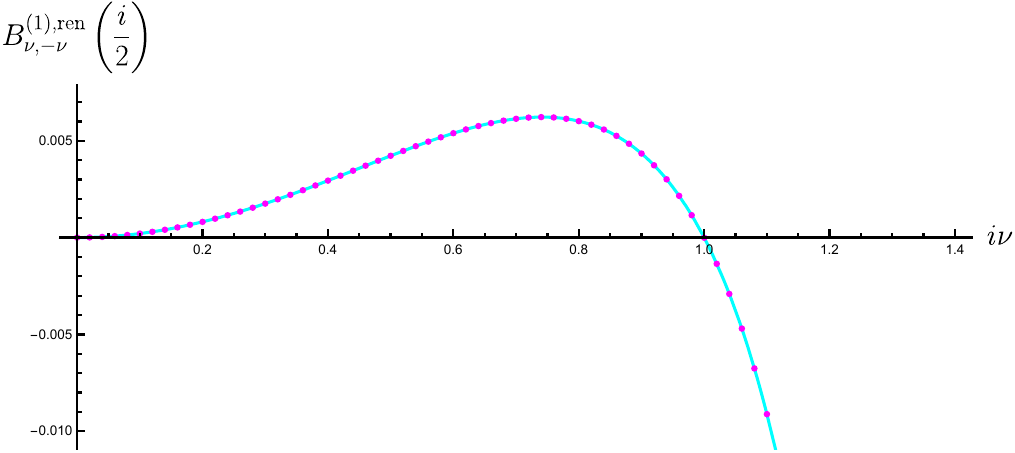}
    \caption{Plot of $B^{(1),\text{ren}}_{\nu, \pm\nu}\left(\pm \frac i 2 \right)$ in \(d = 3\) obtained separately evaluating the finite part of the contributions in \eqref{eq:numMsquared} for several values of \(\nu\) (dotted magenta line), superimposed with the analytic formula for \(M_{\gamma}^2\) in \eqref{eq:phmassphisqd3} (solid turquoise line).}
    \label{fig:AdS-mass-generation}
\end{figure}

We plot the mass-squared of the photon $M^2_\gamma$ in figure \ref{fig:AdS-mass-generation} as a function of the parameter $i\nu$, related to the mass-squared of the scalar as $m^2 = -\frac{9}{4} -\nu^2$. The origin $i\nu = 0$ corresponds to the Breitenlohner-Freedman bound, and we restrict to positive $i\nu$ because the result is symmetric under $\nu\leftrightarrow-\nu$, which corresponds to $\varphi_1 \leftrightarrow \varphi_2$. The dimension of $\varphi_2$ is above the unitarity bound for $i\nu\leq 1$. We see that $M^2_\gamma \geq 0$ precisely in this range, and it turns negative when unitarity is violated $i\nu > 1$. It vanishes for $i\nu= 0$ because for that value $\varphi_1$ and $\varphi_2$ have the same boundary condition and the symmetry is not broken. We observe that it also vanishes for $i\nu =1$. This can be explained as follows: as discussed in \cite{Bashmakov:2016pcg}, by appropriately rescaling the Lagrangian of $\varphi_2$ as we take $i\nu \to 1$, one obtains that the whole bulk dynamics of the field trivializes, and only a free scalar mode at the boundary survives the limit. As a result the photon decouples from the scalars and this explain why there is no correction to the tree-level zero mass.

\begin{figure}[ht]
    \centering
    \includegraphics[width=0.8\textwidth]{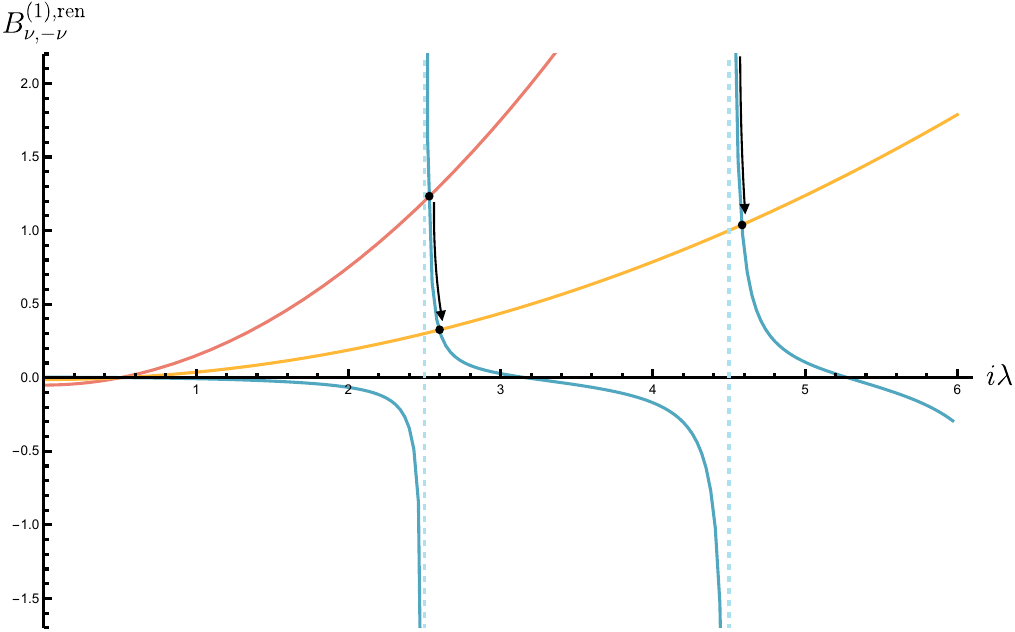}
    \caption{All the curves have the same meaning as in figure \ref{fig:lambda-plot-finite-part-equal-insertions}, with the same values $d=3$, $\nu=0.22 i$, $C=0$ and $e^2_{\text{ren}} = \{5, 20\}$, the difference being that here we have the symmetry-breaking $(\nu,-\nu)$ boundary condition. Due to the mass generation, there is no intersection of the orange/brown curve with the blue one at $i\lambda =1/2$, and consequently no conserved current in the boundary spectrum.
    }
    \label{fig:lambda-plot-finite-part-opposite-insertions}
\end{figure}

Similarly to what we explained above for the $(\nu,\nu)$ boundary condition, it is also interesting to look at the additional zeroes of the resummed one loop propagator of the photon, since they correspond to the dimensions of the operators appearing in the boundary OPE (bOPE) of the photon. In this case, the lightest operator is the current that is broken due to the Higgsing, that gets a positive anomalous dimension. We illustrate the poles in figure \ref{fig:lambda-plot-finite-part-opposite-insertions}.

\subsection{4--point function at 1-loop}

It is interesting to study how the inclusion of the one-loop correction to the photon propagator affects the boundary four-point function of the charged fields. Restricting to the s-channel photon exchange, we derived the expression of the spin 1 conformal partial wave expansion of the four-point function at the tree-level. The result is in \eqref{eq:Dirichlet-current-photon-current-transverse}-\eqref{eq:dirichlet-spectral-density-tree-level} for the case of symmetry-preserving boundary condition, and in \eqref{eq:Dirichlet-current-photon-current-transverse-opposite}-\eqref{eq:dirichlet-spectral-density-tree-level-opposite} for the case of symmetry-breaking boundary condition. It is clear in the structure of \eqref{eq:dirichlet-spectral-density-tree-level}-\eqref{eq:dirichlet-spectral-density-tree-level-opposite} that some factors in the density arise from kinematics, and are unaltered at one-loop, while the one-loop correction does affect the factor arising from the spectral density of the photon propagator. Namely, we need to replace the tree-level spectral density with its one-loop resummed version
\begin{equation}
\frac{1}{\lambda^2 + \left(\frac{d}{2} - 1\right)^2} \longrightarrow \frac{1}{\lambda^2 +(\frac d2 -1)^2  + e^2
        B^{(1)}_{\nu, \pm\nu}(\lambda)}~.
\end{equation}
The consequences of this replacement are:
\begin{itemize}
\item{In the case of symmetry breaking boundary condition, the shift of the photon pole implies that a conserved current operator is no longer exchanged in the four-point function, and it gets replaced by an unprotected spin one operator $j$ with scaling dimension $\Delta_j = d-1 + e^2 \gamma_j +\mathcal{O}(e^4)$, related to the non-zero photon mass in \eqref{eq:phmassphisqd3}, through the equation $M^2_\gamma = (\Delta_j-1)(\Delta_j-d+1) = (d-2)e^2 \gamma_j + \mathcal{O}(e^4)$. In the case of symmetry-preserving boundary condition instead the current remains conserved and regarding this pole the only difference compared to the tree-level result is the value of the residue, encoding the OPE coefficient.}
\item{There is an additional tower of spin 1 operators exchanged in the four-point function. They are associated to the additional poles that appear at one loop in the resummed propagator, illustrated in fig.~\ref{fig:lambda-plot-finite-part-equal-insertions} and fig.~\ref{fig:lambda-plot-finite-part-opposite-insertions}, for the cases of symmetry preserving and symmetry breaking boundary condition, respectively. In the limit $e^2 \to 0$ these operators coincide with the spin 1 ``double-trace'' operators of the GFF on the boundary of the free complex scalar in AdS. Note that, as already remarked above, the location of these poles is only trustworthy as long as the anomalous dimensions of the ``double-trace'' operators are small, because we are including up to $\mathcal{O}(e^2)$  and for larger anomalous dimensions two-loop corrections also matter.}
\end{itemize}

%% file: SSB-and-Higgs-mech-in-AdS.tex
In this section we discuss a different approach for obtaining the formula for the mass of the photon, alternative to the explicit loop calculation presented in the previous section, which is based on the general concept of multiplet recombination and therefore has a wider applicability. This approach builds on the understanding of SSB of continous global symmetries in AdS in terms of the existence of a protected tilt operator, long understood in the context of BCFT \cite{Bray:1977fvl, Herzog:2017xha, Cuomo:2021cnb, Padayasi:2021sik} and recently emphasized for QFT in AdS in \cite{Copetti:2023sya}, and also in \cite{Porrati:2024zvi} where the relation to the bulk Higgs mechanism is also discussed.   

We first collect some general results regarding the spontaneous breaking of a global $U(1)$ symmetry in AdS, like the existence of a tilt operator in the boundary and a universal formula for its two-point function coefficient; we then apply it to the case in which the symmetry is gauged to show that it leads to a recombination of the boundary current with the tilt operator, and to derive a universal formula for the mass-squared of the photon at leading order in the gauge coupling. 

\subsection{SSB and tilt operator}\label{sec:SSBtilt}
Consider a QFT in (Euclidean) AdS$_{d+1}$ with a $U(1)$ symmetry and associated current $J^\mu$. We denote with $\alpha\sim \alpha +2\pi$ the $U(1)$ parameter, and $\delta\alpha$ its infinitesimal version. Given a bulk local operator $O$ we denote its infinitesimal variation as
\begin{equation}
\delta O(x) = \delta\alpha\, O'(x)~,
\end{equation}
so the operator transforms non-trivially as long as the operator $O'(x)$ is non-vanishing. Like in flat space, we say that this $U(1)$ undergoes spontaneous symmetry breaking (SSB)  if there exist such a pair $O(x)$, $O'(x)$ such that the one-point function of $O'(x)$ is non-vanishing
\begin{equation}
\text{SSB:} \quad \langle O'(x) \rangle = a_{O'} \neq 0~.
\end{equation}
By AdS isometries $a_{O'}$ is a constant. In this situation we say that $O'$ is an order parameter.

Promoting $\alpha$ to a function of $x$, the action is not invariant but rather
\begin{equation}\label{eq:varS}
\delta S = -\int d^{d+1}x \sqrt{g(x)}\,\nabla_\mu \delta\alpha(x)\,J^\mu(x)~~\Rightarrow \nabla_\mu J^\mu(x) = \frac{\delta S}{\delta \alpha(x)}~,
\end{equation}
which can be taken to be a definition of the current $J^\mu$. The Ward identity then gives
\begin{align}
\begin{split}\label{eq:WI}
\nabla^{x_1}_\mu \langle J^\mu(x_1) O(x_2) \rangle & = - \frac{\int \mathcal{D}\Phi \, O(x_2)\,\frac{\delta }{\delta \alpha(x_1)} e^{-S} }{\int \mathcal{D}\Phi \,e^{-S}}\\
& = \delta^{d+1}(x_1-x_2) \langle O'(x_2) \rangle~.
\end{split}
\end{align}
Let us now assume that $O'$ is an order parameter for SSB.  Let us integrate both sides of the equation over the whole space in the variable $x_1$. We choose Poincar\'e coordinates $x_i=(z_i,\vec{x}_i)$. We obtain
\begin{equation}\label{eq:WISSB}
-\underset{z_1\to 0}{\lim} \int d^d \vec{x}_1\, z_1^{-d-1} \langle J^z(z_1,\vec{x}_1) O(z_2,\vec{x}_2) \rangle = a_{O'} \neq 0~. 
\end{equation}
In order for this equation to be satisfied there must exist a boundary scalar operator $t(\vec{x})$ appearing in the bOPE  of $J^\mu(z,\vec{x})$ so that
\begin{equation}\label{eq:bOPEJz}
J^z(z,\vec{x})\underset{z\to 0}{\sim} z^{d+1} (t(\vec{x})+\text{descendants}) + \text{other operators}~.
\end{equation}
This boundary scalar operator $t$ is called the {\it tilt} operator. We decide to fix the normalization of the tilt operator by demanding that it appears with unit coefficient in this bOPE. Compatibility of this bOPE with the scaling isometry $(z,\vec{x})\to \lambda (z,\vec{x})$ fixes the scaling dimension of the tilt to be $\Delta_t = d$, i.e. it is a boundary marginal operator. The most general bulk-boundary two-point function between $t$ and $O$ is fixed by the isometries to be
\begin{equation}\label{eq:tO}
\langle t(\vec{x}_1) O(z_2,\vec{x}_2) \rangle = N \left(\frac{z_2}{z_2^2 + (\vec{x}_1-\vec{x}_2)}\right)^d~,
\end{equation}
up to a normalization constant $N$. In appendix \ref{app:boundary-limit-bulk-boundary-propagators} we compute
\begin{equation}\label{eq:integral}
\int d^d \vec{x}_1\,\left(\frac{z_2}{z_2^2 + (\vec{x}_1-\vec{x}_2)}\right)^d =  \frac{2^{1-d}\pi^{\frac{d+1}{2}}}{\Gamma(\tfrac{d+1}{2})}~.
\end{equation}
Plugging \eqref{eq:integral} and \eqref{eq:bOPEJz} in \eqref{eq:WISSB} we obtain
\begin{equation}
N = - \frac{2^{d-1} \Gamma(\tfrac{d+1}{2})}{\pi^{\frac{d+1}{2}}} a_{O'}~.
\end{equation}

Next, we compute $N$ in a second way. Equation \eqref{eq:tO} implies that $t$ must appear in the bOPE of $O$
\begin{equation}
O(z,\vec{x})\underset{z\to 0}{\sim} b_{Ot}\, z^d(t(\vec{x})+\text{descendants}) + \text{other operators}~.
\end{equation}
with a certain bulk-to-boundary OPE coefficient $b_{Ot}$. Plugging this bOPE in the expansion of \eqref{eq:tO} in the limit $z_2\to 0$, and defining the coefficient in the two-point function of $t$ as
\begin{equation}
\langle t(\vec{x}_1) t(\vec{x}_2)\rangle = \frac{C_t}{|\vec{x}_1 - \vec{x}_2|^{2d}}~,
\end{equation}
we obtain that $N = b_{Ot} \, C_t$. 

Comparing with the previous expression for $N$ we finally obtain
\begin{equation}
C_t = - \frac{2^{d-1} \Gamma(\tfrac{d+1}{2})}{\pi^{\frac{d+1}{2}}} \frac{a_{O'}}{b_{O t}}~.
\end{equation}
As a sanity check, note that the normalization of $O$ cancels from this final result. Since $t$ is a marginal operator, with protected scaling dimension and an associated one-dimensional conformal manifold, the positive coefficient $C_t$ has the meaning of the Zamolodchikov metric that measure distances on the space of boundary theories, or equivalently on the space of vacua of the bulk theory.  This relation was discussed previously in the context of BCFT, see for instance appendix A of \cite{Padayasi:2021sik}. The tilt operator for SSB in AdS was discussed in \cite{Copetti:2023sya}, see also the related discussion in \cite{Porrati:2024zvi}. 

From the bOPE \eqref{eq:bOPEJz}, we note that when there is SSB, in order to preserve the condition $\nabla_\mu J^\mu(x) = \frac{\delta S}{\delta \alpha(x)}$ that we used to define $J_\mu$ and derive the Ward identity, we need to take care of boundary terms in the special case that the transformation parameter approaches a non-trivial function at the boundary 
\begin{align}
\begin{split}\label{eq:varSSSB}
\delta\alpha(z,\vec{x})& \underset{z\to 0}{\sim} \delta\alpha_\partial(\vec{x})+\text{subleading} \\
 \Rightarrow \delta S & = -\int d^{d+1}x \sqrt{g(x)}\,\nabla_\mu \delta\alpha(x)\,J^\mu(x) - \int_{z=0} d^d \vec{x} \,\delta\alpha_\partial(\vec{x})\,t(\vec{x})\\
& = \int d^{d+1}x \sqrt{g(x)}\,\delta\alpha(x)\,\nabla_\mu J^\mu(x)~.
\end{split}
\end{align} 

\subsection{Higgs mechanism and photon mass}

In order to gauge the $U(1)$ we add a gauge field $A_\mu$ and modify the action to be
\begin{equation}
\label{eq:maxwell-coupled-to-matter-generic}
S = S_m + \int d^{d+1}x\,\sqrt{g(x)}\,\left( \frac{1}{4}F^{\mu\nu}F_{\mu\nu} + e A_\mu J^\mu + (\text{interactions of}~\mathcal{O}(A_\mu^2))\right)~.
\end{equation}
Here $S_m$ is the action of the un-gauged theory that satisfies \eqref{eq:varS} or its variant \eqref{eq:varSSSB}, therefore this action is guaranteed to be gauge-invariant at the linear level in $\delta\alpha(x)$ as long as
\begin{equation}
\delta A_\mu(x) = \frac{1}{e}\nabla_\mu\delta\alpha(x)~.
\end{equation}
We assume that the higher order interactions involving $A_\mu$ can be fixed in such a way to ensure full gauge invariance at the non-linear level. Besides modifying the action, we also need to change the prescription of the path integral: we only sum over equivalence classes of the fields, including $A_\mu$, under gauge transformations.

At leading order in $e^2$, or equivalently at the linearized/free theory level, the leading boundary modes of the field $A_\mu(x)$ are
\begin{equation}\label{eq:asymptotics}
A_i(z,\vec{x}) \underset{z\to 0}{\sim} z^{d-2} j_i(\vec{x}) + a_i(\vec{x})+\ldots
\end{equation}
Here $i$ runs over the components parallel to the boundary, and the boundary behavior of $A_z$ is also determined using the gauge-fixed equations of motion.  Before turning on the interaction with matter, we choose a Dirichlet boundary condition, namely we take $a_i(\vec{x})$ to be a fixed $c$-number function, possibly vanishing, and as a result $j_i(\vec{x})$ is an operator of the boundary theory. The operator $j_i(\vec{x})$ is required by the scaling isometry to have scaling dimension $\Delta_j = d-1$, and consistently using the equations of motion one obtains that it is a conserved current, i.e. it obeys the operator equation $\partial_i j^i = 0$. The $c$-number function $a_i(\vec{x})$ is a source for the operator $j_i(\vec{x})$. To ensure a good variational principle in the presence of a non-zero source we need to supplement the quadratic action by a boundary term
\begin{equation}\label{eq:boundaryterm}
\int_{z=0} d^d \vec{x} \,(d-2)\,a_i(\vec{x}) j^i(\vec{x})~.
\end{equation}
The Dirichlet boundary condition fixes the gauge invariance at the boundary. ``Large'' gauge transformations with parameter $\alpha(z,\vec{x})$ that approaches a constant at the boundary leave the full action invariant and act as a global $U(1)$ symmetry transformation on the boundary conformal correlators. Relaxing the requirement that $\alpha(z,\vec{x})$ approaches a constant, but rather taking it to be a non-trivial function of the boundary coordinates, then the boundary condition is modified
\begin{align}
\begin{split}
\delta\alpha(z,\vec{x})& \underset{z\to 0}{\sim} \delta\alpha_\partial(\vec{x})+\text{subleading} \\
 \Rightarrow \delta a_i(\vec{x}) & = \frac{1}{e}\partial_i \delta\alpha_\partial(\vec{x})~,
\end{split}
\end{align}
and using the boundary term \eqref{eq:boundaryterm} we get
\begin{equation}
\delta S = \int_{z=0} d^d\vec{x} \,\partial_i \delta\alpha_\partial(\vec{x})\, \frac{1}{e}(d-2) j^i(\vec{x})~.
\end{equation}
This is precisely the condition for $\frac{1}{e}(d-2)j^i$ to be the Noether current, and allows to derive Ward identities in correlation functions with insertions of charged operators. A computation in the free theory \cite{Freedman:1998tz}\footnote{Note that $J_\text{there} = \frac{1}{e}(d-2) j_\text{here}$.} shows that at leading order in $e^2$
\begin{align}
\begin{split}\label{eq:lo2ptj}
\langle j_i(\vec{x}) j_k(0)\rangle & = C^{(0)}_j \frac{I_{ik}(\vec{x})}{|\vec{x}|^{2(d-1)}}(1+\mathcal{O}(e^2))~,~~I_{ik}\equiv  \delta_{ik} - \frac{2 x_i x_k}{\vec{x}^2}~,\\
C^{(0)}_j & =  \frac{\Gamma(d)}{2(d-2)\pi^{\frac{d}{2}}\Gamma(\tfrac{d}{2})}~.
\end{split}
\end{align}

Now, let us specify to the case in which the matter theory described by the action $S_m$ breaks the symmetry spontaneously. Then, using \eqref{eq:varSSSB}, we see that under a gauge transformation that approaches a function at the boundary we have an additional contribution, and the total variation of the action is
\begin{equation}
\delta S = \int_{z=0} d^d\vec{x} \,\delta\alpha_\partial(\vec{x})\,\left(- \frac{1}{e}(d-2)\partial_i j^i(\vec{x}) - t(\vec{x})\right)~,
\end{equation}
This means that on-shell, i.e. as an operator equation, the following recombination identity holds
\begin{equation}\label{eq:rec}
\partial_i j^i(\vec{x}) = -\frac{e}{d-2} t(\vec{x})~.
\end{equation}
Therefore at leading order in perturbation theory the operator $j_i$ stops being a conserved current and gets an anomalous dimension. The form of the correlation function in the interacting theory will thus be, by conformal invariance
\begin{equation}
\langle j_i(\vec{x}) j_k(0)\rangle = C_j(e^2) \frac{I_{ik}(\vec{x})}{|\vec{x}|^{2\Delta_j(e^2)}}~,
\end{equation}
where the normalization $C_j(e^2)$ and the scaling dimension $\Delta_j(e^2)$ are functions of the coupling. Taking the divergence of both currents in the correlator, and using the recombination equation \eqref{eq:rec}, we obtain
\begin{equation}
\langle t(\vec{x}) t(0)\rangle = 4 \frac{(d-2)^2}{e^2} C_j(e^2) (\Delta_j(e^2)-d+1)(\Delta_j(e^2)-\tfrac{d}{2}+1)\frac{1}{|\vec{x}|^{2\Delta_j(e^2)+2}}~.
\end{equation}
Plugging $C_j(e^2) = C_j^{(0)} + \mathcal{O}(e^2)$ and $\Delta_j(e^2) = d-1 + e^2\gamma_j + \mathcal{O}(e^4)$ we see that the two-point function above has a finite limit $e^2\to 0$, in which case we must recover the two-point function of the tilt operator in the matter theory with SSB. We thus obtain the following relation between the leading-order anomalous dimension $\gamma_j$ of the current and the coefficient of the two-point function of the tilt in the un-gauged theory
\begin{equation}
\gamma_j =  \frac{C_t}{2 d (d-2)^2 \, C_j^{(0)}}~.
\end{equation}
We can then use the expression for $C_t$ in terms of the data of the bulk operators $O'$ and $O$, and the expression of $C_j^{(0)}$ in \eqref{eq:lo2ptj} to get
\begin{equation}
\gamma_j =  -\frac{1}{d(d-2)}\frac{a_{O'}}{b_{Ot}}~.
\end{equation}
Equivalently, we can write the mass-squared $M^2_\gamma$ of the higgsed photon using the relation $M^2_\gamma = (\Delta_j(e^2) -1)(\Delta_j(e^2) -d +1) = (d-2) e^2 \gamma_j +\mathcal{O}(e^4)$, obtaining
\begin{equation}\label{eq:phmass}
M^2_\gamma =  -\frac{e^2}{d}\frac{a_{O'}}{b_{Ot}} +\mathcal{O}(e^4)~.
\end{equation}

\paragraph{Check: standard SSB} Take a complex scalar $\Phi =\frac{1}{\sqrt{2}}\left(\varphi_1 + i \varphi_2\right)$ of charge $q$ under a $U(1)$ symmetry, with a potential that spontaneously breaks the $U(1)$ symmetry $\langle \Phi \rangle = v$. With the parametrization
\begin{equation}
\Phi = (v + \delta\rho)e^{i\Omega}~,
\end{equation}
we get that $\Omega$ is a massless scalar in AdS and as a result it will have a marginal operator in its bOPE, which we identity with the tilt, namely
\begin{equation}
\Omega(z,\vec{x}) \underset{z\to 0}{\sim}b_{\Omega t} \,z^d (t(\vec{x})+\text{descendants}) +\text{subleading}~.
\end{equation}
The bulk conserved current is $J_\mu = i q (\Phi^* \nabla_\mu \Phi - \Phi\nabla_\mu \Phi^*) = -2 q v^2 \nabla_\mu \Omega + \dots$ and therefore
\begin{equation}\label{eq:JzclassicalHiggs}
J_z(z,\vec{x}) \underset{z\to 0}{\sim} -2 q v^2 \,b_{\Omega t} \,d \,z^{d-1} (t(\vec{x})+\text{descendants}) +\text{subleading}~\Rightarrow b_{\Omega t} = -\frac{1}{2 q v^2 d}~.
\end{equation}
Now let us take $O = \varphi_2 = \sqrt{2}(v + \delta\rho)\sin\Omega =\sqrt{2}v \Omega + \dots$, so that $b_{O t} = \sqrt{2}v b_{\Omega t} =  -\frac{1}{\sqrt{2} q v d}$, and $O' = q \varphi_1= \sqrt{2} q(v + \delta\rho)\cos\Omega$ so that $a_{O'}= \sqrt{2}qv$. 
Plugging in \eqref{eq:phmass} we obtain
\begin{equation}
M^2_\gamma = -\frac{e^2}{d}(-\sqrt{2} q v d) \sqrt{2}q v +\mathcal{O}(e^4)= 2 e^2 q^2 v^2+\mathcal{O}(e^4)~,
\end{equation}
which is indeed the correct value of the photon mass-squared derived from the Lagrangian.

\paragraph{Higgsing induced by boundary conditions} Now we specify to the setup of subsection \ref{subsec:gauge-breaking-boundary-conditions}, namely a free complex scalar $\Phi = \frac{\varphi_1 + i \varphi_2}{\sqrt{2}}$ with mass-squared in the range of double quantization, and different quantization for $\varphi_1$ and $\varphi_2$, i.e. the asymptotics are 
\begin{align}
\begin{split}
\varphi_1(z,\vec{x})& \underset{z\to 0}{\sim} z^{\frac{d}{2}+i \nu} \left(O_1(\vec{x})+\text{descendants}\right)~,\\
\varphi_2(z,\vec{x}) & \underset{z\to 0}{\sim} z^{\frac{d}{2}-i\nu}  \left(O_2(\vec{x})+\text{descendants}\right)~,
\end{split}
\end{align}
with $0\leq i\nu \leq 1$. Plugging the asymptotics in the conserved current $J_\mu = \varphi_2 \nabla_\mu \varphi_1 - \varphi_1 \nabla_\mu \varphi_2$ we get
\begin{equation}
\label{eq:Jz-asymptotics-Higgsed-phase}
J_z(z,\vec{x}) \underset{z\to 0}{\sim} z^{d-1} 2 i \nu (O_1O_2(\vec{x})+\text{descendants})\Rightarrow t(\vec{x})=2 i \nu\,O_1O_2(\vec{x})~.
\end{equation}

In the notation of subsection \ref{subsec:gauge-breaking-boundary-conditions} we take $O= \mathrm{Im}\,\Phi^2 = \frac12\varphi_1\varphi_2$, so that $b_{Ot}=\frac{1}{4i\nu}$, and $O'=\mathrm{Re}\,\Phi^2 = \frac 12(\varphi_1^2 - \varphi_2^2)$, which gives
\begin{align}
\begin{split}
 a_{O'} & =\frac 12(\langle \varphi_1^2 \rangle - \langle \varphi_2^2 \rangle)\\
 & = \frac 12\underset{x_1\to x_2}{\lim}\left(\Pi^{(0)}_{\frac{d}{2}+i\nu}(x_1,x_2) - \Pi^{(0)}_{\frac{d}{2}-i\nu}(x_1,x_2)\right)\\
 & =-\frac{\sin (\pi  i \nu ) \Gamma \left(\frac{d}{2}+i \nu \right) \Gamma \left(\frac{d}{2}-i \nu \right)}{(4 \pi )^{\frac{d+1}{2}} \Gamma \left(\frac{d+1}{2}\right)}~.
\end{split}
\end{align}
Note that, for any $d\geq 2$ the limit above is a finite non-positive number in the whole range $0\leq i \nu \leq 1 $ of double-quantization, vanishing at $i\nu = 0$, as expected because there is no symmetry breaking in that case. For $d>2$ it vanishes also for $i\nu =1$, in which case $O_2$ hits the unitarity bound and decouples from the bulk \cite{Bashmakov:2016pcg}.

As a result from the formulas above we obtain that upon coupling to a $U(1)$ gauge field 
\begin{equation}
\gamma_j = \frac{4 i \nu \sin (\pi  i \nu ) \Gamma \left(\frac{d}{2}+i \nu \right) \Gamma \left(\frac{d}{2}-i \nu \right)}{(4 \pi )^{\frac{d+1}{2}} d(d-2)\Gamma \left(\frac{d+1}{2}\right)}~,
\end{equation}
and
\begin{equation}\label{eq:phmassphisq}
M^2_\gamma = e^2 \frac{4 i \nu \sin (\pi  i \nu ) \Gamma \left(\frac{d}{2}+i \nu \right) \Gamma \left(\frac{d}{2}-i \nu \right)}{(4 \pi )^{\frac{d+1}{2}} d\,\Gamma \left(\frac{d+1}{2}\right)}+\mathcal{O}(e^4)~.
\end{equation}
From the comments above about $a_{O'}$, we see that $M^2_\gamma\geq 0$ for every $i\nu$ in the range of double-quantization, and that it vanishes when expected. Moreover, this is in agreement with the explicit loop calculation of subsection \ref{subsec:gauge-breaking-boundary-conditions}, providing a non-trivial check of both computations.

%% file: 1-form-symmetries.tex
At the end of section \ref{subsec:wilson-line-prescription-transverse}, we observed that the boundary operator $f_{ij}$, although above the unitarity bound and allowed by representation theory, does not appear in the OPE of the four-point function with Neumann boundary conditions. The main aim of this section is to provide  a generalized-symmetry-based explanation for this decoupling, which also ensures the decoupling of $f_{ij}$ at the non-perturbative level. 

In the process, we clarify several consequences of spontaneously broken one-form symmetries in AdS, including the role of the associated tilt operator and the behavior of Wilson lines at the boundary. Our analysis further offers an alternative perspective on a familiar result in AdS/CFT: the emergence of boundary global symmetries from bulk gauge symmetries. Unlike the standard explanation based on large gauge transformations or asymptotic symmetries, our argument, which holds even in the presence of charged matter, is based purely on generalized symmetries. 

Related analyses of broken one-form symmetries in free Maxwell theory with boundaries appeared in \cite{Arbalestrier:2025jsg}. Spontaneously broken higher-form symmetries have been analyzed in \cite{Lake:2018dqm} and were revisited more recently using boundaries and corners within Symmetry Topological Field Theory in \cite{Bonetti:2025dvm}.
\subsection{One-form symmetries and their spontaneous breaking}
\label{sec:ssb-1-form-symmetries}
In recent years, our understanding of symmetry in quantum field theory has undergone revolution, driven mainly by the realization \cite{Gaiotto:2014kfa} that symmetry operators are special cases of more general topological operators. For an ordinary continuous global symmetry, the symmetry operator can be written as $\exp \left[\alpha \int_{\Sigma^{(D-1)}}\star \mathcal{J}\right]$ where $\mathcal{J}$ is a conserved current and $\Sigma^{(D-1)}$ is a codimension $1$ surface. Current conservation $d\star \mathcal{J}=0$ ensures that this operator is independent of the shape of $\Sigma^{(D-1)}$, making it topological. On the other hand, operators {\it charged} under the symmetry are zero-dimensional and one can measure their charge by surrounding them by the codimension-$1$ symmetry operator.

A natural generalization is provided by {\it higher-form symmetries}, in which symmetry operators are supported on higher-codimension manifolds. Specifically, a 
$p$-form symmetry is generated by an operator defined on a codimension $(p+1)$ submanifold $\Sigma^{(D-p-1)}$ and, when continuous, is associated with a conserved $(p+1)$-form current $\mathcal{J}^{(p)}$. On the other hand, operators charged under the symmetry are now $p$-dimensional and one can measure their charges by ``linking" the symmetry operator around the charged operators \cite{Gaiotto:2014kfa}. In this language, ordinary global symmetries correspond to $p=0$. Simple and important examples of higher-form symmetry are provided by free Maxwell theory and QED, which we discuss in more detail later.

The properties summarized above apply generally to {\it local} quantum field theories, namely those with a conserved stress tensor. In theories without a stress tensor, such as conformal theories realized on the boundary of QFT in AdS, more exotic phenomena can occur  as we see below, including continuous “non-local” symmetries that admit topological symmetry operators but nevertheless are not associated with a conserved current. 

In the following, we examine the physical consequences of spontaneous breaking of higher-form symmetries in AdS space. While the discussion extends straightforwardly to general continuous higher-form symmetries, we mostly focus on one-form symmetries, which are of direct relevance to our analysis.

\paragraph{SSB of one-form symmetry and tilt operator.} Let us first explain the definition of SSB of one-form symmetry in AdS. Following the argument in section \ref{sec:SSBtilt} for ordinary global symmetries, we define the SSB of a continuous $1$-form symmetry $G^{(1)}$ by requiring that the symmetry operator has a finite action when pushed to the boundary (c.f.~\eqref{eq:WISSB}). This translates to the existence of the following contribution in the boundary expansion of the transverse component of the current $\mathcal{J}_{\mu\nu}$:  
\begin{equation}
    \label{eq:ssb-1-form-symmetries-definition}
	\mathcal{J}_{iz}(\vec{x}, z) \underset{z \rightarrow 0}{\sim} z^{d - 3}\tau_{i}(\vec{x}) + \mathcal{O}(z^{d - 2}),
\end{equation}
or equivalently,
\begin{align}
\left(\star^{(D)} \mathcal{J}\right)_{i_1\ldots i_{d - 1}}(\vec{x}, z) \underset{z \rightarrow 0}{\sim}
    \varepsilon_{i_1\ldots {i_{d}}}\tau^{i_{d}}(\vec{x}) + \mathcal{O}(z),
\end{align}
where \(\varepsilon_{i_1\ldots i_{d}}\) is the Levi-Civita symbol.
Here \(\star^{(D)}\) is the Hodge star operator in the bulk, which acts on a generic \(k\)-form \(A\) as
\begin{equation}
    \left(\star^{(D)} A\right)_{i_{k+1} \ldots i_D} = \frac{\sqrt{\vert g \vert}}{(D - k)!} \, \varepsilon_{i_1 \ldots i_D}A^{i_1 \ldots i_k},
\end{equation}
where indices on the right-hand side are contracted with the flat-space metric \(\delta_{ab}\).
The operator $\tau_i (\vec{x})$ is the {\it tilt} operator for one-form symmetry. Unlike the tilt operator for ordinary symmetries, it is a spin-1 operator and its dimension is fixed by the scaling isometry $(z,\vec{x})\mapsto \lambda (z,\vec{x})$ to be $\Delta_{\tau}=d-1$, saturating the unitarity bound. Hence it is a conserved current on the boundary satisfying $\partial_i \tau^{i}=0$.  When applied to the electric one-form symmetry in free Maxwell theory, this is basically a symmetry-based reinterpretation of the familiar statement in AdS/CFT that the gauge symmetry in the bulk implies a conserved current on the boundary. We will also show later how the argument can be generalized to setups with charged matter, in which the one-form symmetry is explicitly broken in the bulk.

Being a conserved current on the boundary, one can use $\tau_i$ to define a topological operator {\it on the boundary}. Indeed, pushing the entire support surface \(\Sigma^{(D - 2)}\) on the boundary, the bulk 1-form symmetry operator morphs into a boundary topolgical operator, given by the integral of the tilt on the boundary surface \(\Sigma'^{(d - 1)}\):
\begin{equation}
    \label{eq:limit-bulk-1-form-bdy-0-form-charge}
	\mathcal{Q}_G = \int_{\Sigma^{(D - 2)}} \star^{(D)} \mathcal{J} \xrightarrow[\Sigma^{(D - 2)} \rightarrow \Sigma'^{(d - 1)} \subseteq \partial \text{AdS}]{} \int_{\Sigma'^{(d - 1)}} \star^{(D)} \mathcal{J} = 
    \int_{\Sigma'^{(d - 1)}} \star^{(d)} \tau.
\end{equation}
Here $\star^{(d)}$ is the standard Hodge star operator on the boundary induced by the {\it flat space} metric. 
Thus, the SSB of one-form symmetry induces a {\it local} \(0\)-form symmetry on the boundary, whose (conserved) current is the tilt \(\tau_i(\vec{x})\).

Conversely, if the bulk \(1\)-form symmetry is unbroken, $\mathcal{J}_{iz}$ goes to zero faster as $z\to 0$, and the limiting procedure in \eqref{eq:limit-bulk-1-form-bdy-0-form-charge} just produces an identically vanishing operator. Thus, when we push the bulk $1$-form symmetry operator to the boundary, a portion of the surface $\Sigma^{(D-2)}$ parallel to the boundary disappears, meaning that we can end the support surface $\Sigma^{(D-2)}$ at the boundary. Such an ``open"  surface ending at the boundary defines a topological operator for the combined bulk-boundary system. However, there is no corresponding local current on the boundary since the topological operator always extends to the bulk. In this sense, this defines a {\it non-local} symmetry from the boundary viewpoint.

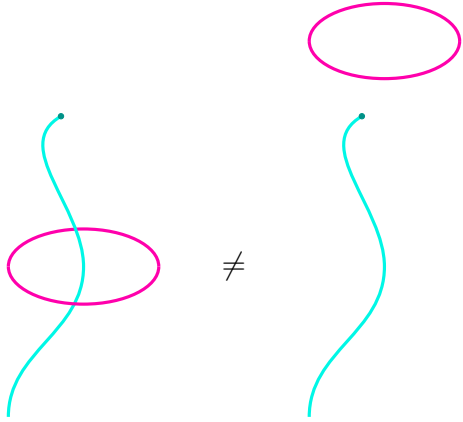
\begin{figure}
    \centering
    \begin{tikzpicture}
        \begin{scope}
            \draw[scale=1, fuchsia, very thick] (0,2) [xscale=2, yscale=1] arc[start angle=180, end angle=0, radius=0.5cm];
            \draw[turquoise, very thick] (0, 0) to[out = 90, in = -90] (1, 2) to[out = 90, in = -150] (0.7, 4);
            \filldraw[black!41!turquoise] (0.7, 4) circle (1pt);
            \draw[scale=1, fuchsia, very thick] (0,2) [xscale=2, yscale=1] arc[start angle=180, end angle=360, radius=0.5cm];
        \end{scope}
        \node[black] at (3,2) {\(\neq\)};
        \begin{scope}[xshift=4cm]
            \draw[scale=1, fuchsia, very thick] (0,{2 + 3}) [xscale=2, yscale=1] arc[start angle=180, end angle=0, radius=0.5cm];
            \draw[turquoise, very thick] (0, 0) to[out = 90, in = -90] (1, 2) to[out = 90, in = -150] (0.7, 4);
            \filldraw[black!41!turquoise] (0.7, 4) circle (1pt);
            \draw[scale=1, fuchsia, very thick] (0,{2 + 3}) [xscale=2, yscale=1] arc[start angle=180, end angle=360, radius=0.5cm];
        \end{scope}
    \end{tikzpicture}
    \caption{If a charged line (turquoise) had an endpoint in the bulk, the charge operator (fuchsia) would not be topological, since carrying it over the line-endpoint \(\Sigma^{D - 2}\) becomes contractible.}
    \label{fig:contractible-topological-operator-line-bulk-endpoint}
\end{figure}

\paragraph{One-form symmetries and endability of  line operators.} Whether a one-form symmetry is spontaneously broken at the boundary has direct consequences for the behavior of charged line operators. Let us first recall the well-known fact \cite{Gaiotto:2014kfa} that, if the one-form symmetry is  unbroken in the bulk, charged line operators cannot have {\it bulk} endpoints; this is simply because the existence of the endpoints is inconsistent with the action of the symmetry operator, see figure \ref{fig:contractible-topological-operator-line-bulk-endpoint}.

\begin{figure}[!h]
    \begin{tikzpicture}
        \begin{scope}
            \draw[black] (0, 0) -- (2, 1) -- (2, 5) -- (0, 4) -- (0, 0);
            \draw[turquoise, thick, dashed] (3.3, 1.1) -- (3.4, 0.5);
            \draw[fuchsia, thick] (3.1, 1.2) ellipse (0.7 and 0.2);
            \draw[fuchsia, thick, ->] (3.1, {1.2-0.2}) -- (3.15, 1);
            \draw[very thick, turquoise] (1, 3) to[out = 0, in = 100]  (3.3, 1.1);
            \filldraw[black!41!turquoise] (1, 3) circle (1.5pt);
        \end{scope}
        \node[black] at (4.5, 3) {\(\longrightarrow\)};
        \begin{scope}[xshift = 6cm]
            \draw[black] (0, 0) -- (2, 1) -- (2, 5) -- (0, 4) -- (0, 0);
            \draw[turquoise, thick, dashed] (3.3, 1.1) -- (3.4, 0.5);
            \draw[fuchsia, thick, ->] (0.5, 1) -- (3.1, 1);
            \draw[fuchsia, thick, ->] (3.1, 1) to[out = 0, in = 0, looseness = 4] (3.3, 1.4);
            \draw[fuchsia, thick, ->] (3.3, 1.4) -- (1.3, 1.4);
            \draw[fuchsia, thick] (1.3, 1.4) -- (0.5, 1);
            \draw[very thick, turquoise] (1, 3) to[out = 0, in = 100]  (3.3, 1.1);
            \filldraw[black!41!turquoise] (1, 3) circle (1.5pt);
        \end{scope}
        \node[black] at (10.5, 3) {\(\longrightarrow\)};
        \begin{scope}[xshift = 12cm]
            \draw[black] (0, 0) -- (2, 1) -- (2, 5) -- (0, 4) -- (0, 0);
            \draw[turquoise, thick, dashed] (3.3, 1.1) -- (3.4, 0.5);
            \draw[fuchsia, thick, ->] (0.5, 1) -- (3.1, 1);
            \draw[fuchsia, thick, ->] (3.1, 1) to[out = 0, in = 0, looseness = 4] (3.3, 1.4);
            \draw[fuchsia, thick, ->] (3.3, 1.4) -- (1.3, 1.4);
            \draw[fuchsia, dashed] (1.3, 1.4) -- (0.5, 1);
            \draw[very thick, turquoise] (1, 3) to[out = 0, in = 100]  (3.3, 1.1);
            \filldraw[black!41!turquoise] (1, 3) circle (1.5pt);
        \end{scope}
        \draw[black, ->] (13, -0.7) -- (13, -1.4);
        \begin{scope}[xshift = 12cm, yshift = -7cm]
            \draw[black] (0, 0) -- (2, 1) -- (2, 5) -- (0, 4) -- (0, 0);
            \draw[turquoise, thick, dashed] (3.3, 1.1) -- (3.4, 0.5);
            \draw[fuchsia, thick, ->] (0.5, {1 + 2.5}) -- (3.1, {1 + 2.5});
            \draw[fuchsia, thick, ->] (3.1, {1 + 2.5}) to[out = 0, in = 0, looseness = 4] (3.3, {1.4 + 2.5});
            \draw[fuchsia, thick, ->] (3.3, {1.4 + 2.5}) -- (1.3, {1.4 + 2.5});
            \draw[very thick, turquoise] (1, 3) to[out = 0, in = 100]  (3.3, 1.1);
            \filldraw[black!41!turquoise] (1, 3) circle (1.5pt);
        \end{scope}
        \node[black] at (10.5, -5) {\(\longleftarrow\)};
        \begin{scope}[xshift = 6cm, yshift = -7cm]
            \draw[black] (0, 0) -- (2, 1) -- (2, 5) -- (0, 4) -- (0, 0);
            \draw[turquoise, thick, dashed] (3.3, 1.1) -- (3.4, 0.5);
            \draw[fuchsia, thick, ->] (0.5, {1 + 2.5}) -- (3.1, {1 + 2.5});
            \draw[fuchsia, thick, ->] (3.1, {1 + 2.5}) to[out = 0, in = 0, looseness = 4] (3.3, {1.4 + 2.5});
            \draw[fuchsia, thick, ->] (3.3, {1.4 + 2.5}) -- (1.3, {1.4 + 2.5});
            \draw[fuchsia, thick] (1.3, {1.4 + 2.5}) -- (0.5, {1 + 2.5});
            \filldraw[turquoise] (1, 3) circle (1pt);
            \draw[very thick, turquoise] (1, 3) to[out = 0, in = 100]  (3.3, 1.1);
            \filldraw[black!41!turquoise] (1, 3) circle (1.5pt);
        \end{scope}
        \node[black] at (4.5, -5) {\(\longleftarrow\)};
        \begin{scope}[yshift = -7cm]
            \draw[black] (0, 0) -- (2, 1) -- (2, 5) -- (0, 4) -- (0, 0);
            \draw[turquoise, thick, dashed] (3.3, 1.1) -- (3.4, 0.5);
            \draw[fuchsia, thick] (3.1, {1.2 + 2.5}) ellipse (0.7 and 0.2);
            \draw[fuchsia, thick, ->] (3.1, {1.2-0.2 + 2.5}) -- (3.15, {1 + 2.5});
            \filldraw[turquoise] (1, 3) circle (1pt);
            \draw[very thick, turquoise] (1, 3) to[out = 0, in = 100]  (3.3, 1.1);
            \filldraw[black!41!turquoise] (1, 3) circle (1.5pt);
        \end{scope}
        \begin{scope}[xshift=1cm, yshift= -1.05cm]
            \node[black, rotate=90] at (0,0) {\(\neq\)};
        \end{scope}
    \end{tikzpicture}
    \caption{
    A line operator charged under the \(1\)-form symmetry can have an endpoint on AdS boundary only when the symmetry is spontaneously broken. A contrapositive of this statement---if the symmetry is unbroken, a line operator cannot have an endpoint at the boundary---can be proven by a sequence of moves depicted above. More concretely, if the line operator had an endpoint on the boundary and there was no tilt operator in the conservation equation, it would be possible to unlink the charged line (turquoise) from the charge operator (fuchsia), as shown above. This contradicts the assumptions that the line operator is charged and  the symmetry operator is topological.
    The key step is that, since the boundary limit of 
    \(\left(\star \, \mathcal{J}\right)_{i_1\ldots i_{d - 1}} \) vanishes, we can remove the boundary segment of the topological operator and freely translate it as shown in the figure (the dashed part in the third figure indicates the segment which we can remove). 
    }
    \label{fig:contractible-topological-operator-line-endpoint}
\end{figure}
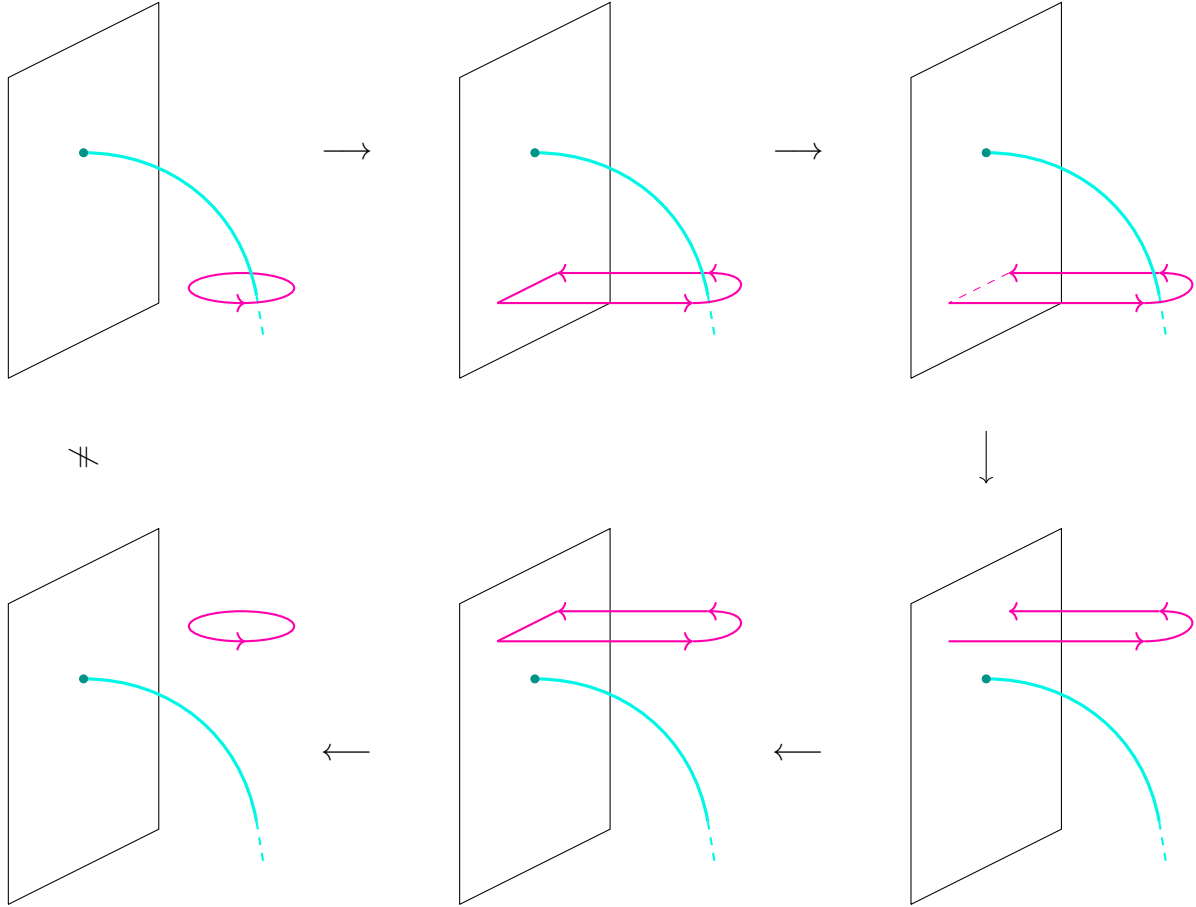

This argument can be generalized to a setup with a boundary: If the one-form symmetry is not (spontaneously) broken at the boundary, the charged line operators cannot have {\it boundary} endpoints. Again, the argument is based on the inconsistency between the action of the symmetry operator and the endpoints, and it can be shown through a sequence of manipulations depicted in figure \ref{fig:contractible-topological-operator-line-endpoint}. On the other hand, if the one-form symmetry is spontaneously broken at the boundary, the line operators can end at the boundary as we can see explicitly in the example of Maxwell theory.

Needless to say, the contrapositives of these statements are also true. Namely, if the line operators can have bulk endpoints, the one-form symmetry is explicitly broken in the bulk while, if they can have boundary endpoints, the one-form symmetry is either explicitly broken in the bulk or spontaneously broken at the boundary.



\subsection{Symmetries of Maxwell theory and QED in AdS}
Let us now examine one-form symmetries and their breakings, discussed above, in the concrete examples of free Maxwell theory\footnote{ See also recent works on free Maxwell theory with a boundary \cite{Arbalestrier:2025jsg}.} and QED in AdS. In what follows, we focus mainly on the case of $D=4$ ($d=3$), where the magnetic symmetry is a $1$-form symmetry\footnote{In a $D$-dimensional bulk, the magnetic symmetry is a $(D-3)$-form symmetry.}. On the other hand, many of the results concerning the electric $1$-form symmetry hold in arbitrary dimensions. We thus keep dimensions $D$ and $d$ general in the formulae below although it should be understood as $D=4$ and $d=3$ for the magnetic symmetries.
\paragraph{Free Maxwell theory.}

\begin{table}[]
    \centering
\begin{tabular}{|c|c|c|}
\hline
&Maxwell \(\mathcal{D}\) & Maxwell \(\mathcal{N}\)  \\
\hline & &  \\
$U(1)^{(1)}_{\mathcal{E}}$
& SB & \checkmark  \\[15pt]
\hline & &  \\
$U(1)^{(1)}_{\mathcal{M}}$& \checkmark  &  SB  \\[15pt]
\hline & & \\
Boundary & $U(1)^{(1)}_{\mathcal{M}}\times \left(U(1)^{(0)}_{\mathcal{E}}\right)_{\text{loc}}$ & $U(1)^{(1)}_{\mathcal{E}}\times \left(U(1)^{(0)}_{\mathcal{M}}\right)_{\text{loc}}$   \\[15pt]
\hline 
\end{tabular}
\caption{Summary of one-form symmetries of free Maxwell theory in AdS with Dirichlet (D) and Neumann (N) boundary conditions. SB stands for Spontaneously Broken, EB for Explicitly Broken, and \protect\checkmark   \, for Unbroken. The boundary symmetries are induced by the bulk ones, as explained in the text. Symmetries with the subscript $\Big( \quad \Big)_{\rm loc}$ are local symmetries generated by a conserved current while the other symmetries are non-local.}
\end{table}

It is well-known that free Maxwell theory in \(4\) dimensions enjoys two distinct \(1\)-form symmetries: the electric and magnetic \(U(1)^{(1)}\) symmetries, whose conserved currents are:
\begin{equation}
    \mathcal{J}_{\mathcal{E}} = \frac{1}{e} F, \quad\quad
    \mathcal{J}_{\mathcal{M}} = \frac{e}{2\pi} \star^{(D)} F.
\end{equation}
The conservation of the electric current $\mathcal{J}_{\mathcal{E}}$ follows from the equation of motion of free Maxwell theory $d\star^{(D)} F=0$ while the conservation of the magnetic current $\mathcal{J}_{\mathcal{M}}$ is ensured by the Bianchi identity $dF=0$.

The fate of these symmetries in the presence of the AdS boundary is determined by the asymptotics of field strength and its dual. For Dirichlet boundary conditions we have (see table \ref{table:F}): 
\begin{equation}
    \begin{split}
        F_{zi} \underset{z \rightarrow 0}{\sim} (d - 2) j_i(\vec{x}) + \mathcal{O}(z), 
        \quad\quad
        \left(\star^{(D)} F\right)_{kz} \underset{z \rightarrow 0}{\sim} \mathcal{O}(z).
    \end{split}
\end{equation}
Comparing them with the expected asymptotics \eqref{eq:ssb-1-form-symmetries-definition} for $d=3$, we find that the electric \(U(1)_{\mathcal{E}}^{(1)}\) symmetry is spontaneously broken with the tilt given by the current  $j_i$, while the magnetic one is unbroken. As discussed above, the SSB of the $1$-form symmetry in the bulk induces the $0$-form symmetry on the boundary, and the charge operator associated with this $0$-form symmetry can be obtained by pushing the entire support surface $\Sigma^{(2)}$ of the $1$-form charge operator to the boundary:
\begin{align}
Q_{\mathcal{E}}=\int_{\Sigma^{(2)}} \star^{(D)}J_{\mathcal{E}} \longrightarrow (d - 2)\int_{\Sigma^{(2)}} \star^{(d)}j\,.
\end{align}

On the other hand, with Neumann boundary conditions we have
\begin{equation}\label{eq:Neumanncurrenttilt}
    \begin{split}
        F_{iz} \underset{z \rightarrow 0}{\sim} \mathcal{O}(z),
        \quad\quad
        \left(\star^{(D)} F\right)_{i z} \underset{z \rightarrow 0}{\sim} \varepsilon_{ijk} \, f_{jk}(\vec{x}) + \mathcal{O}(z).
    \end{split}
\end{equation}
Thus, the magnetic \(U(1)_{\mathcal{M}}^{(1)}\) symmetry is spontaneously broken with the tilt given by (the Hodge dual of) a boundary field strength $\epsilon_{ijk}f_{jk}$, while the electric one remains unbroken. In this case, the charge operator for the induced boundary $0$-form symmetry is given by an integral of the boundary field strength:
\begin{align}
\label{eq:boundary-limit-magnetic-charge}
\begin{aligned}
Q_{\mathcal{M}} &= \int_{\Sigma^{(2)}} \star^{(D)} J_{\mathcal{M}} = \int_{\Sigma^{(2)}} \frac{e}{4\pi} F_{\alpha\beta} \left(\dd x^{\alpha} \wedge \dd x^{\beta}\right)
    =
    \frac{e}{2\pi}\int_{\Sigma^{(2)}} F_{\alpha\beta} \dd x^{\alpha} \dd x^{\beta}\\
&\longrightarrow\frac{e}{2\pi}\int_{\Sigma^{(2)}} f_{ij} \dd x^{i} \dd x^{j}.
\end{aligned}
\end{align}

Note that these patterns of symmetry breakings have consequences on charged line operators, as we discussed above. With Dirichlet boundary conditions, the Wilson lines can have boundary endpoints while they cannot with Neumann boundary conditions (and viceversa for 't Hooft lines).

\paragraph{QED with Neumann photon.}

\begin{table}
    \centering
\begin{tabular}{|c|c|c|c|}
\hline
& QED \(\mathcal{D}\): Coulomb & QED \(\mathcal{D}\): Higgs & QED \(\mathcal{N}\) \\
\hline & & &  \\
$U(1)^{(1)}_{\mathcal{E}}$
& EB \,$\rightarrow$ \,\checkmark$|_{\rm bdy}$\, $\rightarrow$\, SB$|_{\rm bdy}$ & EB & EB\\[15pt]
\hline & & &  \\
$U(1)^{(1)}_{\mathcal{M}}$& \checkmark  &  \checkmark & SB  \\[15pt]
\hline & & &  \\
Boundary & $U(1)^{(1)}_{\mathcal{M}}\times \left(U(1)^{(0)}_{\mathcal{E}}\right)_{\text{loc}}$ & $U(1)^{(1)}_{\mathcal{M}}$  & $\left(U(1)^{(0)}_{\mathcal{M}}\right)_{\text{loc}}$ \\[15pt]
\hline 
\end{tabular}
\caption{
Summary of one-form symmetries for QED in AdS with Dirichlet (D) and Neumann (N) boundary conditions. SB stands for Spontaneously Broken, EB for Explicitly Broken, and \protect\checkmark   \, for Unbroken. The  \(U(1)^{(0) / (1)}_{\mathcal{M}}\) factor of the boundary symmetries is induced by the bulk magnetic one-form symmetries, as explained in the text. In the Coulomb phase, the electric one-form symmetry $U(1)^{(1)}_{e}$ is explicitly broken in the bulk, restored near the boundary and spontaneously broken by the boundary condition as shown in the table.}
\label{tab:bulk-boundary-symmetries-maxwell-matter}
\end{table}

{
We now analyze the symmetries of Maxwell theory coupled to electrically charged matter, which is of direct relevance to the analyses in the preceding sections (see a summary in Table  \ref{tab:bulk-boundary-symmetries-maxwell-matter}). Let us first recall symmetries in the bulk: the matter fields in the bulk transform under the standard $0$-form \(U(1)_{\text{mat}}\) symmetry, which is however gauged and is not a genuine global symmetry of the theory. Among the two one-form symmetries of free Maxwell theory, the electric \(U(1)^{(1)}_{\mathcal{E}}\) symmetry is explicitly broken by the presence of charged matter, since the equation of motion is modified by the matter current, while the magnetic \(U(1)^{(1)}_{\mathcal{M}}\) symmetry remains as an exact global symmetry\footnote{Provided that no magnetically charged matter is introduced.}  since it is a consequence of the Bianchi identity.

We now consider the effect of the AdS boundary. Let us first focus on the photon with Neumann boundary conditions. As in the case of free Maxwell theory, this boundary condition spontaneously breaks the global \(U(1)^{(1)}_{\mathcal{M}}\) symmetry and induces a global $0$-form symmetry. Since the magnetic symmetry follows from the Bianchi identity and is not modified by the presence of charged matter, the $1$-form symmetry current in the bulk and the $0$-form symmetry current on the boundary are both identical to those in free Maxwell theory with Neumann boundary conditions (see \eqref{eq:Neumanncurrenttilt} and \eqref{eq:boundary-limit-magnetic-charge}). In particular, it follows that the conformal dimension of the boundary field strength $f_{ij}$, which is, up to contraction with $\varepsilon$, the tilt operator for $U(1)_{\mathcal{M}}^{(1)}$, is protected to be $\Delta=2$.

On the other hand, the electric $1$-form symmetry $U(1)_{\mathcal{E}}^{(1)}$, which manifests itself as a non-local boundary symmetry in free Maxwell theory with Neumann boundary conditions, ceases to exist in the presence of charged matter since the topological operator for this non-local symmetry always extend to the AdS bulk where the symmetry is explicitly broken.

\paragraph{QED with Dirichlet photon.}
We next consider photons with Dirichlet boundary conditions. In this case, the magnetic $1$-form symmetry \(U(1)^{(1)}_{\mathcal{M}}\) remains unbroken and survives on the boundary as a nonlocal symmetry. By contrast, the fate of the electric $1$-form symmetry and its induced boundary $0$-form symmetry requires more careful analysis and depends on the phase of the theory, as we see below.

As noted above, the electric one-form symmetry is explicitly broken in the bulk because the equation of motion in the presence of charged matter modifies the conservation law to\footnote{Here and below, we use the same normalization for the gauge field as in section \ref{sec:SecHiggs}, see equation \eqref{eq:maxwell-coupled-to-matter-generic}.}
\begin{align}\label{eq:brokencurrentconserv}
 \nabla_{\mu} \left(\mathcal{J}_{\mathcal{E}}\right)^{\mu\nu} = J^{\nu}_{\text{mat}}\,.
\end{align}
Although such explicitly broken symmetries might not seem to have direct physical implications, the presence of an AdS boundary can change this conclusion. Depending on the near-boundary scaling of both sides of this equation, the broken $1$-form symmetry may be effectively restored close to the boundary, inducing an exact $0$-form symmetry in the boundary theory.

To see this explicitly, we set $\nu=z$ in \eqref{eq:brokencurrentconserv} and analyze the boundary operator expansion of both sides. 
In the free limit $e\to 0$, the right hand side vanishes identically, while with Dirichlet boundary conditions the left-hand side gives 
\begin{align}
\nabla_{\mu} \left(\mathcal{J}_{\mathcal{E}}\right)^{\mu z} 
\underset{z\to 0}{\sim
}-\frac{1}{e} z^{d+1}(d-2)\partial_{i}j^{i}(\vec{x})\,~,
\end{align}
reproducing the conservation of the boundary current in free Maxwell theory. When the coupling $e$ is nonzero, the right-hand side can produce a nontrivial cotribution and we instead obtain
\begin{equation}\label{eq:tiltrec}
    \nabla_{\mu} \left(\mathcal{J}_{\mathcal{E}}\right)^{\mu\nu} =  J^{\nu}_{\text{mat}} \quad\quad \Longrightarrow \quad\quad   \partial_i j^i(\vec{x}) = - \frac{e}{d - 2} \lim_{z \rightarrow 0}z^{-(d+1)} J^{z}_{\text{mat}}(\vec{x}, z).
\end{equation}
Thus, if $J^{z}_{\text{mat}}(\vec{x}, z)$ vanishes faster than $z^{d+1}$ near the boundary, the boundary conservation law remains intact despite explicit breaking in the bulk. In this case, the $1$-form symmetry is effectively restored near the boundary and then spontaneously broken by the boundary condition, inducing a conserved boundary current (the tilt operator) associated with a 0-form symmetry. Physically, this corresponds to the {\it Coulomb phase} in the bulk, in which the photon remains massless.
By contrast, if $J^{z}_{\text{mat}}(\vec{x}, z) \sim z^{d+1}$, the boundary current $j^{i}$ is no longer conserved, and the induced $0$-form symmetry is lost. This corresponds to the {\it Higgs phase} in the bulk, with boundary current non-conservation reflecting the photon mass generation. Note that \eqref{eq:tiltrec} precisely matches the equation \eqref{eq:rec} for the recombination that we derived earlier.

We can confirm these predicted scalings of $J^{z}_{\rm mat}$ in different phases of the scalar QED. As discussed in sections \ref{sec:neumann-photon-exchange} and \ref{sec:SecHiggs}, when the two real scalars satisfy the same boundary condition, the $U(1)_{\rm mat}$ symmetry is preserved by the boundary and the theory is in the Coulomb phase with massless photons. The boundary expansion of $J_{{\rm mat},z} \propto \varphi_{[1}\partial_{z}\varphi_{2]}$ can be determined as
\begin{equation}
    \begin{split}
        \varphi_i(\vec{x}, z) &\underset{z \rightarrow 0}{\sim} z^{\frac{d}{2} + i\nu} \mathcal{O}_i(\vec{x}) + \text{ subleading}; \\
        \varphi_1\partial_z\varphi_2(\vec{x}, z) &\underset{z \rightarrow 0}{\sim} z^{d + 2i\nu - 1} 2\Delta_{\nu}\mathcal{O}_1 \mathcal{O}_2(\vec{x}) + \text{ subleading}; \\
        J_{\text{mat}, z}(\vec{x}, z) &\underset{z \rightarrow 0}{\sim} z^{d + 2i\nu + 1} \partial^i(\mathcal{O}_{[1}\partial_i\mathcal{O}_{2]})(\vec{x}) + \text{ subleading}, 
    \end{split}
\end{equation}
Note that the leading power in $ \varphi_{1}\partial_z \varphi_2$, $z^{d+2i\nu-1}$, is cancelled upon antisymmetrization $\varphi_{[1}\partial_z \varphi_{2]}$. Accounting for the metric contraction, we have \( J^{z}_{\text{mat}}(\vec{x}, z) \overset{z\to 0}{\sim} \mathcal{O}\left(z^{d + 3 + 2i\nu}\right)\), and the unitarity bound ensures \(d + 3 + 2i\nu \ge d + 1\). Thus, $J_{\rm mat}^{z}$ vanishes faster than $z^{d+1}$ and $j^{i}$ is still conserved at the boundary.  Note that this argument is valid irrespective of the mass of charged matter as long as the unitarity bound\footnote{Precisely speaking, the boundary conserved current exists even when the unitarity bound is violated, as long as the boundary condition for the scalar does not break the $U(1)$ symmetry: when the unitarity bound is violated, the boundary expansion of $J_{\rm mat,z}$ can produce a term more dominant near the boundary than $j_i$, arising from the left hand side. However such terms generally come with a different power of $z$ and one can still extract a term proportional to $z^{d+1}$ to show the current conservation.} is satisfied and it holds even for massless matter.
As in free Maxwell theory \eqref{eq:limit-bulk-1-form-bdy-0-form-charge}, the boundary $0$-form charge is simply the limit of the bulk $1$-form charge, which is preserved when placed near the boundary;
\begin{equation}
    Q_{\mathcal{E}} = \int_{\Sigma^{(d - 1)} \subseteq \partial \text{AdS}} \star^{(D)} J_{\mathcal{E}}=\int_{\Sigma^{(d - 1)} \subseteq \partial \text{AdS}} \star^{(d)} j.
\end{equation}
This boundary 0-form symmetry is usually attributed to large  gauge transformations in the bulk (i.e.~{\it asymptotic symmetries}). Our analysis shows instead that it can be understood purely as a consequence of the $1$-form symmetry effectively restored near the boundary, without invoking gauge transformations. In this sense, it is similar in spirit to a recent (re-)derivation of soft theorems based on generalized symmetries \cite{Berean-Dutcher:2025ohp}, see also \cite{Lake:2018dqm}.

Instead, if the two scalars satisfy the opposite boundary conditions, the photon is screened and the theory is in the Higgs phase. In this case, the leading power for $\varphi_1 \partial_z \varphi_2$ is modified to $(\frac{d}{2}\pm i\nu) z^{d-1} \mathcal{O}_1\mathcal{O}_2$ and it will not be canceled by the anti-symmetrization. We thus recover the multiplet recombination equation \eqref{eq:rec}, 
\begin{equation}
    \nabla_{\mu} \left(\mathcal{J}_{\mathcal{E}}\right)^{\mu z} = e J^{z}_{\text{mat}} \quad\quad \Longrightarrow \quad\quad  z^{d + 1} (d - 2)\partial^i j_i(\vec{x}) = z^{d + 1} \, e \underbrace{2i\nu \mathcal{O}_1 \mathcal{O}_2(\vec{x})}_{t(\vec{x})}.
\end{equation}

We can also study the standard Higgs phase in which a charged scalar acquires a classical VEV. As discussed around \eqref{eq:JzclassicalHiggs}, the boundary limit of $J_z^{\rm mat}(z,\vec{x}) $ behaves as $J^{\rm mat}_z (z,\vec{x})\propto z^{d-1} v^2   t(\vec{x})$, where $v$ is the classical VEV of the Higgs field and $t(\vec{x})$ is a marginal scalar dual to the Goldstone mode associated with SSB of $U(1)_{\rm mat}$. Thus, $J_{\rm mat}^{z}$ scales as $z^{d+1}$  and the current conservation is violated. 

}

\begin{figure}
	\centering
	\begin{tikzpicture}
	\draw[green, thick] (5, 1.7) to[out = -70, in = -110, looseness = 2] (9, 1.7);
	\node[fuchsia, thick] at (5, 1.7) {\(\times\)};
	\node[fuchsia, thick, below] at (4.9, 1.6) {\(P_1\)};
	\node[fuchsia, thick] at (9, 1.7) {\(\times\)};
	\node[fuchsia, thick, below] at (9.1, 1.6) {\(P_2\)};
	\draw[turquoise, very thick, dashdotted] (5, 1.7) ellipse (2 and 0.7);
	\node[turquoise, thick, below] at (3.2, 1.4) {\(\Sigma^{(2)}\)};
	\draw[black, thick] (0,0) -- (12, 0) -- (15, 3) -- (3, 3) -- (0, 0);
	\end{tikzpicture}
	\caption{Boundary picture of the surface \(\Sigma^{(2)}\) surrounding only the insertion point of operator \(\mathcal{O}_1\). The Wilson line is sent along the bulk geodesic connecting the two points.}
	\label{fig:surface-sigma-single-operator}
\end{figure}
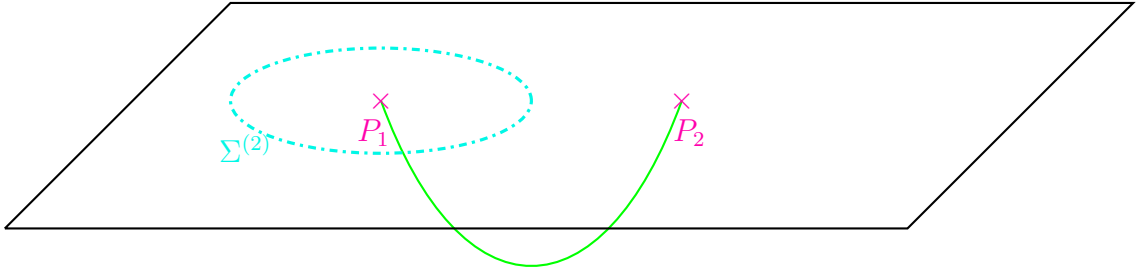

\subsection{Decoupling of boundary field strength}\label{subsec:decoupling_of_F_from_symmetry}
We now provide a non-perturbative explanation for the decoupling of $f_{ij}$ noted at the end of section \ref{sec:neumann-photon-exchange} leveraging on the global symmetries and conformal invariance of the theory. 

Consider the flux through a boundary surface \(\Sigma^{(2)}\) surrounding only the insertion of \(\mathcal{O}_{\Delta_1}\) (see figure \ref{fig:surface-sigma-single-operator}):
\begin{equation}
    \label{eq:magnetic-charge-boundary}
	\int_{\Sigma^{(2)}}
	\langle 
	\mathcal{O}_{\Delta_1}(P_1) W\left[\gamma(P_1, P_2)\right]
	\mathcal{O}_{\Delta_2}(P_2) 
	f_{ij}(P_3)
	\rangle \dd x_3^i \dd x_3^j.
\end{equation}
Notice we had to dress the matter operators with the Wilson line, otherwise the object in  \eqref{eq:magnetic-charge-boundary} would not be gauge invariant.
By the argument in \eqref{eq:boundary-limit-magnetic-charge}, \eqref{eq:magnetic-charge-boundary} computes the charge of the operator \(\mathcal{O}_{\Delta_1}\) under the magnetic boundary $0$-form symmetry. However, we can compute this alternatively by moving the surface $\Sigma^{(2)}$ into the bulk and shrinking it around the bulk Wilson line. This measures the magnetic $1$-form charge of the Wilson line in the bulk, which is zero. We thus conclude the flux integral \eqref{eq:magnetic-charge-boundary} vanishes.

On the other hand, due to the conformal invariance at the boundary, the three-point function in \eqref{eq:magnetic-charge-boundary} admits a single tensor structure; this is because in \(d = 3\) we can write \(f_{ij} = \epsilon_{ijk} \tilde{j}_k\), and the three-point function of two scalars and a (traceless symmetric) spin \(\ell\) operator is proportional to a unique tensor structure (see \cite{Rychkov:2016iqz} among others):
\begin{equation}
	\begin{aligned}
		&\varepsilon_{ijk}\langle \mathcal{O}_{\Delta_1}(P_1) \mathcal{O}_{\Delta_2}(P_2) \tilde{j}_{k}(P_3) \rangle = \frac{c_{\mathcal{O}_{\Delta_1} \mathcal{O}_{\Delta_2} f}}{|P_{12}|^{\Delta_{123}} |P_{13}|^{\Delta_{132}} |P_{23}|^{\Delta_{231}}} 
		\varepsilon_{ijk} R_{k}(P_1, P_2 \vert P_3)~,\\
		&\Delta_{ijk} := \frac{\Delta_i + \Delta_j - \Delta_k}{2}~, \quad
		R_{k}(P_1, P_2 \vert P_3) := \frac{\partial P_3^M}{\partial x^{k}}
		\left[
			\frac{\left(P_1 \cdot P_3\right)P_{2, M} - \left(P_2 \cdot P_3\right)P_{1, M}}
			{\sqrt{\left(P_1 \cdot P_2\right) \left(P_1 \cdot P_3\right) \left(P_2 \cdot P_3\right)}}
		\right]~.
	\end{aligned}
    \label{eq:tensor-structure-current-scalar-scalar}
\end{equation}
Performing the flux integral \eqref{eq:magnetic-charge-boundary} of the right hand side of \eqref{eq:tensor-structure-current-scalar-scalar}, we find a non-zero kinematical factor\footnote{Note that the relevant integral precisely measures the flux for $0$-form symmetries on the boundary and common to any boundary theory with $0$-form symmetry. Hence, it must be nonzero and finite.} times the OPE coefficient $c_{\mathcal{O}_{\Delta_1} \mathcal{O}_{\Delta_2} f}$. We thus conclude that
\(c_{\mathcal{O}_{\Delta_1} \mathcal{O}_{\Delta_2} f} = 0\) at the non-perturbative level;
therefore, neither the field strength \(f_{ij}\), or any of its descendants, can appear in the OPE of \(\mathcal{O}_{\Delta_1}(P_1) W\left[\gamma(P_1, P_2)\right]\mathcal{O}_{\Delta_2}(P_2)\).
Note that, to derive this result, it was important to use the Wilson line sent along the geodesics (rather than arbitrary curves in the bulk) since it enables us to constrain the three-point using the conformal symmetry. 

This discussion was specific to \(D = 3 + 1\) dimensions. In higher dimensions, the electric \(U(1)^{(1)}_{\mathcal{E}}\) is still a $1$-form symmetry while the magnetic symmetry becomes a \((D - 3)\)-form symmetry. The patterns of the symmetry breaking summarized in  
table \ref{table:F} still hold in higher dimensions. However it is less physically interesting for two reasons. First the tilt operator \(f_{ij}\) falls below unitarity bound for $D>4$ and hence the Neumann boundary condition is non-unitary. Second, the decoupling of $f_{ij}$ from scalars is guaranteed by the conformal symmetry alone, since for $d>3$, there is no conformally-invariant tensor structure one can write among two scalar operators and one rank-2 antisymmetric tensor. 


\paragraph{Dirichlet photon.}
{
As a final remark, let us point out that a similar decoupling takes place even for scalar QED with Dirichlet photons in the Coulomb phase, if we dress charged scalar operators with a geodesic Wilson line. From the diagrammatic perspective, the computation to show this is completely analogous to the Neumann one (as Dirichlet and Neumann only differ by the integration contour of the spectral integral). 
Again, this can be explained by leveraging on a global symmetry, namely the boundary $0$-form symmetry \(U(1)_{\mathcal{E}}^{(0)}\).

Similarly to how we treated the Neumann case, we consider the integrated three-point function of the boundary conserved current \(j^i(\vec{x}) \) and two scalars (attached to a bulk geodesic Wilson line).
On one hand, this is governed by the same tensor structure as in  \eqref{eq:tensor-structure-current-scalar-scalar} and is given by a kinematical factor times the OPE coefficient \(c_{\mathcal{O}_{\Delta_1} \mathcal{O}_{\Delta_2} j}\). On the other hand,  the Wilson line dressing makes operators inserted in \(P_1\) and \(P_2\) charge-neutral, since under action of the charge \(U(1)\) Wilson lines end-points transform with charge opposite to those of the respective operator \(\mathcal{O}_{\Delta_i}\). Since the integrated three-point function measures the charge of the object at $P_1$, we get
\begin{equation}
	\label{eq:charge-operator-plus-Wilson-line}
	\int_{\Sigma^{(d-1)}}
	\langle \mathcal{O}_{\Delta_1}(P_1) W\left[\gamma(P_1, P_2)\right]\mathcal{O}_{\Delta_2}(P_2) \left(\star^{(d)} j\right)(P_3) \rangle = 0.
\end{equation}
So again, we infer that the OPE coefficient \(c_{\mathcal{O}_{\Delta_1} \mathcal{O}_{\Delta_2} j}\) vanishes and we conclude that the ``photon multiplet" $j^{i}$ decouples from the OPE of the Wilson line dressed operators.

Note that the decoupling does not happen in the Higgs phase in which the charge-breaking boundary conditions are imposed on the scalar fields ($\nu_1\neq \nu_2$). In this case, the current $j^{i}(\vec{x})$ is no longer conserved as we saw above, and we find that, even if we dress the scalar fields with the geodesic Wilson line, the OPE expansion of 
\begin{equation}
\langle\varphi_1(P_1)W_{\gamma}[P_1, P_2]\varphi_2(P_2)\varphi_1(P_3)W_{\gamma}[P_3, P_4]\varphi_2(P_4)\rangle
\end{equation}
produces a conformal block corresponding to the exchange of the boundary {\it non-conserved} current $j^{i}$, whose dimension in perturbation theory is close to \(\Delta = d - 1\).  (At loop order, its dimension gets corrected since the bulk photon dual to it develops mass given in \eqref{eq:phmassphisqd3}).
}

%% file: conclusions.tex
In this paper we studied the dynamics of abelian gauge theories with matter in the background of Anti-de Sitter space, investigating the effect of boundary conditions, and using the example of scalar QED. In particular, we analyzed in details boundary conditions of the matter fields that break the symmetry arising from the gauge field with Dirichlet boundary condition. We saw that this setup gives rise to a type of Higgs mechanism peculiar to AdS. In addition, we have studied the Neumann boundary condition for the gauge field. In this case we have proposed, and computed perturbatively, a new boundary observable that is the closest analogue to the 4-point function of the matter field that is allowed by gauge-invariance. The observable is obtained by dressing the four-point function with Wilson lines operators, placed along geodesics of the bulk to ensure compatibility with conformal invariance at the boundary. 

Beyond explicit computations, we also developed symmetry-based arguments that do not rely
on perturbation theory. First, we derived the photon mass in AdS directly from Ward
identities and multiplet recombination, providing a general explanation of the AdS Higgs
phenomenon. Second, we analyzed higher-form symmetries in AdS and their spontaneous
breaking. In particular, we identified the tilt operator associated with the bulk electric $1$-form symmetry as a
conserved current for an induced boundary $0$-form symmetry. Importantly, we showed that this
conserved current persists even in the presence of bulk charged matter, which explicitly
breaks the bulk $1$-form symmetry. This is because the $1$-form symmetry is
effectively restored near the AdS boundary. Our argument offers a new generalized-symmetry-based perspective on a
familiar statement in AdS/CFT: a bulk $U(1)$ gauge symmetry implies a conserved
$U(1)$ current on the boundary. Our analysis shows that this connection can be
understood without invoking gauge symmetry, relying instead on generalized global symmetries. Furthermore, for photons with Neumann boundary conditions, we used the magnetic 1-form symmetry to provide a nonperturbative argument for the decoupling of boundary field strengths from Wilson-line–dressed charged scalars. The result provides a nice example of how generalized symmetries can be used to yield concrete dynamical predictions of the theory.

Several avenues remain open for future investigation. One clear direction is to apply these results to gauge theories in de Sitter space. For QFT in dS, unlike AdS, the conformal boundary is the spacelike boundary at future infinity. Once the bulk state is fixed, there is no freedom in picking the behavior of the bulk fields as they approach this boundary. Therefore, while in AdS one can pick between two possible boundary modes for the bulk fields, typically in dS both of those modes are activated at the late time boundary. This observation is at the basis of a well-established connection between quantum field theories in dS and AdS \cite{Sleight:2019hfp, Sleight:2020obc, Sleight:2021plv, di2022analyticity}. The implication for gauge theories is that both the Neumann and the Dirichlet modes of the gauge field are present in dS. Similarly, for the matter fields, both modes are turned on, giving rise also to the symmetry breaking behavior near the boundary. As a result, both of the phenomena studied in this paper are bound to play an important role. In particular, the dressed correlators of charged matter that we considered in AdS are a promising starting point to define conformally-invariant late-time correlators of charged matter fields in dS, which so far remained elusive, see e.g. \cite{Woodard:2004ut, Glavan:2022nrd}. An important motivation is that eventually one would like to extend the definition of these dressed observables to theories with dynamical gravity, and explore the existence of conformally-invariant correlation functions at the late times in dS even in the presence of a dynamical graviton.

In view of this application, but also remaining in the context of AdS with Neumann boundary condition, a compelling direction for the future is also to study in more depth the nature of this dressed observables. It would be nice to better understand how the operators at the endpoint of geodesic Wilson lines fit within the structure of the boundary conformal field theory, for instance whether a sensible OPE can be defined for them, and whether crossing symmetry applies in some form to their correlation functions. At a technical level, the discussion of crossing symmetry requires to include in the dressed correlator also the t-channel exchange diagrams, which we have not considered in this work. It is also possible to consider different configurations of Wilson lines, and possibly their superpositions. It would also be desirable to compare this approach to obtain gauge-invariant observables to other, perhaps more standard, approaches based on gauge-fixing or BRST symmetry, see for instance the recent nice discussion in \cite{Grassi:2024vkb}. 

QFT in AdS can be also applied to flat-space physics by means of the flat-space limit \cite{Paulos:2016fap, Komatsu:2020sag, vanRees:2022zmr}. The latter is however more subtle when massless fields and long-range forces are present, see for instance the review in \cite{Li:2021snj} and references therein. Perhaps relatedly, in this context there are additional difficulties in defining scattering observables in flat space, due to infra-red divergences. It would be interesting to understand, in the spirit of the recent studies for non-abelian gauge theories \cite{Ciccone:2024guw, Gabai:2025hwf, Ciccone:2025dqx, DiPietro:2025ozw}, which one of the possible boundary conditions for scalar QED is more appropriate to describe scattering in the flat space limit. Once this is understood, one could explore the definition of IR-finite scattering observables from the limit of properly dressed correlation functions at the boundary of AdS.

It would also be interesting to develop further the generalized-symmetry-based arguments presented in sections \ref{sec:SecHiggs} and \ref{sec:one-form-symmetries}. Our derivation of a boundary conserved $0$-form current from a spontaneously broken $1$-form symmetry in the bulk shares a conceptual similarity with the derivation of the soft theorem from higher-form symmetries in \cite{Berean-Dutcher:2025ohp}. It would be interesting to see if our discussion can be generalized to other cases discussed in that context, such as the spontaneously-broken $2$-group symmetry. We also expect similar symmetries to play an important role in gauge theories in dS.


%% file: propagators-details.tex
This appendix is dedicated to exposing all the details about the derivation of the propagators described in section \ref{sec:propagators}. For both of the possible boundary condition we will first argue that the proposed solution solves the equation of motion \eqref{eq:equation-of-motion-embedding}, and then we will show that it satisfies the correct boundary conditions.

\subsection{The Dirichlet propagator}
We will derive the Dirichlet propagator in the spectral representation; since it is a spin \(1\) object we need to expand on spin-1 eigenfunctions of the EAdS Laplacian (the orthonormal over-complete functional basis of harmonic functions) and it will be of the form 

\begin{equation}
    \begin{aligned}
        \Pi^{\mathcal{D}}_{d - 1}(X_1, X_2; W_1, W_2) =&
        \int_{-\infty}^{+\infty}\dd \lambda \, \Pi^{\perp}_{\mathcal{D}}(\lambda) \, \Omega^{(1)}_{\lambda}(X_1, X_2; W_1, W_2) \\
        &+\int_{-\infty}^{+\infty}\dd \lambda \, \Pi^{\parallel}_{\mathcal{D}} (\lambda)\,  \left(W_1 \cdot \nabla_1\right)\left(W_2 \cdot \nabla_2\right) \Omega^{(0)}_{\lambda} (X_1, X_2)~.
    \end{aligned}
\end{equation}

In the following, we will refer to the part proportional to \(\Omega^{(1)}_{\lambda}\) and \(\Omega^{(0)}_{\lambda}\) as {\it transverse} and {\it longitudinal} respectively.

\subsubsection{The transverse part}
The spin \(1\) harmonic function satisfies
\begin{align}
    - \nabla_1^2 \Omega^{(1)}_{\lambda}(X_1, X_2; W_1, W_2) &= \left(\lambda^2 + \frac{d^2}{4} + 1\right) \Omega^{(1)}_{\lambda}(X_1, X_2; W_1, W_2)~, \\
    \left(K_1 \cdot \nabla_1\right)\Omega^{(1)}_{\lambda}(X_1, X_2; W_1, W_2) &= 0~.
\end{align}\,
where \(K_{i, A}\) is the projector operator, as defined in  \cite{costa2014spinning}, needed to recover space-time indices contracted with polarization vector \(W_i\):
\begin{equation}
    \label{eq:K-index-projector-definition}
    \begin{aligned}
        K_A &=   \frac{d - 1}{2}\left[\frac{\partial}{\partial W^A} - X_A \left(X \cdot \frac{\partial}{\partial W}\right)\right] 
        + \left(W \cdot \frac{\partial}{\partial W}\right) \frac{\partial}{\partial W^A}\\
        & \quad+ X_A \left(W \cdot \frac{\partial}{\partial W}\right)\left(X \cdot \frac{\partial}{\partial W}\right) 
        - \frac{1}{2}W_A \left[
            \frac{\partial^2}{\partial W \cdot \partial W}
            +\left(X \cdot \frac{\partial}{\partial W}\right)\left(X \cdot \frac{\partial}{\partial W}\right)
        \right]~,
    \end{aligned}
\end{equation}
\(K-\)projector is coming with a \(\frac{1}{J!\left(\frac{d - 1}{2}\right)_J}\) normalization factor when recovering the indices of a generic spin-\(J\) field:
\begin{equation}
    H_{A_1\ldots A_J} = \frac{1}{J!\left(\frac{d - 1}{2}\right)_J}K_{A_1}\ldots K_{A_J} \, H(X, W)~.
\end{equation}

The action of the equation of motion on the transverse part then is
\begin{equation}
    \begin{aligned}
        \left[-\nabla_1^2 - d +\frac{1}{\frac{d - 1}{2}}\right. & \left. \,\left(1 - \frac{1}{\xi}\right)(W_1\cdot\nabla_1)(K_1\cdot\nabla_1)\right] \int_{-\infty}^{+\infty}\dd \lambda \, \Pi^{\perp}_{\mathcal{D}}(\lambda) \, \Omega^{(1)}_{\lambda}(X_1, X_2; W_1, W_2)  \\
        & \mspace{60mu}=  \int_{-\infty}^{+\infty}\dd \lambda \, \left(\lambda^2 + \left(\frac{d}{2} - 1\right)^2\right)\Pi^{\perp}_{\mathcal{D}}(\lambda) \, \Omega^{(1)}_{\lambda}(X_1, X_2; W_1, W_2)
    \end{aligned}
\end{equation}
so, by choosing 
\begin{equation}
    \Pi^{\perp}_{\mathcal{D}}(\lambda) = \frac{1}{\lambda^2 + \left(\frac{d}{2} - 1\right)^2}~,
\end{equation}
we obtain the desired \(\delta-\)function, up to a longitudinal term (this is related to the (over)--completeness relation for \(\Omega^{(1)}_{\lambda}\)):
\begin{equation}
    \begin{aligned}
        \int_{-\infty}^{+\infty}\dd \lambda \, \Omega^{(1)}_{\lambda}(X_1, X_2; W_1, W_2)& = \left(W_1 \cdot W_2\right) \delta^{d + 1}(X_1, X_2) \\
        & \quad - \left(W_1\cdot \nabla_1 \right)\left(W_2\cdot \nabla_2 \right) \int_{-\infty}^{+\infty}\dd \lambda \, \frac{1}{\lambda^2 + \frac{d^2}{2}}\Omega^{(0)}_{\lambda}(X_1, X_2) \,.
    \end{aligned}
\end{equation}

\subsubsection{The longitudinal part}
\label{subsec:Dirichlet-propagator-longitudinal-part}
The spin--0 harmonic function satisfies
\begin{equation}
    \begin{aligned}
        - \nabla_1^2 &\left[\left(W_1\cdot \nabla_1 \right)\left(W_2\cdot \nabla_2 \right)\Omega^{(0)}_{\lambda}(X_1, X_2)\right] \\
        & \qquad= \left(W_1\cdot \nabla_1 \right)\left(W_2\cdot \nabla_2 \right)\left[d - \nabla_1^2\right]\Omega^{(0)}_{\lambda}(X_1, X_2) \\
        & \qquad= \left(\lambda^2 + \frac{d^2}{4} + d\right) \left(W_1\cdot \nabla_1 \right)\left(W_2\cdot \nabla_2 \right)\Omega^{(0)}_{\lambda}(X_1, X_2)~.
    \end{aligned}
\end{equation}
\begin{equation}
    \begin{aligned}
        \left(K_1 \cdot \nabla_1\right)&\left(W_1\cdot \nabla_1 \right)\left(W_2\cdot \nabla_2 \right)\Omega^{(0)}_{\lambda}(X_1, X_2)   \\
        & \qquad= \left(W_2\cdot \nabla_2 \right) \frac{d - 1}{2} \nabla_1^2 \,\Omega^{(0)}_{\lambda}(X_1, X_2) \\
        & \qquad= -\frac{d - 1}{2} \left(\lambda^2 + \frac{d^2}{4}\right)\left(W_2\cdot \nabla_2 \right)\,\Omega^{(0)}_{\lambda}(X_1, X_2)~.
    \end{aligned}
\end{equation}

\textbf{Remark:}
    It is important to note that, even if the embedding space is flat, embedding--derivatives with respect to the same point do not commute.
    In particular, for the Levi-Civita connection
    \[
    \left[\nabla_1^2, (W_1\cdot\nabla_1)\right] = -d \, (W_1\cdot\nabla_1)~.    
    \]

The longitudinal part of the equation of motion then is
\begin{align*}
    0 &= - \frac{1}{\lambda^2 + \frac{d^2}{4}} + \Pi^{\parallel}_{\mathcal{D}} \, \left[\lambda^2 + \frac{d^2}{4} - \left(1 - \frac{1}{\xi}\right)\left(\lambda^2 + \frac{d^2}{4}\right)\right]
\end{align*}
which is solved by
\begin{equation}
    \Pi^{\parallel}_{\mathcal{D}}(\lambda) = \frac{\xi}{\left(\lambda^2 + \frac{d^2}{4}\right)^2}~.
\end{equation}

\subsubsection{The boundary limit}
As remarked in \eqref{eq:Dirichlet-propagator}, we have shown that
\begin{equation}
    \begin{aligned}
        \Pi^{\mathcal{D}}_{d - 1}(X_1, X_2; W_1, W_2) =&
        \int_{-\infty}^{+\infty}\dd \lambda \, \frac{1}{\lambda^2 + \left(\frac{d}{2} - 1\right)^2} \, \Omega^{(1)}_{\lambda}(X_1, X_2; W_1, W_2) \\
        &+\int_{-\infty}^{+\infty}\dd \lambda \, \frac{\xi}{\left(\lambda^2 + \frac{d^2}{4}\right)^2} \,  \left(W_1 \cdot \nabla_1\right)\left(W_2 \cdot \nabla_2\right) \Omega^{(0)}_{\lambda} (X_1, X_2)
    \end{aligned}
\end{equation}
is a particular solution to the equation of motion.
To prove this is the actual Dirichlet propagator, we are left to show that it satisfies the correct boundary conditions. It is convenient to introduce 
\[
u := \frac{(X_1 - X_2)^2}{2} = - (1 + X_1 \cdot X_2)~.
\]
In particular, in Poincaré coordinates
\begin{equation}
    u = \frac{(z_1 - z_2)^2 + \vert \vec{x}_{12}\vert^2}{2z_1 z_2}~,
\end{equation}
and the boundary limit is obtained for \(z_1 \rightarrow 0\), for which \(u \sim \frac{\vert \vec{x}_{12}\vert^2 + z_2^2}{2z_2} \frac{1}{z_1} \rightarrow \infty\).

By EAdS isometries, the propagator will only depend on \(u\), so, to work out the boundary limit it is convenient to move to the tensor basis 
\[
    \Pi^{\mathcal{D}}_{d - 1, \mu\nu} (x_1, x_2)
    = -F_0^{\mathcal{D}} \frac{\partial^2 u}{\partial x_1^{\mu}\partial x_2^{\nu}} 
    + F_1^{\mathcal{D}}\frac{\partial u}{\partial x_1^{\mu}}\frac{\partial u}{\partial x_2^{\nu}}~.
\]
In particular, the harmonic functions can be written as
\begin{align}
    \label{eq:harmonic-function-spin-1}
    \Omega^{(1)}_{\lambda}(X_1, X_2; W_1, W_2) 
    &= \omega_{0,\lambda}(u) (W_1 \cdot W_2)+ \omega_{1,\lambda}(u) (X_1\cdot W_2)(X_2 \cdot W_1), \\
    \label{eq:double-derivative-harmonic-function-spin-0}
    \left(W_1\cdot \nabla_{1}\right)\left(W_2\cdot \nabla_{2}\right)\Omega^{(0)}_{\lambda}(X_1, X_2) 
    &=  - \frac{\partial \Omega^{(0)}_{\lambda}(u)}{\partial u}(W_1 \cdot W_2) 
    + \frac{\partial^2 \Omega^{(0)}_{\lambda}(u)}{\partial u^2}(X_1\cdot W_2)(X_2 \cdot W_1),
\end{align}
where explicitly (see appendix F of \cite{Loparco:2023rug} for example):
\begin{align}
    \label{eq:omega-0-definition}
    &\begin{aligned}
    \omega_{0,\lambda}(u) & = 
        \frac{\lambda\sinh(\pi\lambda)(d^2 + 4\lambda^2)\Gamma\left(\frac{d}{2} - 1 \pm i \lambda\right)}
        {2^{d + 4}\pi^{\frac{d + 3}{2}}\Gamma\left(\frac{d + 3}{2}\right)} \\
        &\hspace{1cm}\left[
            (d + 1)\, {}_2F_1\left(\frac{d}{2} + i \lambda, \frac{d}{2} - i \lambda, \frac{d + 1}{2}; - \frac{u}{2}\right) \right. \\
            &\hspace{2cm}-(1 + u)\left.{}_2F_1\left(\frac{d}{2} + 1 + i \lambda, \frac{d}{2} + 1 - i \lambda, \frac{d + 3}{2}; - \frac{u}{2}\right)
            \right],
    \end{aligned} \\
    \label{eq:omega-1-definition}
    &\begin{aligned}
        \omega_{1,\lambda}(u) & = 
            \frac{\lambda\sinh(\pi\lambda)(d^2 + 4\lambda^2)\Gamma\left(\frac{d}{2} - 1 \pm i \lambda\right)}
            {2^{d + 4}\pi^{\frac{d + 3}{2}}\Gamma\left(\frac{d + 3}{2}\right)} 
            \, \frac{1}{u (2 + u)}\\
            &\hspace{1cm}\left[
                (d + 1)(1 + u)\, {}_2F_1\left(\frac{d}{2} + i \lambda, \frac{d}{2} - i \lambda, \frac{d + 1}{2}; - \frac{u}{2}\right) \right. \\
                &\hspace{2cm}-\left(d + (1 + u)^2\right)\left.{}_2F_1\left(\frac{d}{2} + 1 + i \lambda, \frac{d}{2} + 1 - i \lambda, \frac{d + 3}{2}; - \frac{u}{2}\right)
                \right],
        \end{aligned} \\
    \label{eq:Omega-0-definition}    
    &\begin{aligned}
            \Omega_{\lambda}^{(0)}(u) & = 
                \frac{1}
                {\left(4\pi\right)^{\frac{d + 1}{2}}\Gamma\left(\frac{d + 1}{2}\right)} 
                \frac{\Gamma\left(\frac{d}{2} \pm i \lambda\right)}{\Gamma\left( \pm i \lambda\right)}
                \,
                {}_2F_1\left(\frac{d}{2} + i \lambda, \frac{d}{2} - i \lambda, \frac{d + 1}{2}; - \frac{u}{2}\right).
        \end{aligned}
\end{align}
Taking the pull-back on EAdS from the embedding space
\[
    (W_1 \cdot W_2) \mapsto - \frac{\partial^2 u}{\partial x_1\partial x_2},
    \quad\quad
    (X_1\cdot W_2)(X_2 \cdot W_1) \mapsto \frac{\partial u}{\partial x_1}\frac{\partial u}{\partial x_2}
\]
so
\begin{align}
    F_0^{\mathcal{D}}(u) &= 
    \int_{-\infty}^{+\infty}\dd \lambda \, 
    \frac{1}{\lambda^2 + \left(\frac{d}{2} - 1\right)^2} \, \omega_{0,\lambda}(u) - \frac{\xi}{\left(\lambda^2 + \frac{d^2}{4}\right)^2} \, \frac{\partial \Omega^{(0)}_{\lambda}(u)}{\partial u}, \\
    F_1^{\mathcal{D}}(u) &=
    \int_{-\infty}^{+\infty}\dd \lambda \,
    \frac{1}{\lambda^2 + \left(\frac{d}{2} - 1\right)^2} \, \omega_{1, \lambda}(u)
    + \frac{\xi}{\left(\lambda^2 + \frac{d^2}{4}\right)^2} \, \frac{\partial^2 \Omega^{(0)}_{\lambda}(u)}{\partial u^2}.
\end{align}
At this point we can expand the integrand for \(F_0^{\mathcal{D}}(u)\) in the limit \(u \rightarrow \infty\), and we find
\begin{align}
    \label{eq:F0-integrand-boundary-expansion}
    \begin{split}
         F_0^{\mathcal{D}}(u) 
        \underset{u\to\infty}{\sim} -\frac{1}{2\pi ^{\frac{d + 4}{2}}}\int_{-\infty}^{+\infty}&\dd \lambda  \, 
        \frac{ \Gamma (1+i \lambda )(2 \lambda +i d) \sinh (\pi  \lambda ) \Gamma \left(\frac{d}{2}-i \lambda -1\right)}
        {(d-2)^2+4 \lambda ^2} 
         \left(\frac{1}{2u}\right)^{\frac{d}{2}-i \lambda }\\
        &+
        \frac{\Gamma(1-i \lambda )(2 \lambda -i d) \sinh (\pi  \lambda ) \Gamma \left(\frac{d}{2}+i \lambda -1\right)}{(d-2)^2+4 \lambda ^2}
         \left(\frac{1}{2u}\right)^{\frac{d}{2}+i \lambda }~.
    \end{split}
\end{align}
We can perform the spectral integral in the complex plane by pulling the contour above for the first term and below for the second term, and we find
\begin{equation}
    \begin{aligned}
        F_0^{\mathcal{D}}(u) 
        &\underset{u \to \infty}{\sim} \frac{\Gamma \left(\frac{d + 1}{2}\right)}
        {2\pi ^{\frac{d + 1}{2}}} \,\frac{1}{d - 2} \, \frac{1}{u^{d - 1}} + \mathcal{O}\left(\frac{1}{u^d}\right).
    \end{aligned}
\end{equation}
as reported in \eqref{eq:Dirichlet-propagator-boundary-limit}.

By means of the boundary behaviour of the gauge propagator in Poincaré coordinates\footnote{In order to derive the correct behaviour of the \(A_z\) compontent it is necessary to expand \(F_0^{\mathcal{D}}(u)\) and \(F_1^{\mathcal{D}}(u)\) beyond the leading order:
\begin{equation}
F_0^{\mathcal{D}}(u) 
\underset{u \rightarrow \infty}{\sim}
\frac{\Gamma \left(\frac{d + 1}{2}\right)}
        {2\pi ^{\frac{d + 1}{2}}} \, \frac{1}{d - 2} \, \left(\frac{1}{u^{d - 1}} - \frac{d - 1}{u^d}\right),
\quad\quad
F_1^{\mathcal{D}}(u) 
\underset{u \rightarrow \infty}{\sim}
\frac{\Gamma \left(\frac{d + 1}{2}\right)}
        {2\pi ^{\frac{d + 1}{2}}} \, \frac{1}{d - 2} \, \left(\frac{1}{u^{d}} - \frac{d}{u^{d + 1}}\right).
\end{equation}
}, we can derive the operator limit of the field \(A_{\mu}^{\mathcal{D}}(x, z)\) as \(z \rightarrow 0\):
\begin{equation}
    \begin{aligned}
         \frac{1}{\mathcal{C}^{(1)}_{d - 1}}
        \langle A^{\mathcal{D}}_{i}(x_1, z_1) A^{\mathcal{D}}_{j}(x_2, z_2) \rangle
        & \underset{z_1, z_2 \rightarrow 0}{\sim} z_1^{d - 2} z_2^{d - 2} \, \frac{1}{\vert \vec{x}_{12}\vert^{2(d - 1)}}\left(
            \delta_{ij}
            - 2 \frac{\vec{x}_{12, i} \, \vec{x}_{12, j}}{\vert \vec{x}_{12}\vert^{2}}
        \right)~, \\
        \frac{1}{\mathcal{C}^{(1)}_{d - 1}}
        \langle A^{\mathcal{D}}_{z}(x_1, z_1) A^{\mathcal{D}}_{z}(x_2, z_2) \rangle
        &\underset{z_1, z_2 \rightarrow 0}{\sim} z_1^{d - 1} z_2^{d - 1} \, \left(-2 ( d - 1)\right) \, \frac{1}{\vert \vec{x}_{12}\vert^{2d}} ~,
    \end{aligned}
\end{equation}
so we can identify the boundary operator \(j_i\left(\vec{x}\right)\) 
\begin{equation}
	\begin{split}
		A_i \underset{z\to 0}{\sim} z^{d - 2} \, j_i \left(\vec{x}\right), \quad\quad
        A_z \underset{z\to 0}{\sim} \mathcal{O}\left(z^{d - 1}\right),
	\end{split}
\end{equation}
where \(j_i(\vec{x})\) has the dimension of a conserved current \(\Delta_j = d - 1\), while the \(A_z\) component is subleading. The boundary behaviour of \(A_i\) and \(A_z\) is consistent with what it would be expected from the equations of motion.
\subsection{The Neumann propagator}
As anticipated in section \ref{sec:propagators}, the Neumann propagator is obtained by adding to the Dirichlet propagator the appropriate homogeneous solution\footnote{
    Typically, for a generic Proca propagator, we would simply use the identity
\begin{equation}
    \Omega_{\lambda}^{(1)}(X_1, X_2; W_1, W_2) = \frac{i\lambda}{2\pi} \left[\Pi^{(1)}_{\frac{d}{2} + i\lambda}(X_1, X_2; W_1, W_2)  - \Pi^{(1)}_{\frac{d}{2} - i\lambda}(X_1, X_2; W_1, W_2) \right]~.
\end{equation}
However, for the gauge propagator \(\lambda = \pm i\left(\frac{d}{2} - 1\right)\), which corresponds to a pole of the harmonic function, so this identity does not hold for massless vectors.
In any case, rewriting this identity as a contour integral in the complex plane is inspirational to guess the correct homogeneous solution \eqref{eq:homogeneous-solution}.
}, 
which turns out to be \eqref{eq:homogeneous-solution}.

\subsubsection{Proof of solution homogeneity}
Acting with the equation of motion on \(\eqref{eq:homogeneous-solution}\) we find
\begin{equation}
    \begin{aligned}
        \text{EoM} \left[\eqref{eq:homogeneous-solution}\right] = \int_{\lambda = \pm i\left(\frac{d}{2} - 1\right)}^{\substack{\circlearrowright \\ \circlearrowleft}} &\Omega_{\lambda}^{(1)}(X_1, X_2; W_1, W_2) \, +\\
        &+ \left(W_1 \cdot \nabla_1\right)\left(W_2 \cdot \nabla_2\right) \int_{\lambda = \pm i\frac{d}{2}}^{\substack{\circlearrowright \\ \circlearrowleft}}\frac{1}{\lambda^2 + \frac{d^2}{4}} \Omega_{\lambda}^{(0)}(X_1, X_2)~.  
    \end{aligned}
\end{equation}
This is known to be zero analytically, as in (4.27) of \cite{Loparco:2023rug}, they compute the residues of the spin \(1\) harmonic function around its spurious poles to be:
\begin{equation}
    \begin{aligned}
        i \mspace{1mu} d \, &\underset{\lambda = - i \left(\frac{d}{2} - 1\right)}{\Res}\left[\Omega_{\lambda}^{(1)}(X_1, X_2; W_1, W_2)\right] = 
         \left(W_1 \cdot \nabla_1\right)\left(W_2 \cdot \nabla_2\right) \Omega_{-i \frac{d}{2}}^{(0)}(X_1, X_2)~.
    \end{aligned}
\label{eq:spurious-residue-spin-1}
\end{equation}

\subsubsection{Boundary limit of homogeneous solution}
In this subsection, we derive the boundary limit of the homogeneous solution of the equation of motion defined in \eqref{eq:homogeneous-solution}.

Similarly to the Dirichlet case, we can write the homogeneous solution in terms of the EAdS geodesic distance \(u\), and find the components of the two tensor structures:
\begin{align}
    F_{0}^{\text{Hom}}(u) &= \int_{\lambda = \pm i\left(\frac{d}{2} - 1\right)}^{\substack{\circlearrowright \\ \circlearrowleft}} 
     \dd \lambda
    \frac{1}{\lambda^2 + \left(\frac{d}{2} - 1\right)^2} \, \omega_{0,\lambda}(u) -
    \int_{\lambda = \pm i\frac{d}{2}}^{\substack{\circlearrowright \\ \circlearrowleft}} 
     \dd \lambda\frac{\xi}{\left(\lambda^2 + \frac{d^2}{4}\right)^2} \, \frac{\partial \Omega^{(0)}_{\lambda}(u)}{\partial u}~, \\
    F_1^{\text{Hom}}(u) &=
    \int_{\lambda = \pm i\left(\frac{d}{2} - 1\right)}^{\substack{\circlearrowright \\ \circlearrowleft}} 
    \dd \lambda
    \frac{1}{\lambda^2 + \left(\frac{d}{2} - 1\right)^2} \, \omega_{1,\lambda}(u) +
    \int_{\lambda = \pm i\frac{d}{2}}^{\substack{\circlearrowright \\ \circlearrowleft}}\dd \lambda
    \frac{\xi}{\left(\lambda^2 + \frac{d^2}{4}\right)^2} \, \frac{\partial^2 \Omega^{(0)}_{\lambda}(u)}{\partial u^2}~.
\end{align}

By expanding for large \(u\), we are left with taking the residues of the expression in \eqref{eq:F0-integrand-boundary-expansion} at \(\lambda = \pm i\left(\frac{d}{2} - 1\right)\), and we find 
\begin{equation}
    \begin{aligned}
        F_0^{\text{Hom}}(u) &\underset{u\to\infty}{\sim}
        -\frac{\Gamma \left(\frac{d + 1}{2}\right)}
        {2\pi ^{\frac{d + 1}{2}}(d-2)}   \frac{1}{u^{d - 1}} -  \frac{1}{4 \pi^{\frac{d}{2}}\Gamma\left(2-\frac{d}{2}\right)}
        \left[\left(1 - \frac{d-2}{d}\xi\right)\log\left(\frac{u}{2}\right) + C \right]\frac{1}{u} , \\
        C := & -\frac{d-1}{d-2}+\left(1-\frac{d-2}{d}\xi\right)\left(\log(4)+\gamma_E + \psi^{(0)} \left(-\frac{d}{2}\right)-\frac{2}{d}\right)~.
    \end{aligned}
\end{equation}
as reported in \eqref{eq:homogeneous-solution-boundary-limit}.

%% file: field-strength-2p-function.tex
In this appendix we would like to derive the bulk--to--bulk field strength 2--point function in AdS, and then compute its boundary limit.
Since we derived the bulk two--point function of the gauge field in section \ref{sec:propagators} and appendix \ref{sec:propagators-details}, we can use it to compute the field strength 2--point function. The field strength 2--point function in embedding space is 
\begin{equation}
    \label{eq:field-strength-2p-embedding-from-gauge-propagator}
    \langle F^{AB}(X_1) F_{CD}(X_2) \rangle = 
    \nabla^{1, [A} \nabla_{2, [C} K^{1, B]} K_{2, D]} 
    \left(W_1^M W_2^N \langle A_M(X_1) A_N(X_2) \rangle\right).
\end{equation}
The expression in tensorial basis for \(\left(W_1^M W_2^N \langle A_M(X_1) A_N(X_2) \rangle\right)\) is given in \eqref{eq:Dirichlet-propagator-tensor-basis} for Dirichlet boundary conditions, while \(\nabla^{1, A}\) is the covariant derivative in embedding space, and \(K^{1, A}\) is defined in \eqref{eq:K-index-projector-definition} 

The only two (bi--)tensor structures that are antisymmetric in \([A, B]\) and \([C, D]\), and that are transverse with respect to \(X_1\) (in \(A, B\)) and \(X_2\) (in \(C, D\)) are
\begin{equation}
    \label{eq:field-strength-2p-tensor-structures}
    \begin{aligned}
        \mathbb{T}_{1}^{AB, CD} &= G_{12}^{AC}G_{12}^{BD} - G_{12}^{AD}G_{12}^{BC}, \\
        \mathbb{T}_{2}^{AB, CD} &= V_1^{[A}G_{12}^{B][C}V_2^{D]}~,
    \end{aligned}
\end{equation}
where
\begin{equation}
    G_{12}^{AC} = \eta^{AC} (X_1 \cdot X_2) - X_2^A X_1^C~,
    \quad\quad
    V_1 = X_2 + (X_1 \cdot X_2) X_1~,
    \quad\quad
    V_2 = X_1 + (X_1 \cdot X_2) X_2~.
\end{equation}
We compute \eqref{eq:field-strength-2p-embedding-from-gauge-propagator} and find
\begin{equation}
\label{eq:B4}
    \begin{aligned}
        \langle F^{AB}(X_1) F^{CD}(X_2) \rangle &= 
        \alpha_{FS}(u) \mathbb{T}_{1}^{AB, CD} + \beta_{FS}(u) \mathbb{T}_{2}^{AB, CD}, \\
        \alpha_{FS}(u) &= 
        - \frac{2}{(1 + u)^2} \left(F_1(u) + \frac{\partial F_0(u)}{\partial u}\right), \\
        \beta_{FS}(u) &= 
        \frac{2}{(1 + u)^2} \left(F_1(u) + \frac{\partial F_0(u)}{\partial u}\right) 
        + \frac{1}{1 + u} \left(\frac{\partial F_1(u)}{\partial u} + \frac{\partial^2 F_0(u)}{\partial u^2}\right).
    \end{aligned}
\end{equation}
It is immediately clear that \(\langle F^{AB}(X_1) F^{CD}(X_2) \rangle\) is gauge invariant, a nice cross-check of our computation of the longitudinal part of the gauge field 2--point function.
Moreover, we can now extract the boundary limit of the field strength 2--point function. Let us do it separately for Dirichlet and Neumann boundary conditions.

\subsection{Dirichlet boundary conditions}
\label{subsec:field-strength-2p-dirichlet}

In the limit of \(X_1\) approaching the boundary, we have that \(u \rightarrow +\infty\), and the field strength 2--point function components have the following asymptotics
\begin{equation}
    \begin{aligned}
         \alpha_{FS}^{\mathcal{D}}(u) & \underset{u \to \infty}{\sim}
        \frac{\Gamma \left(\frac{d+1}{2}\right)}{\pi^{\frac{d + 1}{2}}}  \frac{1}{u^{d+2}}, \\
         \beta_{FS}^{\mathcal{D}}(u) &\underset{u \to \infty}{\sim} \frac{d - 2}{2}  \frac{\Gamma \left(\frac{d+1}{2}\right)}{\pi^{\frac{d + 1}{2}}} \frac{1}{u^{d+2}}~.
    \end{aligned}
\end{equation}
Notice that, interestingly, despite the gauge field two--point function is singular in \(d = 2\), the field strength two--point function has a regular limit.

\subsection{Neumann boundary conditions}
\label{subsec:field-strength-2p-neumann}

The contribution to the field strength 2--point from the homogeneous solution of the gauge field equation of motion, which corresponds to the difference between Neumann and Dirichlet boundary conditions, is 

\begin{equation}
    \begin{aligned}
        \alpha_{FS}^{\text{Hom}}(u) 
        &\underset{u\to\infty}{\sim}-\frac{\Gamma \left(\frac{d+1}{2}\right)}{\pi^{\frac{d + 1}{2}}} \frac{1}{u^{d+2}}
        + \frac{\Gamma\left(\frac{d - 2}{2}\right)}{2\pi^{\frac{d + 2}{2}}}\sin \left(\frac{\pi d}{2}\right)  \frac{1}{u^4}~, \\
        \beta_{FS}^{\text{Hom}}(u) 
        &\underset{u\to\infty}{\sim} -\frac{d - 2}{2}  \frac{\Gamma \left(\frac{d+1}{2}\right)}{\pi^{\frac{d + 1}{2}}} \frac{1}{u^{d+2}}
        +\frac{3 \Gamma\left(\frac{d - 4}{2}\right) }{4\pi^{\frac{d + 2}{2}}} \sin \left(\frac{\pi  d}{2}\right) \frac{1}{u^6}~.
    \end{aligned}
\end{equation}
Notice that both these expressions are also regular in \(d = 2\), and that the \(u^{-4}\) and \(u^{-5}\) orders in \(\beta_{FS}^{\text{Hom}}(u)\), which we would a priori expect, exactly cancel.

\subsection{Boundary limit of tensor structures}
In order to perform the boundary limit of the field strength 2--point function, we need to compute the boundary limit of the tensor structures \(\mathbb{T}_{1}^{AB, CD}\) and \(\mathbb{T}_{2}^{AB, CD}\) in \eqref{eq:field-strength-2p-tensor-structures}. This can be done remembering the pull--back of fundamental tensor structures in terms of the geodesic distance:
\begin{equation}
    \begin{aligned}
        \frac{\partial X^A_1}{\partial x^{\alpha}_1} \frac{\partial X^B_2}{\partial x^{\beta}_2} \eta_{AB} 
        &= \frac{\partial^2}{\partial x^{\alpha}_1 \partial x^{\beta}_2}\left(X_1 \cdot X_2\right)
        = -\frac{\partial^2 u}{\partial x^{\alpha}_1 \partial x^{\beta}_2}~, \\
        \frac{\partial X^A_1}{\partial x^{\alpha}_1}X_{2, A}
        &= \frac{\partial}{\partial x^{\alpha}_1}\left(X_1 \cdot X_2\right) = - \frac{\partial u}{\partial x^{\alpha}_1}~.
    \end{aligned}
\end{equation}
or by computing the pull-back maps \(\frac{\partial X^A}{\partial x^{\alpha}}\) already in some charts of coordinates.
We choose to specify Poincaré coordinates, and we find:
\begin{align}
    &\begin{aligned}
        \mathbb{T}_{1}^{0j, 0k} &= \frac{\left(\vec{x}_1 - \vec{x}_2\right)^2 +z_1^2+z_2^2 }{2 z_1^3 z_2^3} \, \delta_{j,k},\\
        \mathbb{T}_{2}^{0j, 0k} &=
        \frac{
            \left(\vec{x}_1 - \vec{x}_2\right)^2+z_1^2+z_2^2}{8 z_1^5 z_2^5} \,
            \left[\left(\left(\vec{x}_1 - \vec{x}_2\right)^4 - \left(z_1^2-z_2^2\right)^2\right) \delta_{j,k} \right.\\
            &\hspace{4.6cm}\left. 
            -2 \left(\left(\vec{x}_1 - \vec{x}_2\right)^2+z_1^2+z_2^2\right)\left(x_{1,j}-x_{2,j}\right) \left(x_{1,k}-x_{2,k}\right)\right],
    \end{aligned}
    \\
    &\begin{aligned}
        &\lim_{z_1 \rightarrow 0}\mathbb{T}_{1}^{ab, cd} = \mathcal{O}\left(\frac{1}{z_1^4}\right), \\
        &\mathbb{T}_{2}^{ab, cd} = -\frac{\left(\vec{x}_1 - \vec{x}_2\right)^2+z_1^2+z_2^2}{2 z_1^4 z_2^4}
        \left[\left(\vec{x}_{1,d}-\vec{x}_{2,d}\right)\left(\vec{x}_{1,b}-\vec{x}_{2,b}\right) \delta_{a,c}- \left(\vec{x}_{1,d}-\vec{x}_{2,d}\right)\left(\vec{x}_{1,a}-\vec{x}_{2,a}\right) \delta_{b,c}\right. \\
        &\hspace{5.2cm}\left.
        -\left(\vec{x}_{1,b}-\vec{x}_{2,b}\right) \left(\vec{x}_{1,c}-\vec{x}_{2,c}\right) \delta_{a,d}+\left(\vec{x}_{1,a}-\vec{x}_{2,a}\right) \left(\vec{x}_{1,c}-\vec{x}_{2,c}\right) \delta_{b,d}\right].
    \end{aligned}
\end{align}
Substitutimg these expansions in \eqref{eq:B4} we get the results of appendix \ref{table:F}

%% file: boundary-limit-bulk-to-boundary-prop.tex
The aim of this appendix is to compute the leading order of scalar bulk to boundary propagators in the limit \(z \rightarrow 0\).
First of all, we should fix the constant in front of the \(\delta-\)function given in formula (2.39) of \cite{witten1998anti}.

Indeed, 
\begin{align*}
    \int_{\R^d} \dd^d \vec{x} \frac{z^{d + 2\alpha}}{(z^2 + \vec{x})^{d + \alpha}} 
    &= z^{-d} \int_{\R^d} \dd^d \vec{x} \frac{1}{(1 + \frac{\vec{x}^2}{z^2})^{d + \alpha}} \\
    &= \int_{\R^d} \dd^d \vec{x} \frac{1}{(1 + \vec{x}^2)^{d + \alpha}} \\
    &= \Omega_{d - 1} \int_{0}^{+\infty} \dd x \, \frac{x^{d - 1}}{(1 + x^2)^{d + \alpha}} \\
    &= \pi^{\frac{d}{2}} \frac{\Gamma\left(\frac{d}{2} + \alpha\right)}{\Gamma\left(d + \alpha\right)},
\end{align*}
where the volume of the \(d\)-dimensional sphere \(\Omega_d = \frac{2\pi^{\frac{d + 1}{2}}}{\Gamma\left(\frac{d + 1}{2}\right)}\).

Now, clearly \(\frac{z^{d + 2\alpha}}{(z^2 + \vec{x})^{d + \alpha}}\) integrated over the real space doesn't depend on \(z\), and in the limit \(z \rightarrow 0\) only has support around \(\vec{x} \sim 0\), hence, at leading order
\begin{equation}
    \lim_{z\rightarrow 0} \frac{z^{d + 2\alpha}}{(z^2 + \vec{x})^{d + \alpha}} = \pi^{\frac{d}{2}} \frac{\Gamma\left(\frac{d}{2} + \alpha\right)}{\Gamma\left(d + \alpha\right)} \delta^d\left(\vec{x}\right) = \tilde{\kappa}_{\nu}\delta^d\left(\vec{x}\right).
\end{equation}

\subsection{Contact term for scalar propagators}
Applying the previous limit to the scalar propagator \(\Pi^{(0)}_{\nu}(X, P)\), we get:
\begin{equation}
    \begin{aligned}
    \Pi^{(0)}_{\frac{d}{2} + i\nu}(X, P)
    &=  \sqrt{\mathcal{C}_{\nu}^{(0)}} 
    \left(\frac{z}{z^2 + \vec{x}_{ij}^2}\right)^{\frac{d}{2} + i \nu}\\
    &=  \sqrt{\mathcal{C}_{\nu}^{(0)}} 
    \frac{z^{d + 2\left(i\nu - \frac{d}{2}\right)}}{\left(z^2 + \vec{x}_{ij}^2\right)^{d +  \left(i\nu - \frac{d}{2}\right)}} \, z^{\frac{d}{2} - i\nu} \\
    &\underset{z\to 0}{\sim} \sqrt{\mathcal{C}_{\nu}^{(0)}} \, z^{\frac{d}{2} - i\nu} 
    \, \pi^{\frac{d}{2}} \frac{\Gamma\left( i\nu\right)}{ \Gamma\left(\frac{d}{2} + i\nu\right)} \delta^d\left(\vec{x}_{ij}\right).
    \end{aligned}
\end{equation}
Similarly,
\begin{equation}
    \begin{aligned}
        \frac{\partial}{\partial z}\Pi^{(0)}_{\frac{d}{2} + i\nu}(X, P) 
        &=  \sqrt{\mathcal{C}_{\nu}^{(0)}} 
        \left(\frac{d}{2} + i\nu\right) 
        \left[\frac{z^{\frac{d}{2} + i \nu - 1}}{\left(z^2 + \vec{x}_{ij}^2\right)^{d +  \left(i\nu - \frac{d}{2}\right)}} 
        - 2\frac{z^{\frac{d}{2} + i \nu + 1}}{\left(z^2 + \vec{x}_{ij}^2\right)^{d +  \left(i\nu + 1 - \frac{d}{2}\right)}}\right] \\
        &= \sqrt{\mathcal{C}_{\nu}^{(0)}}
        \left(\frac{d}{2} + i\nu\right) z^{\frac{d}{2} - i\nu - 1}
        \left[\frac{z^{d + 2\left(i\nu - \frac{d}{2}\right)}}{\left(z^2 + \vec{x}_{ij}^2\right)^{d +  \left(i\nu - \frac{d}{2}\right)}} 
        - 2\frac{z^{d + 2 \left(i\nu + 1 - \frac{d}{2}\right)}}{\left(z^2 + \vec{x}_{ij}^2\right)^{d +  \left(i\nu + 1 - \frac{d}{2}\right)}}\right] \\
        &\underset{z\to 0}{\sim} \sqrt{\mathcal{C}_{\nu}^{(0)}}
        \left(\frac{d}{2} + i\nu\right) z^{\frac{d}{2} - i\nu - 1} \delta^d(\vec{x}_{ij}) \pi^{\frac{d}{2}}
        \left[\frac{\Gamma\left(i\nu\right)}{\Gamma\left(\frac{d}{2} + i\nu\right)} - 2\frac{\Gamma\left(1 + i\nu\right)}{\Gamma\left(1 + \frac{d}{2} + i\nu\right)} \right] \\
        &= \sqrt{\mathcal{C}_{\nu}^{(0)}}
        z^{\frac{d}{2} - i\nu - 1} \delta^d(\vec{x}_{ij}) \pi^{\frac{d}{2}} \left(\frac{d}{2} - i\nu\right) \frac{\Gamma\left(i\nu\right)}{\Gamma\left(\frac{d}{2} + i\nu\right)}.
    \end{aligned}
\end{equation}
This provides a derivation of the leading terms in the boundary asymptotics of the propagator and of its derivative, proportional to Dirac deltas. However the expansion contains also subleading terms. We have also included those in the main text, in equations \eqref{eq:boundary-limit-bulk-boundary-propagator}-\eqref{eq:boundary-limit-nabla-bulk-boundary-propagator}. Plugging these expansions, also including the subleading terms, in the vertex structure, we get 
\begin{equation}
    \begin{aligned}
        &T_{12}^z\left(P_1, P_2, X_1\right)\\
        & \underset{z\to 0}{\sim} z^{d + 1}\mathcal{C}_{\nu}^{(0)} \tilde{\kappa}_{\nu}
       \left(
            \left(\frac{d}{2} + i\nu\right)\frac{\delta^d(\vec{x}_{1\tilde{1}})}{\left(\vec{x}_{2\tilde{1}}^2\right)^{\frac{d}{2} + i\nu}} + \left(\frac{d}{2} - i\nu\right)\frac{\delta^d(\vec{x}_{2\tilde{1}})}{\left(\vec{x}_{1\tilde{1}}^2\right)^{\frac{d}{2} + i\nu}} - (1 \leftrightarrow 2) 
            +\mathcal{O}\left(z^2\right)
            \right)\\
        &+ z^{d + 1}z^{2i\nu}\mathcal{C}_{\nu}^{(0)}
        \left(
            \left(\frac{1}{\vec{x}_{1\tilde{1}}^2\vec{x}_{2\tilde{1}}^2}\right)^{\frac{d}{2} + i\nu}
            - (1 \leftrightarrow 2)
            +\mathcal{O}\left(z^2\right)
        \right)\\
        &\underset{z\to 0}{\sim} z^{d + 1}
        \left(
            \frac{\delta^d(\vec{x}_{1\tilde{1}})}{\left(\vec{x}_{2\tilde{1}}^2\right)^{\frac{d}{2} + i\nu}}
            - \frac{\delta^d(\vec{x}_{2\tilde{1}})}{\left(\vec{x}_{1\tilde{1}}^2\right)^{\frac{d}{2} + i\nu}}
            +\mathcal{O}\left(z^2\right) +\mathcal{O}\left(z^{2(1 + i\nu)}\right)
            \right)~,
    \end{aligned}
\end{equation}
where we used \(2i\nu\,\mathcal{C}_{\nu}^{(0)}\tilde{\kappa}_{\nu} = 1 \). Going from the second to the third expression, we used that some terms vanish when antiymmetrized in $1\leftrightarrow 2$. We neglected all contact terms \(\delta^d(\vec{x}_{12})\) since we assume the insertion points to be separated.

%% file: parallel-current-bdy-expansion.tex
In this appendix, we report the details of the boundary expansion of the parallel part of the Dirichlet current exchange, as follows from the procedure explained in subsection \ref{subsec:longitudinal-dirichlet-photon-exchange}.

We examine in detail the case of equal matter insertions (where \(\nu_1 = \nu_2 = \nu_3 = \nu_4 = \nu\)); the case of opposite matter insertions, where \(\nu_1 = \nu_3 = \nu\) and \(\nu_2 = \nu_4 = -\nu\) can be treated analogously. Moreover this second case is bounded by the former: out of two fields with shadow-related dimensions, one of them is always falling off faster at the AdS-boundary, so the convergence properties of the diagrams cannot be worse than equal matter insertions of conformal dimension close to unitarity bound.

\label{subsec:equal-matter-insertions}

Inserting the boundary expansions \eqref{eq:boundary-limit-bulk-boundary-propagator} and \eqref{eq:boundary-limit-nabla-bulk-boundary-propagator} into the boundary integral of variable \(\tilde{X}_1\) in \eqref{eq:LongitudinalContributionBoundary} we obtain, at quadratic order in \(z\)  
\begin{align}
   & \int_{\partial AdS} \frac{\dd^d \vec{x}_{\tilde{1}}}{z^{d + 1}}  
    T_{12}^A\left(P_1, P_2, X_1\right) \Pi^{(0)}_{\frac{d}{2} + i\lambda}(X_1, P_5)
    \Big|_{X_1 = \left(z, \vec{x}_{\tilde{1}} \right)} 
    \underset{z\to 0}{\sim} \sqrt{\mathcal{C}_{\lambda}^{(0)}} \int_{\R^d}\dd^d \vec{x}_{\tilde{1}} \\
    & \left[\tilde{\kappa}_{\lambda}\delta^d\left(\vec{x}_{5\tilde{1}}\right) z^{\frac{d}{2}-i\lambda} 
    + z^{\frac{d}{2}+i\lambda}\left(
        \left(\frac{1}{\vec{x}_{5\tilde{1}}^2}\right)^{\frac{d}{2}+i\lambda} 
        +\mathcal{O}\left(z^2\right)
    \right)\right] 
    \\
    &\left[
            \frac{\delta^d(\vec{x}_{1\tilde{1}})}{\left(\vec{x}_{2\tilde{1}}^2\right)^{\frac{d}{2} + i\nu}}
            - \frac{\delta^d(\vec{x}_{2\tilde{1}})}{\left(\vec{x}_{1\tilde{1}}^2\right)^{\frac{d}{2} + i\nu}}
        +\mathcal{O}\left(z^2\right) +\mathcal{O}\left(z^{2(1 + i\nu)}\right)
        \right]
        \\
    & = 
    \label{eq:parallel-current-equal-matter-Ia}
    \sqrt{\mathcal{C}_{\lambda}^{(0)}} \tilde{\kappa}_{\lambda}\,  
    z^{\frac{d}{2}-i\lambda} \,
    \left(
        \frac{\delta^d(\vec{x}_{15})}{\left(\vec{x}_{25}^2\right)^{\frac{d}{2} + i\nu}}
        - \frac{\delta^d(\vec{x}_{25})}{\left(\vec{x}_{15}^2\right)^{\frac{d}{2} + i\nu}}
        +\mathcal{O}\left(z^2\right)
    \right)  \\
    \label{eq:parallel-current-equal-matter-IIa}
    &
    +\mathcal{O}\left(z^{\frac{d}{2}-i\lambda + 2(1 + i\nu)}\right)
    \,  \\
    \label{eq:parallel-current-equal-matter-IIIa}
    &+ \sqrt{\mathcal{C}_{\lambda}^{(0)}} \, z^{\frac{d}{2}+i\lambda} \,
    \frac{1}{\left(\vec{x}_{12}^2\right)^{\frac{d}{2}+i\nu}}
    \left(
        \frac{1}{\left(\vec{x}_{15}^2\right)^{\frac{d}{2}+i\lambda}} -  \frac{1}{\left(\vec{x}_{25}^2\right)^{\frac{d}{2}+i\lambda}}
    \right)  \\
    \label{eq:parallel-current-equal-matter-IVa}
    &
    +\mathcal{O}\left(z^{\frac{d}{2}+i\lambda + 2(1 + i\nu)}\right)
\end{align}
The shadow related integral (of variable \({x_{\tilde{2}}}\)) also has an analogous expansion in four power series:
\begin{align}
    \int_{\partial AdS} \frac{\dd^d \vec{x}_{\tilde{2}}}{z^{d + 1}} &\, 
    T_{34}^A\left(P_3, P_4, X_2\right) \Pi^{(0)}_{\frac{d}{2} - i\lambda}(X_2, P_5)
    \\
    \label{eq:parallel-current-equal-matter-Ib}
    & \underset{z\to 0}{\sim}  \sqrt{\mathcal{C}_{-\lambda}^{(0)}} \tilde{\kappa}_{-\lambda}\,  
    z^{\frac{d}{2}+i\lambda} \,
    \left(
        \frac{\delta^d(\vec{x}_{35})}{\left(\vec{x}_{45}^2\right)^{\frac{d}{2} + i\nu}}
        - \frac{\delta^d(\vec{x}_{45})}{\left(\vec{x}_{35}^2\right)^{\frac{d}{2} + i\nu}}
        +\mathcal{O}\left(z^2\right)
    \right) \,  \\
    \label{eq:parallel-current-equal-matter-IIb}
    &+ 
    \mathcal{O}\left(z^{\frac{d}{2}+i\lambda + 2(1 + i\nu)}\right)
    \\
    \label{eq:parallel-current-equal-matter-IIIb}
    &+ \sqrt{\mathcal{C}_{-\lambda}^{(0)}} \, z^{\frac{d}{2}-i\lambda} \,
    \frac{1}{\left(\vec{x}_{34}^2\right)^{\frac{d}{2}+i\nu}}
    \left(
        \frac{1}{\left(\vec{x}_{35}^2\right)^{\frac{d}{2}-i\lambda}} -  \frac{1}{\left(\vec{x}_{45}^2\right)^{\frac{d}{2}-i\lambda}}
    \right) \,  \\
    \label{eq:parallel-current-equal-matter-IVb}
    &+ 
    \mathcal{O}\left(z^{\frac{d}{2}-i\lambda + 2(1 + i\nu)}\right)
\end{align}

The combination of these two terms gives an expansion of \(\mathcal{A}^{\parallel}_{\mathcal{D}}\) in six families of power series in \(z\):
\begin{equation}
    \label{eq:equal-matter-insertions-parallel-current-z-expansion}
    \begin{aligned}
        \mathcal{A}^{\parallel}_{\mathcal{D}}(z) \underset{z\to 0}{\sim} \,
        (-ie)^2\frac{\xi}{\pi} 
        \int_{-\infty}^{+\infty} \dd \lambda
        \frac{\lambda^2 \mathcal{C}_{\lambda}^{(0)}\mathcal{C}_{-\lambda}^{(0)}}{\left(\lambda^2 + \frac{d^2}{4}\right)^2}
        \biggl(
        &\# z^d + \# z^{d + 2} + \ldots \\
        +& \# z^{d + 2 + 2i\nu} + \# z^{d + 4 + 2i\nu} + \ldots \\
        +& \# z^{d + 4 + 4i\nu} + \# z^{d + 6 + 4i\nu} + \ldots \\
        +& \# z^{d \pm 2i\lambda} + \# z^{d + 2 \pm 2i\lambda} + \ldots \\
        +& \# z^{d \pm 2i\lambda + 2(1 + i\nu)} + \# z^{d + 2 \pm 2i\lambda + 2(1 + i\nu)} + \ldots \\
        +& \# z^{d \pm 2i\lambda + 4(1 + i\nu)} + \# z^{d + 2 \pm 2i\lambda + 4(1 + i\nu)} + \ldots
        \biggr).
    \end{aligned}
\end{equation}

Let us comment briefly on the behaviour of each these six families in the limit \(z \to 0\), and explicit their leading term.

\begin{itemize}
    \item 
    \(\boxed{z^d}\): 
    this comes from the product of \eqref{eq:parallel-current-equal-matter-Ia} with \eqref{eq:parallel-current-equal-matter-Ib}, and \eqref{eq:parallel-current-equal-matter-IIIa} with \eqref{eq:parallel-current-equal-matter-IIIb}. This first family of terms goes to zero as \(z \to 0\) for any \(d > 0\).
    The leading piece is:
    \begin{equation}
        \begin{aligned}
         (-ie)^2\frac{\xi}{\pi} \, z^d \,
        & \int_{-\infty}^{+\infty} \dd \lambda
        \frac{\lambda^2 \mathcal{C}_{\lambda}^{(0)}\mathcal{C}_{-\lambda}^{(0)}}{\left(\lambda^2 + \frac{d^2}{4}\right)^2} 
        \frac{1}{\left(\vec{x}_{12}^2 \vec{x}_{34}^2\right)^{\frac{d}{2}+i\nu}}
        \biggl[ 
            \tilde{\kappa}_{\lambda}\tilde{\kappa}_{-\lambda}
            \left(
                \delta^d(\vec{x}_{13}) - \delta^d(\vec{x}_{14}) - \delta^d(\vec{x}_{23}) + \delta^d(\vec{x}_{24})
            \right) \\
          &  + \int_{\R^d} \dd^d x_5 
            \left(\frac{1}{\left(\vec{x}_{15}^2\right)^{\frac{d}{2}+i\lambda}} -  \frac{1}{\left(\vec{x}_{25}^2\right)^{\frac{d}{2}+i\lambda}}\right) 
            \left(\frac{1}{\left(\vec{x}_{35}^2\right)^{\frac{d}{2}-i\lambda}} -  \frac{1}{\left(\vec{x}_{45}^2\right)^{\frac{d}{2}-i\lambda}}\right)
            \,\left(1 + \mathcal{O}\left(z^2\right)\right)
        \biggr].
        \end{aligned}
    \end{equation}
    \item \(\boxed{z^{d + 2 + 2i\nu}}\): 
    this comes from the product of \eqref{eq:parallel-current-equal-matter-Ia} with \eqref{eq:parallel-current-equal-matter-IIb} and \eqref{eq:parallel-current-equal-matter-IIa} with \eqref{eq:parallel-current-equal-matter-Ib}, \eqref{eq:parallel-current-equal-matter-IIIa} with \eqref{eq:parallel-current-equal-matter-IVb} and \eqref{eq:parallel-current-equal-matter-IVa} with \eqref{eq:parallel-current-equal-matter-IIIb}. This second family of terms goes to zero as \(z \to 0\) due to AdS unitarity bound for scalar operators \(\Delta_{\nu} \ge \frac{d - 2}{2} \); indeed \(d + 2 + 2i\nu > 0\) is true in any \(d > 0\) if \(i\nu \ge -1\).
    The leading piece is given by the sum of the four contributions mentioned above, and it is regular -- except for some potential coincident points divergences (in the \(x_{\tilde{2}}\) integral), which give non worrysome contact terms. There are no IR divergences in these spatial integrals.
\item \(\boxed{z^{d + 4 + 4i\nu}}\): 
    this comes from the product of \eqref{eq:parallel-current-equal-matter-IIa} with \eqref{eq:parallel-current-equal-matter-IIb} and \eqref{eq:parallel-current-equal-matter-IVa} with \eqref{eq:parallel-current-equal-matter-IVb}. Again this vanishes due to AdS unitarity bound for scalar operators: in any \(d > 0\), \(d + 4 + 4i\nu > 0 \iff i\nu \ge -1\).
    The spatial integrals are regular except for some potential contact terms.
    \item \(\boxed{z^{d \pm 2i\lambda}}\):
    they come from \eqref{eq:parallel-current-equal-matter-Ia} times \eqref{eq:parallel-current-equal-matter-IIIb} and \eqref{eq:parallel-current-equal-matter-IIIa} times \eqref{eq:parallel-current-equal-matter-Ib}. The leading piece is:
    \begin{equation}
    \label{eq:equal-matter-insertions-lambda-dominant-contribution}
    \begin{aligned}
        \mathcal{A}^{\parallel}_{\mathcal{D}} &\underset{z\to 0}{\sim} \,
        (-ie)^2\frac{\xi}{\pi}
        \int_{-\infty}^{+\infty} \dd\lambda 
        \frac{\lambda^2}{\left(\lambda^2 + \frac{d^2}{4}\right)^2} \,\mathcal{C}_{-\lambda}^{(0)} \mathcal{C}_{\lambda}^{(0)}
        \, \frac{1}{\vec{x}_{12}^{2 \left(\frac{d}{2}+i \nu \right)} \vec{x}_{34}^{2 \left(\frac{d}{2} + i \nu \right)}}\\
        \,& \left[z^{d - 2 i\lambda} \tilde{\kappa}_{\lambda}
        \left(
            \frac{1}{\vec{x}_{13}^{2\left(\frac{d}{2} - i\lambda\right)}} 
            - \frac{1}{\vec{x}_{14}^{2\left(\frac{d}{2} - i\lambda\right)}} 
            - \frac{1}{\vec{x}_{23}^{2\left(\frac{d}{2} - i\lambda\right)}} 
            + \frac{1}{\vec{x}_{24}^{2\left(\frac{d}{2} - i\lambda\right)}}
            \right) \right. \\
        +&\left.
            z^{d + 2 i\lambda} \tilde{\kappa}_{-\lambda}
            \left(
                \frac{1}{\vec{x}_{13}^{2\left(\frac{d}{2} + i\lambda\right)}} 
                - \frac{1}{\vec{x}_{14}^{2\left(\frac{d}{2} + i\lambda\right)}} 
                - \frac{1}{\vec{x}_{23}^{2\left(\frac{d}{2} + i\lambda\right)}} 
                + \frac{1}{\vec{x}_{24}^{2\left(\frac{d}{2} + i\lambda\right)}}\right) 
        \right].
    \end{aligned}
\end{equation}
The integrand is even in \(\lambda\), so we can fold the first term inside the square brackets onto the second, and, recalling from the end of appendix \ref{app:boundary-limit-bulk-boundary-propagators} that \(\mathcal{C}_{-\lambda}^{(0)} \, \tilde{\kappa}_{-\lambda} = \frac{i}{2\lambda}\), and that \(\mathcal{C}_{\lambda}^{(0)} = \frac{1}{2\pi^{d/2}}\frac{\Gamma\left(\frac{d}{2} + i\lambda\right)}{\Gamma(1 + i\lambda)}\) we have a \(\lambda-\)integral that looks like
\begin{equation}
    \label{eq:lambda-integral-longitudinal-contributions}
    (-ie)^2\frac{\xi}{\pi}\,
    \frac{i}{4\pi^{\frac{d}{2}}}
    \int_{-\infty}^{+\infty} \dd \lambda 
    \frac{\lambda}{\left(\lambda^2 + \frac{d^2}{4}\right)^2} \, \frac{\Gamma \left(\frac{d}{2} + i \lambda\right)}{\Gamma(1 + i\lambda)}\,  \frac{1}{\vec{x}_{i,j}^{2\left(\frac{d}{2} + i\lambda\right)}} \, z^{d + 2i\lambda}.
\end{equation}
This integral can be performed by pulling the contour in the lower half-plane (since \(z^{2i\lambda} = e^{-2i\lambda \vert \log z \vert}\) and \(\log z < 0\)  given that \(z\rightarrow 0\)), which picks up the pole at 
\[
    \lambda = -i \frac{d}{2}  \implies \eqref{eq:equal-matter-insertions-lambda-dominant-contribution} \sim z^{d + 2i\lambda}\Big|_{\lambda = -i \frac{d}{2}} = z^{2d} \longrightarrow 0.
\]
\textbf{Remark:} Were we to quantize the photon with alternate boundary condition (Neumann), the integration contour would be modified, so we would pick up the pole at \(\lambda = +i \frac{d}{2}\). 
This is what would give rise to a finite longitudinal part in the exchange of a Neumann photon.
\item \(\boxed{z^{d \pm 2i\lambda + 2(1 + i\nu)}}\) and \(\boxed{z^{d \pm 2i\lambda + 4(1 + i\nu)}}\):
    As in the previous case, we first need to perform the \(\lambda-\)integral, which would pick up a pole at \(\lambda = \mp i \frac{d}{2}\) and so 
    \[
        \mathcal{A}^{\parallel}_{\mathcal{D}}(z) \supseteq z^{d + 2i\lambda + 2n(1 + i\nu)}\Big|_{\lambda = -i \frac{d}{2}} = z^{2d + 2n(1 + i\nu)}  \le z^{2d} \longrightarrow 0.
    \]
    The extra power of \(z^{2n(1 + i\nu)}\), where \(n = 1, 2\), ensures that even in case of the exchange of a Neumann photon, the contribution from these families would vanish due to AdS unitarity bound, in analogy to the second and third familes above.
\end{itemize}



Finally, notice that we can promote \(\xi\) to any generic function of \(\lambda\), and the same conclusion holds if \(\xi(\lambda)\) decays fast enough at infinity so that the integration contour can be closed. Moreover, any antisymmetric part of the function would simply cancel with the shadow contribution, so we can always assume \(\xi(\lambda) = \xi(-\lambda)\).

%% file: spin-1-simple-bubble.tex
\begin{figure}[!ht]
    \centering
    \begin{tikzpicture}
        \node at (0,0) {\(\text{(Simple bubble)}(X_1, X_2; W_1, W_2) = \)};
        \begin{scope}[yshift = 0 cm, xshift = 3.7cm, scale = 1.4]
                \draw[goldenyellow, ultra thick, decorate,decoration={snake, amplitude= 2mm}]  (-0.3, 0) -- (0.5, 0);
                \draw[fuchsia, ultra thick] (1.2, 0) circle (0.7);
                \draw[goldenyellow, ultra thick, decorate,decoration={snake, amplitude= 2mm}]  (1.9, 0) -- (2.7, 0) ;
                \node[fuchsia, above] at (1.2, 0.7) {\(\varphi_1\)};
                \node[fuchsia, below] at (1.2, -0.7) {\(\varphi_2\)};
                \filldraw[turquoise] (0.5, 0) circle (0.05) node[right, black] {\(X_1\)};
                \filldraw[turquoise] (1.9, 0) circle (0.05) node[left, black] {\(X_2\)};
        \end{scope}
    \end{tikzpicture}
    \caption{The diagram named {\it{simple bubble}} contributing to the 1PI photon self-energy.}
    \label{fig:simple-bubble-AdS}
\end{figure}
In this appendix we discuss how to implement dimensional regularization to the one-loop diagrams correcting the photon propagator that we encountered in section \ref{sec:1-loop-self-energy}.

The first diagram is shown in figure \ref{fig:simple-bubble-AdS}. Its expression as an integral of AdS propagators in position space is
\begin{equation}
    \label{eq:simple-bubble-AdS-definition}
    \begin{aligned}
        \text{(Simple bubble)}(X_1, X_2; W_1, W_2) 
        =& \, 2\Pi_{\frac{d}{2} + i\nu_1}^{(0)}(X_1, X_2)\left(W_1\cdot\nabla_1\right)\left(W_2\cdot\nabla_2\right)\Pi_{\frac{d}{2} + i\nu_2}^{(0)}(X_1, X_2) \\
        &2\Pi_{\frac{d}{2} + i\nu_2}^{(0)}(X_1, X_2)\left(W_1\cdot\nabla_1\right)\left(W_2\cdot\nabla_2\right)\Pi_{\frac{d}{2} + i\nu_1}^{(0)}(X_1, X_2) \\
        &- \left(W_1\cdot\nabla_1\right)\left(W_2\cdot\nabla_2\right) \left[\Pi_{\frac{d}{2} + i\nu_1}^{(0)}(X_1, X_2)\Pi_{\frac{d}{2} + i\nu_2}^{(0)}(X_1, X_2)\right]~.
     \end{aligned}
 \end{equation}
We parametrize its spectral representation as
\begin{equation}
    \begin{aligned}
        \text{(Simple bubble)}(X_1, X_2; W_1, W_2)  
        & = -\int_{-\infty}^{+\infty} \dd\lambda \, \langle JJ \rangle_{\nu_1, \nu_2}(\lambda) \Omega_{\lambda}^{(1)}(X_1, X_2; W_1, W_2) \\
        &- \int_{-\infty}^{+\infty} \dd\lambda \, \langle JJ \rangle_{\nu_1, \nu_2}^{\parallel}(\lambda) \left(W_1\cdot \nabla_{1}\right)\left(W_2\cdot \nabla_{2}\right)\Omega_{\lambda}^{(0)}(X_1, X_2)~,
    \end{aligned}
\end{equation}
in terms of two functions, the transverse component $\langle JJ \rangle_{\nu_1, \nu_2}(\lambda)$ and the longitudinal one $\langle JJ \rangle_{\nu_1, \nu_2}^{\parallel}(\lambda)$. We recall that $\Omega_{\lambda}^{(1)}(X_1, X_2; W_1, W_2)$ is divergence-less even at coincidence points \cite{costa2014spinning}. On the other hand, the divergence of $\left(W_1\cdot \nabla_{1}\right)\left(W_2\cdot \nabla_{2}\right)\Omega_{\lambda}^{(0)}(X_1, X_2)$ gives a laplacian acting on $\Omega_{\lambda}^{(0)}(X_1, X_2)$, and for general $\langle JJ \rangle_{\nu_1, \nu_2}^{\parallel}(\lambda)$ this results in a non-vanishing contribution. If, however, $\langle JJ \rangle_{\nu_1, \nu_2}^{\parallel}(\lambda)\propto (\lambda^2 + \frac{d^2}{4})^{-1}$ then the divergence becomes a Laplacian acting on the propagator of a massless scalar, and  one gets only a contact term contribution, a derivative of a Dirac delta. As a result, current conservation at separated points forces $\langle JJ \rangle_{\nu_1, \nu_2}^{\parallel}(\lambda)$ to be a $\lambda$-independent prefactor times $(\lambda^2 + \frac{d^2}{4})^{-1}$. We will first concentrate on the transverse component, and return to the longitudinal one when discussing the tadpole diagram. We reported in the main text, in equation \eqref{eq:simple-bubble-AdS}, the expression for $\langle JJ \rangle_{\nu_1, \nu_2}(\lambda)$ as an infinite sum obtained in \cite{ShenMasterThesis}. Taking $d<1$ this sum is convergent and a direct summation gives the following shadow-symmetric combination of generalized hypergeometric functions of argument $z=1$
\begin{equation}\label{eq:7F6}
    \begin{split}
    \langle JJ \rangle_{\nu_1, \nu_2}(\lambda)  &=
    -\frac{2 \pi^{-\frac{d}{2}} \left(d+i\nu_1+i \nu_2+2\right)_{-\frac{d}{2}}}
    {\left(\frac{d}{2}+ i \nu_1 + i \nu_2 + 1\right) \left(\frac{d}{2}+i \nu_1+1\right)_{\frac{d}{2}} \left(\frac{d}{2}+i \nu_2+1\right)_{-\frac{d}{2}}} \Biggl[ \\
    &\hspace{-2cm}\frac{\, _7F_6\left(
        \begin{array}{c}
        \frac{d + 2}{2},\frac{d + 2}{2}+i \nu_1,\frac{1 + \left(\nu_1 + \nu_2\right)}{2},1 + i\frac{\nu_1 + \nu_2}{2},\frac{d + 2}{4}+\frac{i \lambda }{2}+ i\frac{\nu_1 + \nu_2}{2},\frac{d + 2}{2}+i \nu_2,\frac{d + 2}{2}+i \nu_1+i \nu_2 \\
        1 + i \nu_1,\frac{d + 2}{2}+ i\frac{\nu_1 + \nu_2}{2},\frac{d + 3}{2}+ i\frac{\nu_1 + \nu_2}{2},\frac{d}{4}+\frac{i \lambda }{2}+ i\frac{\nu_1 + \nu_2}{2}+\frac{3}{2},1 + i \nu_2,1 + i \nu_1+i \nu_2
        \end{array}
        ;1\right)}
        {\frac{d}{2}+ i\lambda + i\nu_1 + i\nu_2 + 1} \\
        &+ (\lambda\to-\lambda)\Biggr]~. \\
    \end{split}
\end{equation}
For the two relevant cases of \(\nu_1 = \nu_2\) and \(\nu_1 = -\nu_2\), it reduces to the expressions reported in \eqref{eq:resummed-simple-bubble}.
We recall that the convergence of the hypergeometric sum $ {}_{q + 1}F_q \left(
\begin{array}{c} a_1, \ldots, a_{q + 1} \\ b_1, \ldots, b_q \end{array}; z \right) $ evaluated at argument $z=1$ is determined by the parameter \(Z = \sum_{i = 1}^{q + 1} a_i - \sum_{i = 1}^{q} b_i\), namely it is convergent for $\mathrm{Re}[Z]<0$. We see that the sums in \eqref{eq:7F6} and their simplified form in \eqref{eq:resummed-simple-bubble} are both convergent only in the region $d<1$, and our goal in this appendix is to determine their analytic continuation in $d$ to $d>1$.

To do so, we use the approach of  \cite{cacciatori2024loops} based on recursion relations of hypergeometric functions. The recursion relation in (C.7) of appendix C of \cite{cacciatori2024loops} gives (assuming that the $b_i$'s are all different from each other, and none of them is a negative integer)
\begin{align}
\begin{split}
    \label{eq:dim-reg-hypergeometric-recursion-relation}
   & {}_{q + 1}F_q \left(
        \begin{array}{c}
        a_1, \ldots, a_{q + 1} \\
        b_1, \ldots, b_q
        \end{array}; 1
        \right) 
       \\
       & = - \frac{1}{Z} \sum_{j = 1}^q 
        \frac{\Pi_{k = 1}^{q + 1} \left(b_j - a _k\right)}{b_j \Pi_{k = 1, k \neq j}^{q} \left(b_j - b _k\right)}
        {}_{q + 1}F_q \left(
        \begin{array}{c}
        a_1, \ldots, a_{q + 1} \\
        b_1, \ldots, b_j + 1, \ldots, b_q
        \end{array}; 1
        \right)~.
\end{split}        
\end{align}
This equation has the virtue that on the right-hand side we have factored a pole at $Z=0$, which turns out to be the motivation for the convergence on the left-hand side only in the region $\mathrm{Re}[Z]<0$. Indeed this pole  multiplies a different hypergeometric function with a shift $Z\to Z-1$ which improves the convergence. We can then readily take the right-hand side to be the extension of the left-hand side to the larger region $\mathrm{Re}[Z]<1$.

In the context of one loop Witten diagrams, in which $Z$ is determined by the spacetime dimension $d$, this gives a concrete strategy to implement dimreg. In practice, \eqref{eq:dim-reg-hypergeometric-recursion-relation} furnishes the analytic continuation of a hypergeometric sum that converges only up to dimension \(d < \delta\), expressed as a combination of hypergeometric sums that converge for \(d < \delta + 1\), with coefficients that are simple poles at \(d=\delta\). By applying it \(3\) times to the functions in equation \eqref{eq:resummed-simple-bubble} we can obtain expressions of the form
\[
\begin{aligned}
    \langle JJ \rangle_{\nu, \nu}(\lambda) &= \frac{1}{(d-1) (d - 2)(d - 3)}  \langle\widetilde{JJ}\rangle_{\nu, \nu}(\lambda), \\
    \langle JJ \rangle_{\nu, -\nu}(\lambda) &= \frac{1}{(d-1) (d - 2)(d - 3)} \langle\widetilde{JJ}\rangle_{\nu, -\nu}(\lambda),
\end{aligned}
\]
where \(\langle\widetilde{JJ}\rangle_{\nu, \nu}(\lambda)\) and \(\langle\widetilde{JJ}\rangle_{\nu, -\nu}(\lambda)\) are linear combinations of hypergeometric functions absolutely convergent for \(d < 4\).
This approach can be readily implemented algorithmically to obtain an expression for $\langle\widetilde{JJ}\rangle_{\nu, \nu}(\lambda)$ and $\langle\widetilde{JJ}\rangle_{\nu, -\nu}(\lambda)$. However these expressions turn out to be too lengthy to be simplified analytically, therefore we resorted to evaluating them numerically. We find that the spectral density is real in both cases \(\nu_2 = \pm \nu_1\), and more generally that the real part is quadratic in \(\lambda\). Evaluating the coefficient of $\lambda^2$ for several values of \(\nu\) we find
\begin{equation}
    \langle JJ \rangle_{\nu_1, \nu_2}(\lambda)\underset{d\to 3}{\sim} \frac{-\frac{1}{24\pi^2}\lambda^2 + c(\nu_1, \nu_2)}{d - 3}+ \overline{\langle JJ \rangle}_{\nu,\pm\nu} (\lambda) +\mathcal{O}(d-3)~,
\end{equation}
where $c(\nu_1, \nu_2)$ and $\overline{\langle JJ \rangle}_{\nu,\pm\nu} (\lambda)$ are $\lambda$ and $d$ independent functions of of $\nu_1, \nu_2$ that can be readily evaluated numerically. The coefficient $-\frac{1}{24\pi^2}$ of the divergent term quadratic in $\lambda$ is what determines the running of the gauge coupling at one-loop, and reproduces the known result for the beta function as we showed in the main text.

%% file: tadpole.tex
\begin{figure}[!ht]
    \centering
    \begin{tikzpicture}
        \node at (0,0) {\(T_{\nu}(X_1, X_2; W_1, W_2) = \)};
        \begin{scope}[yshift = 0 cm, xshift = 3.3cm, scale = 1.4]
                \draw[goldenyellow, ultra thick, decorate,decoration={snake, amplitude= 2mm}]  (-0.3, 0) -- (0.5, 0);
                \draw[fuchsia, ultra thick] (0.5, 0) 
                to[out = 140, in = 180, looseness = 2] (0.5, 1)  
                to[out = 0, in = 40, looseness = 2] (0.5, 0);
                \draw[goldenyellow, ultra thick, decorate,decoration={snake, amplitude= 2mm}]  (0.5, 0) -- (1.3, 0);
                \node[fuchsia, below] at (0.5, 1) {\(\varphi_1\)};
                \filldraw[turquoise] (-0.3, 0) circle (0.05) node[below, black] {\(X_1\)};
                \filldraw[turquoise] (1.3, 0) circle (0.05) node[below, black] {\(X_2\)};
                \filldraw[turquoise] (0.5, 0) circle (0.05) ;
            \end{scope}
    \end{tikzpicture}
    \caption{The \(1\)-loop tadpole contribution to the photon self-energy.}
    \label{fig:tadpole-dS}
\end{figure}

Next, we discuss the tadpole contribution, shown in figure \ref{fig:tadpole-dS}. The tadpole amounts to a contact term, which in the spectral representation translates into a constant. In more details, we have
\begin{equation}
T_{\nu}(X_1, X_2; W_1, W_2) = (W_1\cdot W_2)\, \mathcal{T}_\nu\, \delta^{d+1}(X_1,X_2)~,
\end{equation}
where $\mathcal{T}_\nu$ is a function of $\nu$ to be determined below. Using the completeness relation that relates the harmonic functions of spin 1 and spin 0 - see equation (91) of \cite{costa2014spinning} - we can rewrite it as
\begin{align}
\begin{split}
T_{\nu}(X_1, X_2; W_1, W_2) &= \mathcal{T}_\nu \left(\int_{-\infty}^{+\infty} \dd\lambda \, \Omega_{\lambda}^{(1)}(X_1, X_2; W_1, W_2)\right.\\
&\left.+\int_{-\infty}^{+\infty} \dd\lambda \, \frac{1}{\lambda^2 + \frac{d^2}{4}}\left(W_1\cdot \nabla_{1}\right)\left(W_2\cdot \nabla_{2}\right)\Omega_{\lambda}^{(0)}(X_1, X_2)\right)~,
\end{split}
\end{align}
namely, the transverse component of the spectral representation is a constant in $\lambda$ given precisely by $\mathcal{T}_\nu$, while the longitudinal component is given by the same constant $\mathcal{T}_\nu$ multiplying $(\lambda^2 + \frac{d^2}{4})^{-1}$. The fact that we need to include the tadpole is also the reason why so far we have ignored the longitudinal contribution $\langle JJ \rangle_{\nu_1, \nu_2}^{\parallel}(\lambda)$ to the current two-point function. As we recalled above, by current conservation this function is also a constant times $(\lambda^2 + \frac{d^2}{4})^{-1}$. We then fix it simply by requiring that it cancels with the longitudinal part of the tadpole diagram, leaving a one-loop 1PI diagram that is divergence-less also at coincident points, as is appropriate for a current that couples to a dynamical gauge field. 

One way to evaluate the tadpole, that we used in the main text, is by taking the coincident-points limit of the scalar propagator. Doing so, we obtain a finite answer in \(d\le 1\), reported in equation \eqref{eq:tadpole-dim-reg}. In this appendix we show an alternative approach based on the same dimensional regularization technique used above for the current two-point function, and verify that we obtain the same result.

When \(\mathrm{Im}(\nu) < 0\) the coincident point limit fo the scalar propagator can be computed through its spectral representation. This gives
\begin{equation}
    \mathcal{T}_\nu = \lim_{X_4 \rightarrow X_3} \Pi_{\frac{d}{2} + i \nu}^{(0)}(X_4, X_3) = \int_{-\infty}^{\infty} d\nu' \frac{1}{\nu'^2 - \nu^2} \Omega^{AdS}_{\nu'}(x,x)~,
    \label{eq:tadpole-definition}
\end{equation}
where we can substitute the coincident-points limit of the harmonic function
\begin{equation}
    \Omega^{(0)}_{\nu}(x,x) = \frac{\Gamma\left(\frac{d}{2}\right)}{4\pi^{\frac{d}{2} + 1}\Gamma(d)}\frac{\Gamma\left(\frac{d}{2}+ i\nu\right)\Gamma\left(\frac{d}{2}- i\nu\right)}{\Gamma( i\nu)\Gamma(- i\nu)}~.
\end{equation}
Closing the contour with an arc at infinity the result can be written as a sum of the residues at the poles at \(\nu' = \nu\) and \(\nu' = i\left(\frac{d}{2} + n\right)_{n\in\N}\).\footnote{The condition \(d < 1\) is needed to drop the arc at infinity. This can be checked quickly by looking at the asymptotic behavior of the integrand: using \(\Gamma(a \pm i b) \underset{b \rightarrow \infty}{\sim} 2\pi e^{-\pi b} b^{2a - 1}\) valid for \(\vert \mathrm{Arg}(b)\vert \le \pi - \epsilon\), one gets that at infinity in $\nu'$ complex plane the integrand behaves like $\nu'^{d - 2}$.} 
When \(d < 1\) the integral is convergent and the sum over poles can be performed analytically, giving 
\begin{equation}
    \label{eq:tadpole-pole-resummation}
    \begin{aligned}
       & \mathcal{T}_\nu = \frac{\Gamma \left(\frac{d}{2}\right)}{4 \pi ^{\frac{d}{2}+1} \Gamma (d)} 
        \left[-i \sinh (\pi  \nu ) \Gamma \left(\frac{d}{2}-i \nu \right) \Gamma \left(\frac{d}{2}+i \nu \right) \right.  \\
      & + \frac{2 i \pi}{\nu \left(d^2+4 \nu ^2\right)} 
        \frac{\Gamma(d)}{\Gamma \left(-\frac{d}{2}\right) \Gamma \left(\frac{d}{2}\right)}
        \left((d+2 i \nu ) \, _3F_2\left(\frac{d}{2}+1,d,\frac{d}{2}-i \nu ;\frac{d}{2},\frac{d}{2}-i \nu +1;1\right) \right.\\
        & -\left. \left.(d-2 i \nu ) \, _3F_2\left(\frac{d}{2}+1,d,\frac{d}{2}+i \nu ;\frac{d}{2},\frac{d}{2}+i \nu +1;1\right)\right)\right].
    \end{aligned}
\end{equation}
For \(\mathrm{Im}(\nu) > 0\) formula \eqref{eq:tadpole-pole-resummation} still holds, since
\begin{equation}
    \begin{aligned}
        \Pi^{(0)}_{\frac{d}{2} -  i \nu}(X_4, X_3) &= \Pi^{(0)}_{\frac{d}{2} + i \nu}(X_4, X_3) + \frac{2 i\pi}{\nu} \Omega^{(0)}_{\nu}(x_4, x_3), \\
    \implies \lim_{X_4\rightarrow X_3}\Pi^{(0)}_{\frac{d}{2} -  i \nu}(X_4, X_3) 
    &= \mathcal{T}_\nu + \frac{\Gamma \left(\frac{d}{2}\right)}{4 \pi ^{\frac{d}{2}+1} \Gamma (d)} \left[2i \sinh (\pi  \nu ) \Gamma \left(\frac{d}{2} \pm i \nu \right)\right] \\
    &= \mathcal{T}(-\nu) .
    \end{aligned}
\end{equation}
The formula \eqref{eq:tadpole-pole-resummation} is of the same form as the functions in \eqref{eq:resummed-simple-bubble}, so the same regularization algorithm based on the recursion relation \eqref{eq:dim-reg-hypergeometric-recursion-relation} can be applied, and we analytically continue \(\mathcal{T}_\nu\)  to \(d \ge 1\). 
As for the current two-point function, this procedure gives an explicit expression in the form of a simple pole at $d=3$, and whose residue is a linear combination of hypergeometric functions. The resulting expression is also too lengthy to simplify it analytically, so we again compute it numerically. Doing so, we can determine the form of the dimensionally-regularized tadpole to be
\begin{equation}
    \label{eq:tadpole-expansion}
    \mathcal{T}_\nu \underset{d\to 3}{\sim} \frac{c_T(\nu)}{3 - d}  + \text{ finite part} + \mathcal{O}(d-3)~,
\end{equation}
with
\begin{equation}
    \label{eq:tadpole-analytic-residue}
    c_T(\nu) = \frac{\cosh(\pi\nu)}{8\pi^3}\Gamma\left(\frac{3}{2} \pm i\nu\right) = \frac{\nu^2 + \frac{1}{4}}{8\pi^2}~.
\end{equation}
This coincides with the expression for the residue derived in \eqref{eq:tadpole-dim-reg}. The numerical evaluations of the finite part in \eqref{eq:tadpole-expansion} are also consistent with the expression of \(\overline{\mathcal{T}}_\nu\) in  \eqref{eq:tadpole-dim-reg}. This provides an additional check of the regularization procedure developed in \cite{cacciatori2024loops}.

%% file: bibliography.bib
@article{Lake:2018dqm,
    author = "Lake, Ethan",
    title = "{Higher-form symmetries and spontaneous symmetry breaking}",
    eprint = "1802.07747",
    archivePrefix = "arXiv",
    primaryClass = "hep-th",
    month = "2",
    year = "2018"
}

@article{Chakraborty:2025izq,
    author = "Chakraborty, Tuneer and H, Ashik and Raju, Suvrat",
    title = "{Cosmological correlators in gravitationally-constrained de Sitter states}",
    eprint = "2507.15926",
    archivePrefix = "arXiv",
    primaryClass = "hep-th",
    month = "7",
    year = "2025"
}

@article{Herzog:2017xha,
    author = "Herzog, Christopher P. and Huang, Kuo-Wei",
    title = "{Boundary Conformal Field Theory and a Boundary Central Charge}",
    eprint = "1707.06224",
    archivePrefix = "arXiv",
    primaryClass = "hep-th",
    doi = "10.1007/JHEP10(2017)189",
    journal = "JHEP",
    volume = "10",
    pages = "189",
    year = "2017"
}

@article{Cuomo:2021cnb,
    author = "Cuomo, Gabriel and Mezei, M{\'a}rk and Raviv-Moshe, Avia",
    title = "{Boundary conformal field theory at large charge}",
    eprint = "2108.06579",
    archivePrefix = "arXiv",
    primaryClass = "hep-th",
    doi = "10.1007/JHEP10(2021)143",
    journal = "JHEP",
    volume = "10",
    pages = "143",
    year = "2021"
}

@article{Bray:1977fvl,
    author = "Bray, A. J. and Moore, M. A.",
    title = "{Critical behaviour of semi-infinite systems}",
    doi = "10.1088/0305-4470/10/11/021",
    journal = "J. Phys. A",
    volume = "10",
    number = "11",
    pages = "1927",
    year = "1977"
}

@article{Maldacena:2002vr,
    author = "Maldacena, Juan Martin",
    title = "{Non-Gaussian features of primordial fluctuations in single field inflationary models}",
    eprint = "astro-ph/0210603",
    archivePrefix = "arXiv",
    doi = "10.1088/1126-6708/2003/05/013",
    journal = "JHEP",
    volume = "05",
    pages = "013",
    year = "2003"
}

@article{Gabai:2025hwf,
    author = "Gabai, Barak and Gorbenko, Victor and Qiao, Jiaxin",
    title = "{Yang-Mills Flux Tube in AdS}",
    eprint = "2508.08250",
    archivePrefix = "arXiv",
    primaryClass = "hep-th",
    month = "8",
    year = "2025"
}

@article{Aharony:2012jf,
    author = "Aharony, Ofer and Berkooz, Micha and Tong, David and Yankielowicz, Shimon",
    title = "{Confinement in Anti-de Sitter Space}",
    eprint = "1210.5195",
    archivePrefix = "arXiv",
    primaryClass = "hep-th",
    reportNumber = "WIS-14-12-AUG-DPPA",
    doi = "10.1007/JHEP02(2013)076",
    journal = "JHEP",
    volume = "02",
    pages = "076",
    year = "2013"
}

@article{Callan:1989em,
    author = "Callan, Jr., Curtis G. and Wilczek, Frank",
    title = "{INFRARED BEHAVIOR AT NEGATIVE CURVATURE}",
    reportNumber = "IASSNS-HEP-90-4, PUPT-1168",
    doi = "10.1016/0550-3213(90)90451-I",
    journal = "Nucl. Phys. B",
    volume = "340",
    pages = "366--386",
    year = "1990"
}

@article{vanRees:2022zmr,
    author = "van Rees, Balt C. and Zhao, Xiang",
    title = "{Quantum Field Theory in AdS Space instead of Lehmann-Symanzik-Zimmerman Axioms}",
    eprint = "2210.15683",
    archivePrefix = "arXiv",
    primaryClass = "hep-th",
    doi = "10.1103/PhysRevLett.130.191601",
    journal = "Phys. Rev. Lett.",
    volume = "130",
    number = "19",
    pages = "191601",
    year = "2023"
}

@article{Komatsu:2020sag,
    author = "Komatsu, Shota and Paulos, Miguel F. and Van Rees, Balt C. and Zhao, Xiang",
    title = "{Landau diagrams in AdS and S-matrices from conformal correlators}",
    eprint = "2007.13745",
    archivePrefix = "arXiv",
    primaryClass = "hep-th",
    reportNumber = "CPHT-RR119.122020",
    doi = "10.1007/JHEP11(2020)046",
    journal = "JHEP",
    volume = "11",
    pages = "046",
    year = "2020"
}

@article{Paulos:2016fap,
    author = "Paulos, Miguel F. and Penedones, Joao and Toledo, Jonathan and van Rees, Balt C. and Vieira, Pedro",
    title = "{The S-matrix bootstrap. Part I: QFT in AdS}",
    eprint = "1607.06109",
    archivePrefix = "arXiv",
    primaryClass = "hep-th",
    reportNumber = "CERN-TH-2016-162",
    doi = "10.1007/JHEP11(2017)133",
    journal = "JHEP",
    volume = "11",
    pages = "133",
    year = "2017"
}

@article{Sleight:2020obc,
    author = "Sleight, Charlotte and Taronna, Massimo",
    title = "{From AdS to dS exchanges: Spectral representation, Mellin amplitudes, and crossing}",
    eprint = "2007.09993",
    archivePrefix = "arXiv",
    primaryClass = "hep-th",
    doi = "10.1103/PhysRevD.104.L081902",
    journal = "Phys. Rev. D",
    volume = "104",
    number = "8",
    pages = "L081902",
    year = "2021"
}

@article{Sleight:2021plv,
    author = "Sleight, Charlotte and Taronna, Massimo",
    title = "{From dS to AdS and back}",
    eprint = "2109.02725",
    archivePrefix = "arXiv",
    primaryClass = "hep-th",
    doi = "10.1007/JHEP12(2021)074",
    journal = "JHEP",
    volume = "12",
    pages = "074",
    year = "2021"
}

@article{Schaub:2023scu,
    author = "Schaub, Vladimir",
    title = "{Spinors in (Anti-)de Sitter Space}",
    eprint = "2302.08535",
    archivePrefix = "arXiv",
    primaryClass = "hep-th",
    doi = "10.1007/JHEP09(2023)142",
    journal = "JHEP",
    volume = "09",
    pages = "142",
    year = "2023"
}

@article{Sleight:2019hfp,
    author = "Sleight, Charlotte and Taronna, Massimo",
    title = "{Bootstrapping Inflationary Correlators in Mellin Space}",
    eprint = "1907.01143",
    archivePrefix = "arXiv",
    primaryClass = "hep-th",
    reportNumber = "PUPT-2590",
    doi = "10.1007/JHEP02(2020)098",
    journal = "JHEP",
    volume = "02",
    pages = "098",
    year = "2020"
}

@article{MdAbhishek:2025dhx,
    author = "Abhishek, Md. and Sleight, Charlotte and Taronna, Massimo",
    title = "{Cosmological Correlators in Gauge Theory and Gravity from EAdS}",
    eprint = "2509.09536",
    archivePrefix = "arXiv",
    primaryClass = "hep-th",
    month = "9",
    year = "2025"
}

@article{Sleight:2025dmt,
    author = "Sleight, Charlotte and Taronna, Massimo",
    title = "{(Non-)Conserved Currents and Cosmological Correlators}",
    eprint = "2509.18888",
    archivePrefix = "arXiv",
    primaryClass = "hep-th",
    month = "9",
    year = "2025"
}

@article{Freedman:1998tz,
    author = "Freedman, Daniel Z. and Mathur, Samir D. and Matusis, Alec and Rastelli, Leonardo",
    title = "{Correlation functions in the CFT(d) / AdS(d+1) correspondence}",
    eprint = "hep-th/9804058",
    archivePrefix = "arXiv",
    reportNumber = "MIT-CTP-2727",
    doi = "10.1016/S0550-3213(99)00053-X",
    journal = "Nucl. Phys. B",
    volume = "546",
    pages = "96--118",
    year = "1999"
}

@article{Padayasi:2021sik,
    author = "Padayasi, Jaychandran and Krishnan, Abijith and Metlitski, Max A. and Gruzberg, Ilya A. and Meineri, Marco",
    title = "{The extraordinary boundary transition in the 3d O(N) model via conformal bootstrap}",
    eprint = "2111.03071",
    archivePrefix = "arXiv",
    primaryClass = "cond-mat.stat-mech",
    doi = "10.21468/SciPostPhys.12.6.190",
    journal = "SciPost Phys.",
    volume = "12",
    number = "6",
    pages = "190",
    year = "2022"
}

@article{Sadekov:2023ivd,
    author = "Sadekov, Damir",
    title = "{Effective graviton mass in de Sitter space}",
    eprint = "2311.11053",
    archivePrefix = "arXiv",
    primaryClass = "hep-th",
    doi = "10.1103/PhysRevD.109.085001",
    journal = "Phys. Rev. D",
    volume = "109",
    number = "8",
    pages = "085001",
    year = "2024"
}

@article{ShenMasterThesis,
    author = "Shen, Cong",
    title = "{Photon Propagator in de Sitter Spacetime}",
    journal = "Master Thesis, EPFL",
    pages = "Supervisors: J. Penedones and M. Loparco",
    year = "2023"
}

@article{Marotta:2024sce,
    author = "Marotta, Raffaele and Skenderis, Kostas and Verma, Mritunjay",
    title = "{Flat space spinning massive amplitudes from momentum space CFT}",
    eprint = "2406.06447",
    archivePrefix = "arXiv",
    primaryClass = "hep-th",
    doi = "10.1007/JHEP08(2024)226",
    journal = "JHEP",
    volume = "08",
    pages = "226",
    year = "2024"
}

@article{DHoker:1999bve,
    author = "D'Hoker, Eric and Freedman, Daniel Z. and Mathur, Samir D. and Matusis, Alec and Rastelli, Leonardo",
    title = "{Graviton and gauge boson propagators in AdS(d+1)}",
    eprint = "hep-th/9902042",
    archivePrefix = "arXiv",
    reportNumber = "MIT-CTP-2824, UCLA-99-TEP-1",
    doi = "10.1016/S0550-3213(99)00524-6",
    journal = "Nucl. Phys. B",
    volume = "562",
    pages = "330--352",
    year = "1999"
}

@article{DHoker:1998bqu,
    author = "D'Hoker, Eric and Freedman, Daniel Z.",
    title = "{Gauge boson exchange in AdS(d+1)}",
    eprint = "hep-th/9809179",
    archivePrefix = "arXiv",
    reportNumber = "UCLA-98-TEP-33, MIT-CTP-2781",
    doi = "10.1016/S0550-3213(98)00852-9",
    journal = "Nucl. Phys. B",
    volume = "544",
    pages = "612--632",
    year = "1999"
}

@article{Ciccone:2024guw,
    author = "Ciccone, Riccardo and De Cesare, Fabiana and Di Pietro, Lorenzo and Serone, Marco",
    title = "{Exploring Confinement in Anti-de Sitter Space}",
    eprint = "2407.06268",
    archivePrefix = "arXiv",
    primaryClass = "hep-th",
    month = "7",
    year = "2024"
}

@article{Allen:1985wd,
    author = "Allen, Bruce and Jacobson, Theodore",
    title = "{Vector Two Point Functions in Maximally Symmetric Spaces}",
    reportNumber = "UCSB-TH-4-1985",
    doi = "10.1007/BF01211169",
    journal = "Commun. Math. Phys.",
    volume = "103",
    pages = "669",
    year = "1986"
}

@article{witten1998anti,
  title={Anti de Sitter space and holography},
  author={Witten, Edward},
  journal={arXiv preprint hep-th/9802150},
  year={1998}
}

@article{costa2014spinning,
    author = "Costa, Miguel S. and Gon\c{c}alves, Vasco and Penedones, Jo\~ao",
    title = "{Spinning AdS Propagators}",
    eprint = "1404.5625",
    archivePrefix = "arXiv",
    primaryClass = "hep-th",
    doi = "10.1007/JHEP09(2014)064",
    journal = "JHEP",
    volume = "09",
    pages = "064",
    year = "2014"
}

@inproceedings{Woodard:2004ut,
    author = "Woodard, R. P.",
    title = "{de Sitter breaking in field theory}",
    booktitle = "{Deserfest: A Celebration of the Life and Works of Stanley Deser}",
    eprint = "gr-qc/0408002",
    archivePrefix = "arXiv",
    reportNumber = "UFIFT-QG-04-2",
    pages = "339--351",
    month = "7",
    year = "2004"
}

@article{Glavan:2022nrd,
    author = "Glavan, Dra{\v{z}}en and Prokopec, Tomislav",
    title = "{Even the photon propagator must break de Sitter symmetry}",
    eprint = "2212.13997",
    archivePrefix = "arXiv",
    primaryClass = "hep-th",
    doi = "10.1016/j.physletb.2023.137928",
    journal = "Phys. Lett. B",
    volume = "841",
    pages = "137928",
    year = "2023"
}

@article{Grassi:2024vkb,
    author = "Grassi, Pietro Antonio and Porrati, Massimo",
    title = "{Local operator algebras of charged states in gauge theory and gravity}",
    eprint = "2411.08865",
    archivePrefix = "arXiv",
    primaryClass = "hep-th",
    doi = "10.1007/JHEP03(2025)175",
    journal = "JHEP",
    volume = "03",
    pages = "175",
    year = "2025"
}

@article{Liu:1998ty,
    author = "Liu, Hong and Tseytlin, Arkady A.",
    title = "{On four point functions in the CFT / AdS correspondence}",
    eprint = "hep-th/9807097",
    archivePrefix = "arXiv",
    reportNumber = "IMPERIAL-TP-97-98-060",
    doi = "10.1103/PhysRevD.59.086002",
    journal = "Phys. Rev. D",
    volume = "59",
    pages = "086002",
    year = "1999"
}

@article{Fitzpatrick:2011ia,
    author = "Fitzpatrick, A. Liam and Kaplan, Jared and Penedones, Joao and Raju, Suvrat and van Rees, Balt C.",
    title = "{A Natural Language for AdS/CFT Correlators}",
    eprint = "1107.1499",
    archivePrefix = "arXiv",
    primaryClass = "hep-th",
    reportNumber = "SLAC-PUB-14506, HRI-ST-1107",
    doi = "10.1007/JHEP11(2011)095",
    journal = "JHEP",
    volume = "11",
    pages = "095",
    year = "2011"
}

@article{Fitzpatrick:2011hh,
    author = "Fitzpatrick, A. Liam and Shih, David",
    title = "{Anomalous Dimensions of Non-Chiral Operators from AdS/CFT}",
    eprint = "1104.5013",
    archivePrefix = "arXiv",
    primaryClass = "hep-th",
    doi = "10.1007/JHEP10(2011)113",
    journal = "JHEP",
    volume = "10",
    pages = "113",
    year = "2011"
}

@article{DHoker:1999mqo,
    author = "D'Hoker, Eric and Freedman, Daniel Z. and Rastelli, Leonardo",
    title = "{AdS / CFT four point functions: How to succeed at z integrals without really trying}",
    eprint = "hep-th/9905049",
    archivePrefix = "arXiv",
    reportNumber = "MIT-CTP-2857, UCLA-99-TEP-16",
    doi = "10.1016/S0550-3213(99)00526-X",
    journal = "Nucl. Phys. B",
    volume = "562",
    pages = "395--411",
    year = "1999"
}

@article{Faulkner:2012gt,
    author = "Faulkner, Thomas and Iqbal, Nabil",
    title = "{Friedel oscillations and horizon charge in 1D holographic liquids}",
    eprint = "1207.4208",
    archivePrefix = "arXiv",
    primaryClass = "hep-th",
    reportNumber = "NSF-KITP-12-122",
    doi = "10.1007/JHEP07(2013)060",
    journal = "JHEP",
    volume = "07",
    pages = "060",
    year = "2013"
}

@article{Marolf:2006nd,
    author = "Marolf, Donald and Ross, Simon F.",
    title = "{Boundary Conditions and New Dualities: Vector Fields in AdS/CFT}",
    eprint = "hep-th/0606113",
    archivePrefix = "arXiv",
    reportNumber = "DCPT-06-15",
    doi = "10.1088/1126-6708/2006/11/085",
    journal = "JHEP",
    volume = "11",
    pages = "085",
    year = "2006"
}

@article{DiPietro:2025ozw,
    author = "Di Pietro, Lorenzo and Kousvos, Stefanos R. and Meineri, Marco and Piazza, Alessandro and Serone, Marco and Vichi, Alessandro",
    title = "{A Bootstrap Study of Confinement in AdS}",
    eprint = "2512.00150",
    archivePrefix = "arXiv",
    primaryClass = "hep-th",
    month = "11",
    year = "2025"
}

@article{Ciccone:2025dqx,
    author = "Ciccone, Riccardo and De Cesare, Fabiana and Di Pietro, Lorenzo and Serone, Marco",
    title = "{QCD in AdS}",
    eprint = "2511.04752",
    archivePrefix = "arXiv",
    primaryClass = "hep-th",
    month = "11",
    year = "2025"
}

@article{Li:2021snj,
    author = "Li, Yue-Zhou",
    title = "{Notes on flat-space limit of AdS/CFT}",
    eprint = "2106.04606",
    archivePrefix = "arXiv",
    primaryClass = "hep-th",
    doi = "10.1007/JHEP09(2021)027",
    journal = "JHEP",
    volume = "09",
    pages = "027",
    year = "2021"
}

@article{di2022analyticity,
    author = "Di Pietro, Lorenzo and Gorbenko, Victor and Komatsu, Shota",
    title = "{Analyticity and unitarity for cosmological correlators}",
    eprint = "2108.01695",
    archivePrefix = "arXiv",
    primaryClass = "hep-th",
    reportNumber = "CERN-TH-2021-118",
    doi = "10.1007/JHEP03(2022)023",
    journal = "JHEP",
    volume = "03",
    pages = "023",
    year = "2022"
}

@article{hijano2016witten,
    author = "Hijano, Eliot and Kraus, Per and Perlmutter, Eric and Snively, River",
    title = "{Witten Diagrams Revisited: The AdS Geometry of Conformal Blocks}",
    eprint = "1508.00501",
    archivePrefix = "arXiv",
    primaryClass = "hep-th",
    doi = "10.1007/JHEP01(2016)146",
    journal = "JHEP",
    volume = "01",
    pages = "146",
    year = "2016"
}

@article{Chen:2017yia,
    author = "Chen, Heng-Yu and Kuo, En-Jui and Kyono, Hideki",
    title = "{Anatomy of Geodesic Witten Diagrams}",
    eprint = "1702.08818",
    archivePrefix = "arXiv",
    primaryClass = "hep-th",
    reportNumber = "KUNS-2664",
    doi = "10.1007/JHEP05(2017)070",
    journal = "JHEP",
    volume = "05",
    pages = "070",
    year = "2017"
}

@article{Carmi:2018qzm,
    author = "Carmi, Dean and Di Pietro, Lorenzo and Komatsu, Shota",
    title = "{A Study of Quantum Field Theories in AdS at Finite Coupling}",
    eprint = "1810.04185",
    archivePrefix = "arXiv",
    primaryClass = "hep-th",
    doi = "10.1007/JHEP01(2019)200",
    journal = "JHEP",
    volume = "01",
    pages = "200",
    year = "2019"
}

@article{Loparco:2023rug,
    author = "Loparco, Manuel and Penedones, Joao and Salehi Vaziri, Kamran and Sun, Zimo",
    title = {{The K\"all\'en-Lehmann representation in de Sitter spacetime}},
    eprint = "2306.00090",
    archivePrefix = "arXiv",
    primaryClass = "hep-th",
    doi = "10.1007/JHEP12(2023)159",
    journal = "JHEP",
    volume = "12",
    pages = "159",
    year = "2023"
}

@article{simmons2018spacetime,
    author = "Simmons-Duffin, David and Stanford, Douglas and Witten, Edward",
    title = "{A spacetime derivation of the Lorentzian OPE inversion formula}",
    eprint = "1711.03816",
    archivePrefix = "arXiv",
    primaryClass = "hep-th",
    doi = "10.1007/JHEP07(2018)085",
    journal = "JHEP",
    volume = "07",
    pages = "085",
    year = "2018"
}

@article{karateev2019harmonic,
    author = "Karateev, Denis and Kravchuk, Petr and Simmons-Duffin, David",
    title = "{Harmonic Analysis and Mean Field Theory}",
    eprint = "1809.05111",
    archivePrefix = "arXiv",
    primaryClass = "hep-th",
    reportNumber = "CALT-TH 2018-036",
    doi = "10.1007/JHEP10(2019)217",
    journal = "JHEP",
    volume = "10",
    pages = "217",
    year = "2019"
}

@article{ankur2023scalar,
    author = "Ankur and Carmi, Dean and Di Pietro, Lorenzo",
    title = "{Scalar QED in AdS}",
    eprint = "2306.05551",
    archivePrefix = "arXiv",
    primaryClass = "hep-th",
    doi = "10.1007/JHEP10(2023)089",
    journal = "JHEP",
    volume = "10",
    pages = "089",
    year = "2023"
}

@article{cacciatori2024loops,
    author = "Cacciatori, Sergio L. and Epstein, Henri and Moschella, Ugo",
    title = "{Loops in de Sitter space}",
    eprint = "2403.13145",
    archivePrefix = "arXiv",
    primaryClass = "hep-th",
    doi = "10.1007/JHEP07(2024)182",
    journal = "JHEP",
    volume = "07",
    pages = "182",
    year = "2024"
}

@article{Porrati:2001db,
    author = "Porrati, M.",
    title = "{Higgs phenomenon for 4-D gravity in anti-de Sitter space}",
    eprint = "hep-th/0112166",
    archivePrefix = "arXiv",
    reportNumber = "NYU-TH-01-12-03",
    doi = "10.1088/1126-6708/2002/04/058",
    journal = "JHEP",
    volume = "04",
    pages = "058",
    year = "2002"
}

@article{Rattazzi:2009ux,
    author = "Rattazzi, Riccardo and Redi, Michele",
    title = "{Gauge Boson Mass Generation in AdS4}",
    eprint = "0908.4150",
    archivePrefix = "arXiv",
    primaryClass = "hep-th",
    doi = "10.1088/1126-6708/2009/12/025",
    journal = "JHEP",
    volume = "12",
    pages = "025",
    year = "2009"
}

@article{Copetti:2023sya,
    author = "Copetti, Christian and Di Pietro, Lorenzo and Ji, Ziming and Komatsu, Shota",
    title = "{Taming Mass Gaps with Anti\textendash{}de Sitter Space}",
    eprint = "2312.09277",
    archivePrefix = "arXiv",
    primaryClass = "hep-th",
    doi = "10.1103/PhysRevLett.133.081601",
    journal = "Phys. Rev. Lett.",
    volume = "133",
    number = "8",
    pages = "081601",
    year = "2024"
}

@article{Porrati:2024zvi,
    author = "Porrati, M. and Zaffaroni, A.",
    title = "{A Universal Feature for the Higgs Phenomenon in Anti de Sitter Space}",
    eprint = "2408.16919",
    archivePrefix = "arXiv",
    primaryClass = "hep-th",
    month = "8",
    year = "2024"
}

@article{Bashmakov:2016pcg,
    author = "Bashmakov, Vladimir and Bertolini, Matteo and Di Pietro, Lorenzo and Raj, Himanshu",
    title = "{Scalar Multiplet Recombination at Large N and Holography}",
    eprint = "1603.00387",
    archivePrefix = "arXiv",
    primaryClass = "hep-th",
    doi = "10.1007/JHEP05(2016)183",
    journal = "JHEP",
    volume = "05",
    pages = "183",
    year = "2016"
}

@book{Rychkov:2016iqz,
    author = "Rychkov, Slava",
    title = "{EPFL Lectures on Conformal Field Theory in D\ensuremath{>}= 3 Dimensions}",
    eprint = "1601.05000",
    archivePrefix = "arXiv",
    primaryClass = "hep-th",
    reportNumber = "CERN-TH-2016-012",
    doi = "10.1007/978-3-319-43626-5",
    isbn = "978-3-319-43625-8, 978-3-319-43626-5",
    series = "SpringerBriefs in Physics",
    month = "1",
    year = "2016"
}

@article{Giombi:2013yva,
    author = "Giombi, Simone and Klebanov, Igor R. and Pufu, Silviu S. and Safdi, Benjamin R. and Tarnopolsky, Grigory",
    title = "{AdS Description of Induced Higher-Spin Gauge Theory}",
    eprint = "1306.5242",
    archivePrefix = "arXiv",
    primaryClass = "hep-th",
    reportNumber = "MIT-CTP-4471",
    doi = "10.1007/JHEP10(2013)016",
    journal = "JHEP",
    volume = "10",
    pages = "016",
    year = "2013"
}

@article{Compere:2008us,
    author = "Compere, Geoffrey and Marolf, Donald",
    title = "{Setting the boundary free in AdS/CFT}",
    eprint = "0805.1902",
    archivePrefix = "arXiv",
    primaryClass = "hep-th",
    doi = "10.1088/0264-9381/25/19/195014",
    journal = "Class. Quant. Grav.",
    volume = "25",
    pages = "195014",
    year = "2008"
}

@article{Gaiotto:2014kfa,
    author = "Gaiotto, Davide and Kapustin, Anton and Seiberg, Nathan and Willett, Brian",
    title = "{Generalized Global Symmetries}",
    eprint = "1412.5148",
    archivePrefix = "arXiv",
    primaryClass = "hep-th",
    doi = "10.1007/JHEP02(2015)172",
    journal = "JHEP",
    volume = "02",
    pages = "172",
    year = "2015"
}

@article{Berean-Dutcher:2025ohp,
    author = "Berean-Dutcher, Jonah and Derda, Maria and Parra-Martinez, Julio",
    title = "{Soft Theorems from Higher Symmetries}",
    eprint = "2505.03566",
    archivePrefix = "arXiv",
    primaryClass = "hep-th",
    reportNumber = "CALT-TH 2025-013",
    month = "5",
    year = "2025"
}

@article{Arbalestrier:2025jsg,
    author = "Arbalestrier, Adrien and Argurio, Riccardo and Galati, Giovanni and Paznokas, Elise",
    title = "{4d Maxwell on the Edge: Global Aspects of Boundary Conditions and Duality}",
    eprint = "2510.19551",
    archivePrefix = "arXiv",
    primaryClass = "hep-th",
    month = "10",
    year = "2025"
}

@article{Bonetti:2025dvm,
    author = "Bonetti, Federico and Del Zotto, Michele and Minasian, Ruben",
    title = "{SymTFT for Continuous Symmetries: Non-linear Realizations and Spontaneous Breaking}",
    eprint = "2509.10343",
    archivePrefix = "arXiv",
    primaryClass = "hep-th",
    month = "9",
    year = "2025"
}
